\documentclass[11pt,a4paper,superscriptaddress]{article}
\pdfoutput=1
\usepackage{jcappub}
\usepackage{rotating} 
\usepackage{graphicx,epsfig}
\usepackage{amsmath}
\usepackage {amssymb}
\usepackage{subfigure}
\usepackage{color}
\usepackage{amsfonts}
\usepackage{booktabs}
\usepackage{siunitx}
\usepackage{hypcap} % fix the links
\usepackage{cleveref}
\usepackage{physics}
\usepackage{natbib}
\usepackage{relsize}
\usepackage{a4wide}

\def\be{\begin{equation}}
\def\ee{\end{equation}}
\def\bea{\begin{eqnarray}}
\def\eea{\end{eqnarray}}
\def\be{\begin{equation}}
\def\ee{\end{equation}}

\def\nno{\nonumber}
\def\bse{\begin{subequations}}
\def\ese{\end{subequations}}

\def\sphi{\phi_{\ast}}

\graphicspath{{./figs/}}

\begin{document}

\title{(P)reheating after minimal Plateau Inflation and constraints from CMB}

\author{Debaprasad Maity and}

\author{Pankaj Saha}
 
\affiliation{Department of Physics,\\Indian Institute of Technology Guwahati,\\ Guwahati, Assam, India 781039}
 
% \affiliation[b]{Indian Institute of Technology Guwahati, Guwahati 781039, India}
 
\emailAdd{debu@iitg.ac.in}

\emailAdd{pankaj.saha@iitg.ac.in}

\abstract{In this paper we have investigated the preheating phase for a class of plateau inflationary model considering the four-legs interaction term $(1/2)g^2\phi^2\chi^2$ between the inflaton $(\phi)$ and reheating field $(\chi)$. We specifically focus on the effects of our model parameter $\phi_*$ which controls inflationary dynamics. For $\phi_* < M_p$, the departure of the inflaton potential from the usual power-law behavior $\phi^n$ significantly modifies the microscopic behavior of the preheating dynamics. We analyze and compare the efficiency of production, thermalization behavior and the final equation of states of the system for different values of  $n=2,4,6$ considering two different values of $\phi_*$. Most importantly as we increase $n$, or decrease $\phi_*$, the preheating occurs very efficiently with the final equation of state to be that of the radiation, $w=1/3$. However, for $n=2$, the final equation of state turned out to be $w\simeq 0.2$. In the non-perturbative framework complete decay of inflaton could not be achieved with the four-legs interaction for any model under consideration. Therefore, in order to complete the reheating process, we perform the perturbative analysis for the second stage of the reheating phase. With the appropriate initial condition set by the non-perturbative dynamics, we solved a set of homogeneous Boltzmann equations for both the fields supplemented by the constraints coming from the subsequent entropy conservation. In so doing, we calculated the reheating temperature which was otherwise ill-defined right after the end of preheating. The temperature can be uniquely fixed for a given inflaton decay constant and the CMB temperature. We also compare our results with the conventional reheating constraint analysis and discuss the limits on the inflaton decay constant from the field theory perspective.}

%\pacs{98.80.-k, 95.36.+x, 04.50.Kd}
\maketitle

\newpage
%{\fontfamily{pbk}\selectfont 
\section{Introduction}
Inflation is a period in the early universe when the universe expanded exponentially in a quasi-vacuum like state void of any entropy or particles\cite{Starobinsky:1980te,Kazanas:1980tx,Sato:1980yn, Guth:1980zm, Linde:1981mu, Albrecht:1982wi}. This exponential expansion during the inflation is the reason for the remarkable homogeneity of the cosmic microwave background(CMB). The inflationary mechanism is also responsible for generating the seed for all the large-scale structure in the universe. This period of inflation needs a state of negative pressure that can be easily achieved by a scalar field. Despite significant progress on the model independent analysis from effective field theories\cite{Cheung:2007st,Weinberg:2008hq,Boyanovsky:2015tba,Boyanovsky:2015jen}, inflation is still largely a model dependent phenomena. In principle, any scalar field potential satisfying the well-known \textit{`slow-roll'} conditions can yield inflation. The list of models for inflationary cosmology is thus practically inexhaustible\cite{Martin:2013tda}. Nonetheless, the measurements of CMB anisotropies act as a probe for inflationary models. Even though over the years observations of the PLANCK\cite{Planck:2013jfk, Ade:2015lrj, Ade:2015tva} and Keck Array, and BICEP2 Collaborations\cite{Array:2015xqh}, on the inflationary parameters such as the scalar spectral index as $n_s = 0.968 \pm 0.006$, and the scale dependence of scalar spectral index $\dd n_s/\dd \ln {\rm k} =-0.003 \pm 0.007$, and the bound on the tensor-to-scalar ratio, $ r_{0.05} < 0.07 \text{(95 \% CL)}$ are increasingly pointing towards the inflationary paradigm, it turned out that these results rule out a large class of simple and prominent models. One of such model is the well known, single field chaotic inflation models. The models predicting smaller values of tensor-to-scalar ratio are now favored. One such model from super-gravity which also unifies a large class of inflationary models is known as the $\alpha$-attractor model. The potentials of these models are characterized by an infinite plateau for large field value with the minimum at the origin. The shape of the plateau can be controlled by a parameter called $\alpha$ which can be chosen to reproduce the inflationary observables within the PLANCK limit. In our recent work\cite{Maity:2019ltu} we have proposed a class of inflationary model which belong to a different class as compared to $\alpha$-attractor. The form of the potential resembles that of the power-law chaotic models with a non-minimal modification. The inflationary predictions and dynamics of those models have been discussed in details in the work. The reheating constraints on the model through CMB and dark matter abundance has also been studied\cite{Maity:2018exj} considering the perturbative reheating phenomena. The study of perturbative reheating with a phenomenological term with inflaton decay constant $\Gamma_{\phi}$ is simple and convenient to understand several important aspects of the reheating phase especially the connection between CMB and the present dark matter abundance through the reheating temperature. However, this treatment relies on several assumptions on the nature of the reheating processes: (i) The first assumption is the absence of non-perturbative phenomena such as parametric resonance during reheating phase. Although the parametric resonance can be suppressed by choosing small enough value of the coupling parameter. However, it would be more appropriate to consider the perturbative reheating as a final stage of the whole reheating process with the initial conditions set by preheating stage. (ii) The second assumption is that we have also ignored the phenomena of inflaton fragmentation and the growth of inhomogeneity in the inflaton sector. With this the inflaton equation of state for the whole period of reheating has been described by that of the homogeneous inflaton condensate. For potentials $V(\phi) \sim \phi^n$ near the  minimum, using the virial theorem, the average equation of state are found to be $w_{\phi} = (n-2)/(n+2)$. 
It is needless to say that in any realistic reheating scenario the above assumptions have limited applicability. In this paper we have studied the non-linear effects of preheating and thermalization and then look into by perturbative reheating for the class of minimal plateau models we proposed recently. 
%It will be our focus to study the effects of the controlling scale $\sphi$ on the preheating and thermalization process.

The effects of parametric resonance is well studied in classical mechanics\cite{Landau:1982}. The importance of resonant particle production at the initial stage of reheating has been recognized and studied extensively for the first time in the seminal works of Kofman, Linde, and Starobinsky\cite{Kofman:1994rk,Kofman:1997yn}. The structure of resonance in the class of conformally invariant theories such as $(\lambda/4)\phi^4 + (g^2/2)\phi^2\chi^2$ has been developed in \cite{Greene:1997fu}. It has been understood that the analytic study of preheating does not capture the full non-linear dynamics of this phase. A full non-linear study of non-perturbative field theory is needed for. However, if the occupation number of the different species is much larger than one we may study the system by ignoring their quantum nature\cite{Khlebnikov:1996mc,Prokopec:1996rr,Micha:2002ey,Micha:2004bv} and solving the appropriate classical wave equations. This fact serves as the basis for studying the preheating phase in $3 + 1$ dimensional lattice. With the advent of different lattice codes\cite{Felder:2000hq,Frolov:2008hy,Easther:2010qz,Huang:2011gf}, lattice simulation study of preheating becomes quite frequent in the literature (see refs.(\cite{Felder:2007ef,Allahverdi:2010xz,Amin:2014eta,Figueroa:2016wxr}) and references therein). Another import aspect of preheating is the study of self resonance after inflation. In this case the inflaton quanta may became unstable due to small spatial perturbations even without the couplings to other fields. The self-resonance is found to be inefficient for the case of chaotic models\cite{Greene:1997fu}, however they can be efficient for multifield inflation\cite{Hertzberg:2014iza,Hertzberg:2014jza} or in the case of plateau type potentials\cite{Lozanov:2016hid,Lozanov:2017hjm}.

In this work we have studied the preheating and subsequent thermalization after minimal plateau inflation introduced in our earlier work\cite{Maity:2019ltu}. These class of models are parameterized by the power $n$ of the inflaton field $\phi$ and most importantly a scale $\sphi$ that controls the energy scale of inflation, similar to the $\alpha$ parameter for $\alpha$-attractor models. The study of self-resonance phenomena with $\alpha$-attractor type potentials has been performed in \cite{Lozanov:2016hid,Lozanov:2017hjm}. However, in any realistic model of preheating the inflaton coupled with other matter fields must be incorporated. The study of preheating for the archetypal plateau inflation model viz the Starobinsky-Higgs inflation has been done in\cite{Bezrukov:2008ut,GarciaBellido:2008ab}. However, the study of preheating when other matter fields are coupled has not been done in the literature for general plateau type of potentials with a controlling parameter such as $\sphi$. Therefore, in this paper we will consider the class of minimal plateau inflation models mentioned before. We will study, in detail, how the particle production and the subsequent thermalization process depend on the scale $\sphi$ and the power of the potential $n$. The end of preheating can be identified around the scale factor where average value of all the energy components tend to become stationary. Interestingly for $n>2$, all the models lead to effective equation of state equal to that of the radiation $w=1/3$.

It is well known that the inflaton decay is not complete with a four-legs interaction\cite{Podolsky:2005bw}. It has been found here that this situation prevails even if we change the parameter $\sphi$ for all the models considered. Therefore, as a logical next step we study the perturbative reheating process considering a phenomenological inflaton decay term into the Boltzmann equations for the inflaton and the radiation component. The phase of perturbative reheating enables us to connect the reheating phase with the current CMB date in terms of the primordial spectral index of the inflaton fluctuation. For a particular inflation model, the preheating dynamics turned out to be insensitive to the inflationary e-folding number which is a function of scalar spectral index. As a consequence in determining the reheating temperature the effect of preheating appears only though its e-folding number. Finally we have commented on the range of value of the coupling parameter that will set the value of the above perturbative inflaton decay term. Although, this work concerns with the above plateau potentials, the general conclusions in this work will be applicable to any other class of plateau potential having a controlling scale similar to $\sphi$ or $\alpha$.  

We have structured this paper as follows. After briefly describing the minimal plateau inflationary models in \Cref{sec:minimal}, we have described the requisite equations and introduced various quantities of interest in \Cref{sec:resonance}. In \Cref{sec:pre_analytic} we have studied the analytic behavior of parametric resonance with the help of instability chart associated with the Mathieu/Hill type differential equations. \Cref{sec:results} describes the results of the lattice simulation. The perturbative reheating and CMB constrains on reheating phase has been described in \Cref{sec:pert}. Finally we conclude in \Cref{sec:conclusion}.

We will consider $\hbar = c = 1$ unless otherwise stated. We have denoted $m_p( = 1/\sqrt{G})$ as the Planck constant and $M_p( = 1/\sqrt{8\pi G})$ as the reduced Planck constant. We will take the usual Friedmann-Le\^{i}matre-Roberson-Walker (FLRW) metric as our background metric $\dd s^2= \dd t^2 -a^2(t)(\dd x^2+\dd y^2+\dd z^2) $ for deriving our equations. With $a(t)$ is the scale factor and $t$ representing the cosmic time.

\section{{\label{sec:minimal}}A Brief Introduction to the Minimal Inflation Model}
\begin{figure}
 \includegraphics[scale=1]{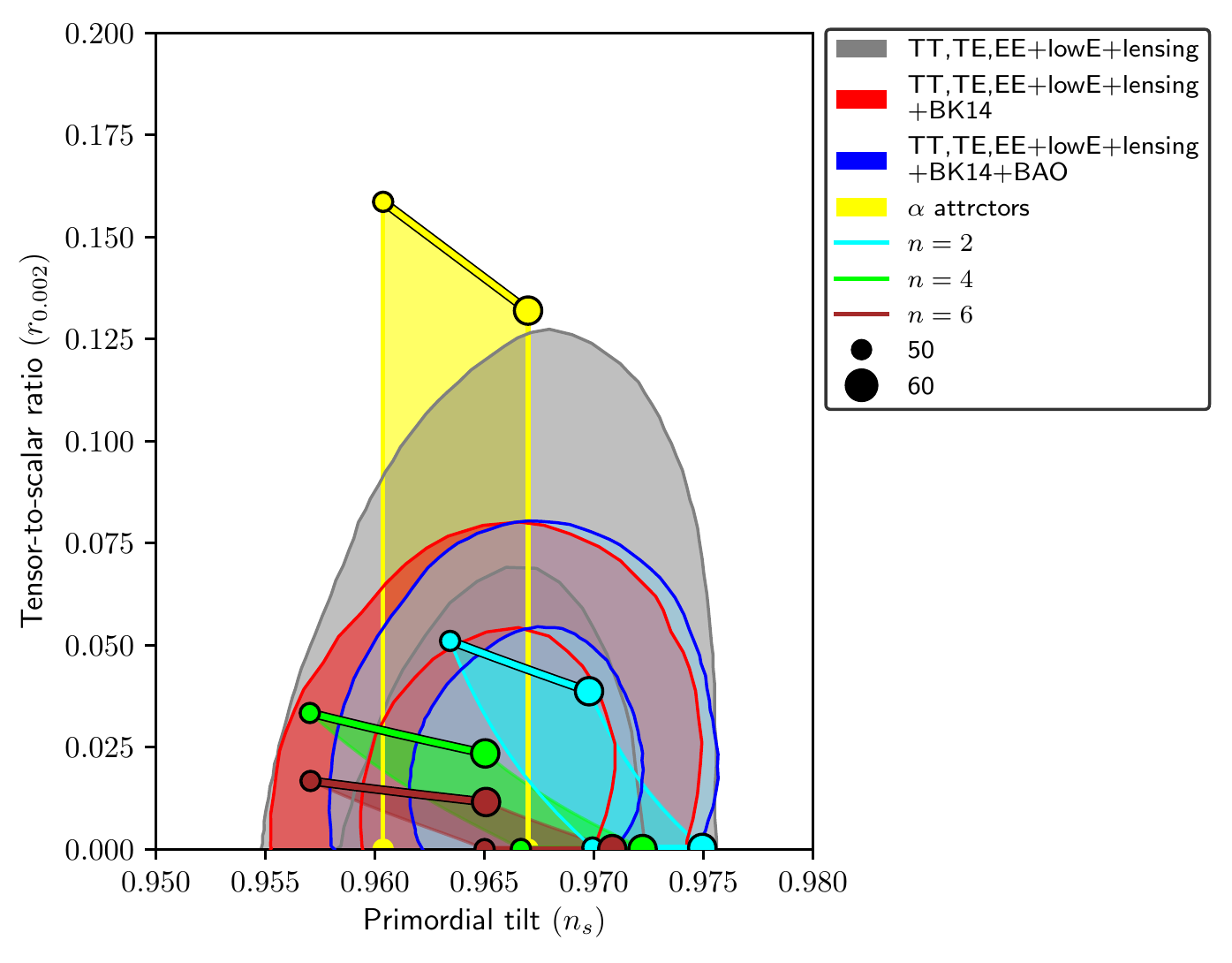}
 \caption{\scriptsize The  $n_s$ and $r$ plot of the model on the marginalized joint $68\%$  and $95\%$  CL regions for $n_s$ and $r$ at $k = 0.002 Mpc^{−1}$ from Planck alone and in combination with BK14 or BK14 plus BAO data. Dotted line corresponds to $\phi_* = 10M_p$ and the solid line is for $\phi_* = 0.1M_p$}
 \label{fig:planckplot}
\end{figure}

In this section we will briefly describe the minimal plateau inflationary model and its characteristics introduced in\cite{Maity:2019ltu}. We have already explained in our previous work, considering a simple power-law potential $\phi^n$, we can obtain our general class of non-polynomial form of the potential given as 
\begin{equation}
V(\phi) = \frac{1}{n}\frac{\lambda ~m^{4-n} \phi^n}{1 + \left(\frac{\phi}{\phi_*}\right)^n}
\label{eq:modelpot}
\end{equation}
either by using a conformal transformation in a certain class of non-minimal scalar-tensor theory, or from a supergravity construction. The scale $\sphi$ can be identified with the non-minimal coupling in scalar-tensor theory, and inflationary energy scale in the supergravity potential. We take $\lambda=1$ for $n \neq 4$ and the values of $\lambda$ or, $m$ is fixed from WMAP normalization. The inflationary predictions of this model has been studied extensively in the original work, here we present the $n_s$ and $r$ plot in fig.(\ref{fig:planckplot}) on the latest Planck data\cite{Akrami:2018odb}. After introducing the scalar potential, next we will introduce the parametric resonance after the end of inflation.

\section{{\label{sec:resonance}}Preheating: Parametric resonance}
\subsection{The Model and equations}
In this subsection we will first describe the main equations and methodology for our subsequent discussions. 
The exponential growth of a dynamical field coupled with an oscillating classical background for certain ranges of parameters is known as the parametric resonance\cite{Landau:1982}. In their seminal work, Kofman, Linde and Starobinsky\cite{Kofman:1994rk,Kofman:1997yn}(see also\cite{Shtanov:1994ce}), introduced and work out the idea of non-perturbative resonance production of particles after inflation based on the idea of parametric resonance.
The fields under consideration could be the fluctuations of the inflaton or any other daughter field or both, which will experience the homogeneous oscillating background inflaton for the present case. The daughter fields coupled with the inflaton could be any other scalar fields or usual standard model particles(usually bosons \footnote{the presence of bosons are important for parametric resonance, for preheating with fermions see\cite{Greene:1998nh,Greene:2000ew}.  we will here only consider the bosonic case}). The energy density of the universe just after the inflation is in the form of the homogeneous inflaton field. This energy starts decaying into fluctuations of the inflaton and other fields at the onset of preheating. The initial stage of preheating is marked by exponential growth of decay product due to resonance effects. The analysis of this initial stage could be treated analytically. However, the expeditious draining of energy due to exponential particle production affects the dynamics of the inflaton field. This back-reaction of the produced particle leads to a fragmentation of the homogeneous inflaton and eventually shuts down the particle production. The rich dynamical nature of the combined inflaton and the produced fields call for a numerical simulation of the system on a lattice. The decay of inflaton by studying the classical field equation can be faithfully done as long as the occupation number of the fields are much larger that unity\cite{Khlebnikov:1996mc}. In the alter part of this work we will do the simulation with the heavy-duty \texttt{LATTICEEASY}\cite{Felder:2000hq} and its parallelized version \texttt{CLUSTEREASY}\cite{Felder:2007nz}. 
The inflaton potential for the model is given in \eqref{eq:modelpot}. Throughout this work we will work exclusively with the four-legged interaction given as
\be
\mathcal{L}_{int} = - \frac{1}{2}g^2 \phi^2 \chi^2
\ee    
It has been noted that this four-legged interaction term will be dominant over the three-legged interaction viz $g^2 \sigma \phi \chi^2$ for the initial stages of preheating when the amplitude of the homogeneous inflaton oscillation is large\cite{Podolsky:2005bw}. This interaction does not lead to any tree level decay of the inflaton and particle production will solely due to non-perturbatives processes. The four-legged interaction is usually not able to complete the decay of inflaton for general chaotic inflationary scenario\cite{Podolsky:2005bw}. Therefore, after the preheating three-legged interaction will be dominating in the perturbative decay process and may complete the reheating dynamics. In this work we will exclusively consider the aforementioned interaction for the  preheating stage. At the end we will discuss about the perturbative reheating and connection with CMB. Nevertheless, the full potential for our lattice simulation is 
\begin{equation}
V(\phi, \chi) = \frac{1}{n} \frac{\lambda~m^{4-n}~\phi^n}{\left[1 + \left(\frac{\phi}{\phi_*}\right)^n\right]} + \frac{1}{2}g^2 \phi^2 \chi^2 .
\end{equation}
With this potential \texttt{LATTICEEASY} will solve the following classical scalar field equations
\begin{align}\label{eq:eoms0}
	\ddot{\phi} + 3 H \dot{\phi} - \frac{1}{a^2} \nabla^2 \phi + \frac{\partial }{\partial \phi}V(\phi, \chi) = 0,\\
	\label{eq:eqchik}
	\ddot{\chi} + 3 H \dot{\chi} - \frac{1}{a^2} \nabla^2 \chi + \frac{\partial }{\partial \chi}V(\phi, \chi) = 0.
\end{align}
While the Hubble parameter $H$ is calculated self-consistently from the Friedmann equations. The redundancy of the Friedmann equations is used in \texttt{LATTICEEASY} to monitor the energy conservation. In all of our results we have kept the energy conservation at the $\mathcal{O}(10^{-4}-10^{-2})$ level. Denoting the preheating fields as by a generic symbol $f(t,\vec{x})$ and its Fourier transform as $f_k(t)$, the (comoving) occupation number of the particles are with this generic symbol $f$ are given by\cite{Felder:2000hq, Felder:2000hr}
\begin{align}
	\label{eq:nk}
	n_k(t) \equiv \frac{1}{\omega_k}|\dot{f}_k|^2 + \frac{\omega_k}{2}|f_k|^2, ~~ 
	\text{with,}\quad \omega_k \equiv \sqrt{k^2 + m^2_{eff}},~~
	\text{and,}\quad m^2_{eff} \equiv 
	\frac{\partial^2 V}{\partial f^2}
\end{align}
The evolution of various energy components such as kinetic, gradient and interaction part contain important information about the thermalization process and the growth of inhomogeneities.
We will study in detail the evolution of those individual components defined below
\begin{equation}
\rho \equiv E_t = (E_{\phi}^K + E_{\chi}^K + E_{\phi}^G + E_{\chi}^G + E_{\phi}^P + E_{\phi\leftrightarrow\chi}^I)
\label{eq:total_energy}
\end{equation}
Where,
\begin{align}
	E_{\phi}^K &= \frac{1}{2}\dot{\phi}^2;\quad\quad\quad~~ E_{\chi}^K = \frac{1}{2}\dot{\chi}^2;\\
	E_{\phi}^G &= \frac{1}{2a^2}(\nabla \phi)^2, \quad E_{\chi}^G = \frac{1}{2a^2}(\nabla \phi)^2;\\
	E_{\phi}^P &= V(\phi), \quad\quad E_{\phi\leftrightarrow\chi}^I = \frac{1}{2}g^2\phi^2\chi^2 .
\end{align}
Where, the subscript $(K,G,P,I)$ stand for kinetic, gradient, potential and interaction component of the total energy respectively. However, for evolution of the scale factor it is the total energy density $\rho$ that plays the important role. The energy total energy density may be expressed as \begin{equation}
\rho \approx \frac{1}{(2\pi)^3a^4}\int d^3k \omega_k n_k,
\end{equation}
and the total (comoving) number density of the $f$ field is expressed as
\begin{equation}
\label{eq:ntotal}
n_f(t) \equiv \frac{1}{2\pi^2} \int d^3 k~ n_k(t).
\end{equation}
Before describing the numerical results, we will analyze the resonance structure analytically by studying the stability-instability diagram of the above equation with the well known method of Floquet theory\cite{McLachlan:1947,Magnus:1966}.     
%\clearpage

\subsection{{\label{sec:pre_analytic}} Parametric resonance: Instability chart}
The instability of a mode is intimately tied with the violation of adiabatic condition that controls the particle production measured in terms of its growth over time. However this growth of a mode depends upon the coupling parameter and the value of the momentum. The exponential growth of a mode function depending upon the parameter values is termed as parametric resonance. Ignoring the gradient term during the initial linear regime, we can study this parametric resonance with the help of Mathieu/Hill equation. To this end, we first do the following re-scaling $X_k(t) \equiv a^{3/2}(t)\chi_k(t)$ to the $\chi$ field Eq.(\ref{eq:eqchik}) as
\be
\ddot{X}_k + \omega_k^2 X_k = 0,
\ee
where
\bea
\label{Hilleq0}
\omega_k^2 \equiv \frac{k^2}{a^2} + g^2 \phi^2(t) +  \Delta,~~~~~ \Delta \equiv -\frac{3}{4}(3H^2 + 2H).
\eea
During preheating, we will set $\Delta=0$. To study the resonance phenomenon in the context of Floquet theory, we will first ignore the expansion of the universe. In that case eq.(\ref{Hilleq0}) can be identified as a form of Hill's differential equation\cite{Magnus:1966}
\begin{figure}[t]
	\centering
	\subfigure[$n=2$]{\includegraphics[scale=0.3]{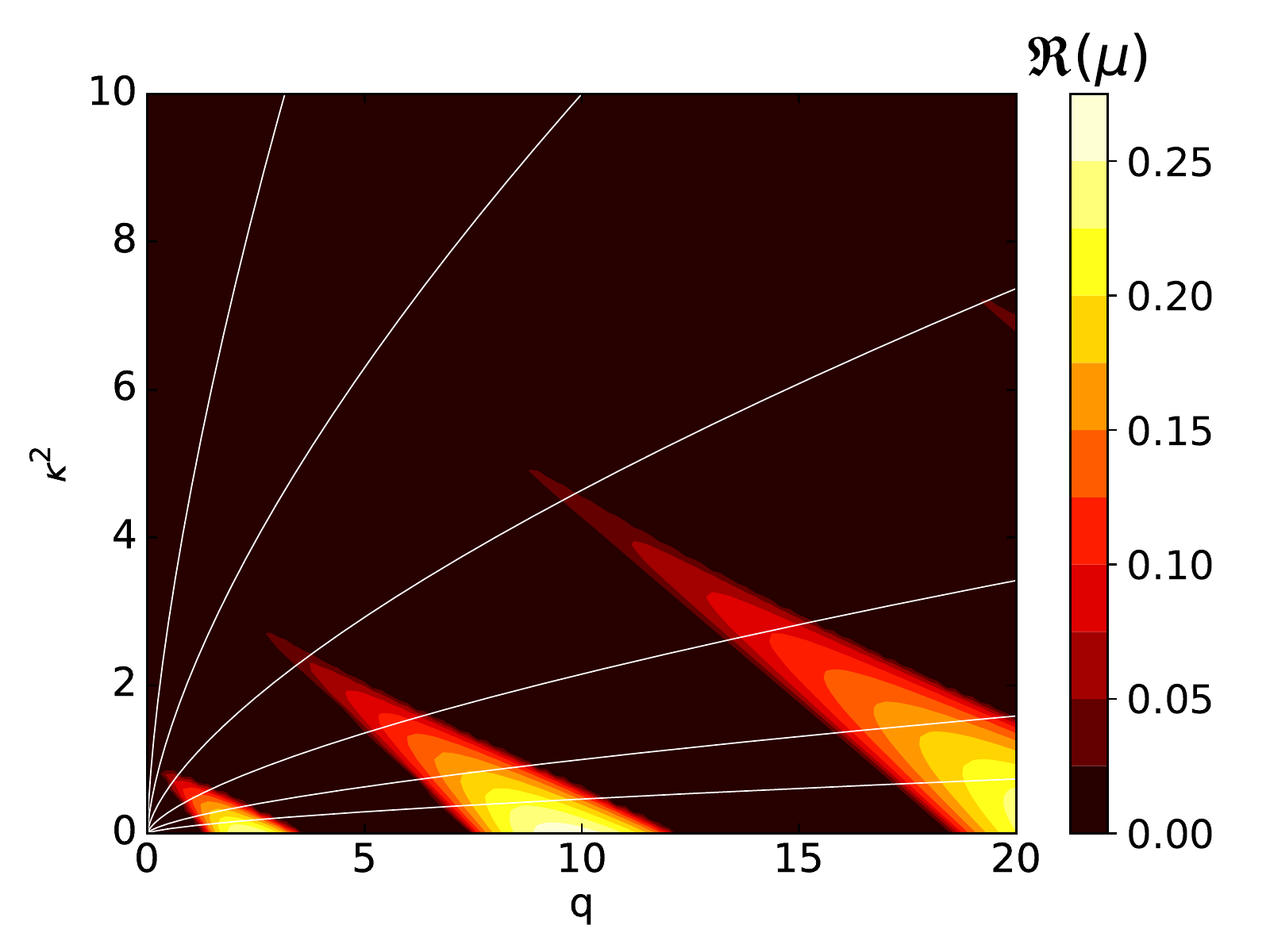}}
	\subfigure[$n=4$]{\includegraphics[scale=0.3]{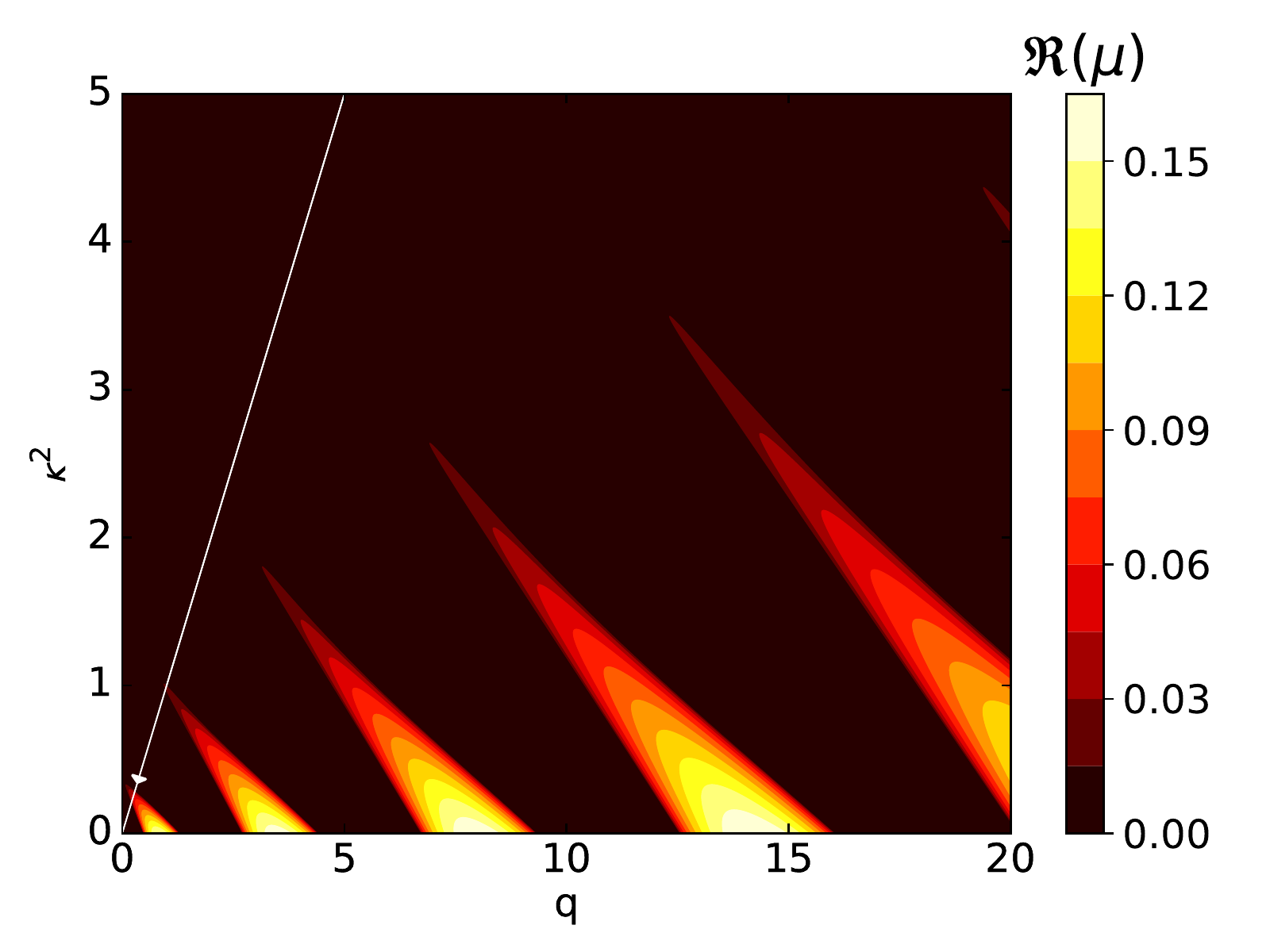}}
	\subfigure[$n=6$]{\includegraphics[scale=0.3]{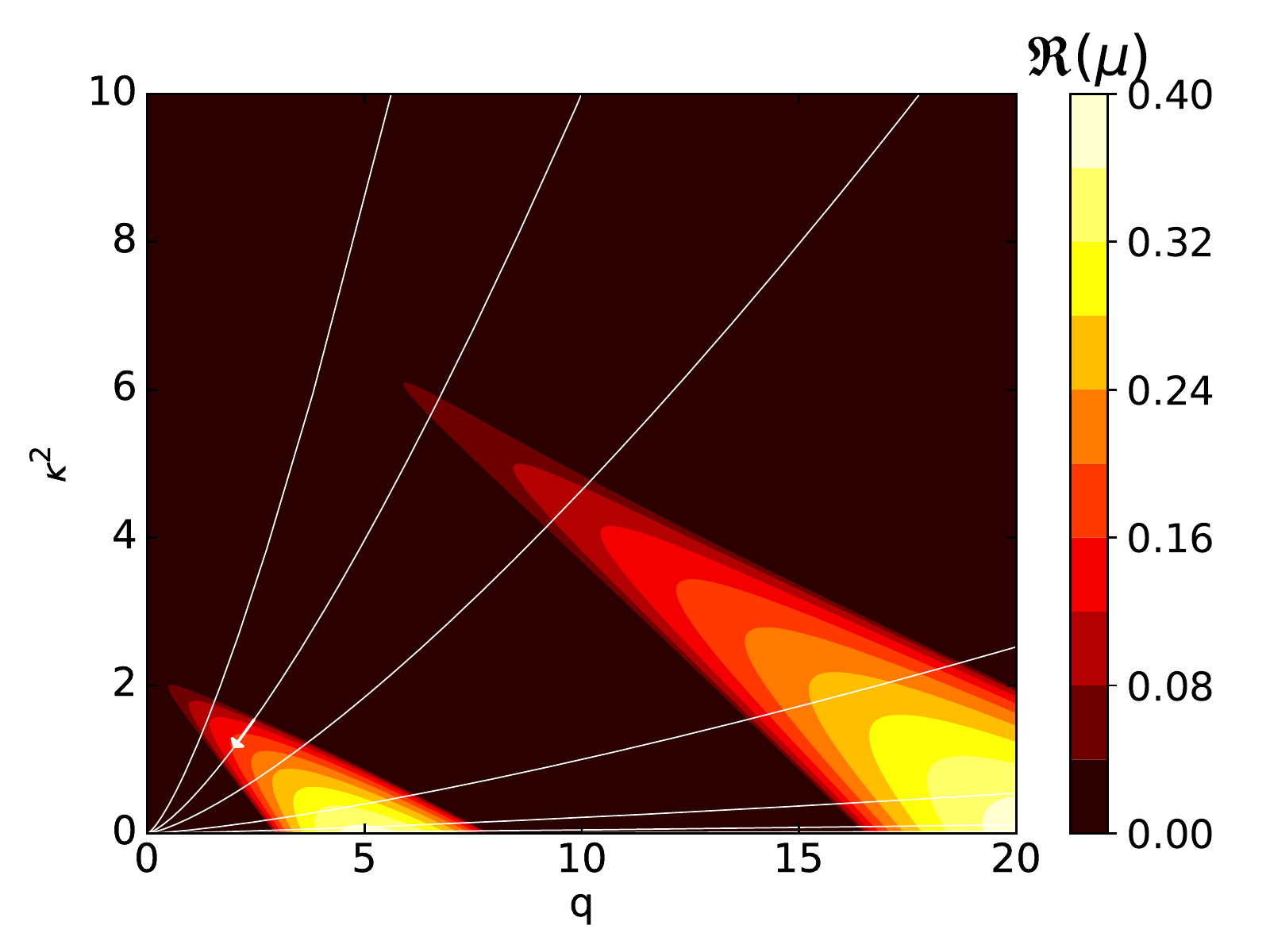}}
	\caption{\scriptsize Instability regions for the models}
	\label{fig:stability_chart}
\end{figure}

\be
X_k'' + \left(\kappa + q \varphi^2(t)\right)X_k = 0,
\label{Hilleq1}
\ee
with $a=1$ for no expansion, the coefficients $\kappa$ and $q$ given by $\kappa = k^2/a^2$ and $q = (g^2 \Phi^2)/\mathcal{B}^2$ are time-independent. The time will be measured in unit of $\mathcal{B}$ with $\mathcal{B}=\phi_0^{n/2-1}\sqrt{\lambda m^{4-n}}$(~$\phi_0$ being the initial inflaton amplitude) set the natural time scale of the systems. Writing the homogeneous oscillatory background solution as $\phi(t) = \Phi \varphi(t)$. $\Phi$ is the amplitude of oscillation. In reality, the amplitude is also decaying with time i.e., $\Phi \equiv \Phi(t)$. However, without expansion, for the present discussion we can take it to be constant. We will introduce the time dependence quantitatively later. The solution of the eq.(\ref{Hilleq1}) is of the form $X_k \propto exp(\mu_k t)$. Where $\mu_k$ is known as the Floquet exponent in the field of differential equations that set the nature of the solution. If $\mathfrak{R}(\mu_{\kappa,q}) > 0$ for certain values of the parameters $(\kappa, q)$, the solution shows exponential growth which is identified as the particle production. The contours of $\mathfrak{R}(\mu_{\kappa,q}) = 0$ in the $(\kappa, q)$ plane, known as the instability bands, help us to understand qualitatively the region of parameter space as well as the strength of the resonance during preheating. In figs.(\ref{fig:stability_chart}), we present the stability-instability chart for three different values of $n~=(2,4,6)$ for $\sphi=10M_p$. The shaded regions are the instability regions while the color-code value shows the strength of resonance. The effect of decreasing $\sphi$ for a particular $n$(not shown in the figure) is to shift the instability regions towards higher values of $q$. This is expected as decreasing $\sphi$ will result in reducing the initial amplitude of inflaton field. Hence, we will naturally need a higher value of the coupling constant $g$ to get the resonance. Also for a particular $q$(or, $g$), decreasing $\sphi$ will lead to resonance only for the higher momentum modes. We will describe other important effects of the scale on preheating and thermalization as we go along. Till now, we have ignored the effect of expansion on the resonance phenomena. A striking difference between the parametric resonance in an expanding universe with the normal resonance is that the parameters $(\kappa, q)$ in an expanding universe are not constants. They depend on time via the scale factor and the time-dependent amplitude $\Phi(t)$ of the inflaton oscillation. As a result of the expanding universe all the momentum modes will be red-shifted while the back reaction of produced particles will eventually shut-down the resonance. Nonetheless, we may incorporate the effect of expansion in the instability analysis by noting including the fact that the amplitude of the inflaton oscillation will decay as $\Phi \propto a^{-6/(n+2)}$. Hence, a particular mode residing in an instability band in an expanding universe will not have indefinite growth rather it will travel through different instability and stability regions(when the solution is oscillatory indicating an absence of particle production) with time. The white flow-lines with the arrow direction shows the trajectory of a mode through different bands. 
We will close this section with an estimate of the maximum allowed momenta range that can be produced during broad parametric resonance. This maximum value has important consequence on setting up the lattice parameters as we will discuss later. To this end, we first note that due to large effective mass(most of the time interval during inflaton oscillation we have $m_{\rm eff}^{\chi} \sim g\phi \simeq g\Phi \gg m_{\phi}$), the preheating field mode $\chi_k$ will oscillate much faster than the inflaton.  This implies that the oscillation frequency $\omega_k$ mostly changes adiabatically. As a result the occupation number $n_k$ with variables frequency $\omega_k$ can be treated as an adiabatic invariant.
	\begin{align}
	n_k &\sim \frac{1}{2}\omega_k^2 \frac{\abs{\chi_k^2}}{\omega_k} \simeq {\rm constant.}\\
	\implies \abs{\chi_k} &\sim \omega_k^{-1/2}
	\end{align}
	The violation of adiabatic condition signals particle production i.e.,
	\begin{equation}
	\dot{\omega} \gtrsim \omega^2 \quad \implies {\rm Particle~production}
	\label{eq:adiabatic}
	\end{equation}
	Now, we will approximate $\dot{\phi}\sim \mathcal{B}\phi_0$. Where $\cal B$ sets, as mentioned before, the natural time scale of inflaton oscillation. With $\omega^2 = k^2 + g^2\phi^2$ and $\dot{\omega} \simeq g^2\phi_0\mathcal{B}\phi$, the adiabatic condition in \eqref{eq:adiabatic} provides the range of excited momenta $k$ as
	\begin{equation}
	0 \leq k^2 \lesssim (g^2\phi_0\mathcal{B}\phi)^{2/3} - g^2\phi^2
	\end{equation}
	Momentum will be maximum when
	\begin{equation}
	\phi \equiv \tilde{\phi} \simeq \frac{1}{2}\Bigg(\frac{\mathcal{B}\phi_0}{g}\Bigg)^{1/2}
	\end{equation}
	This estimate the range of typical momenta $k$ of the particles that are produced in the broad resonance regime as
	\begin{equation}
	0 \leq 2k \lesssim k_{\rm max} \equiv (g\mathcal{B}\phi_0)^{1/2}
	\label{eq:kmax}
	\end{equation}
	Where $k_{\rm max}$ gives the maximum momentum scale that can be reached during preheating. The above estimation of $k_{\rm max}$ for particular cases can be found in \cite{Kofman:1997yn,Figueroa:2016wxr} for $n=2$ and in \cite{Greene:1997fu,Figueroa:2016wxr} for $n=4$.
With these results, we will describe the results of our lattice simulations in the next Section.

\section{{\label{sec:results}}Preheating: Lattice Simulation}
In this section we will describe the results of our lattice simulation for three particular class of models from the potential given in \ref{eq:modelpot}. The details of specific model and its parameters as well as various lattice specification will be mentions in their respective places. However, before diving into details of the specific models we will first mention some general features of the study which is almost universal across all the class of models.

\subsection{The back-reaction and the  emergence of non-linearity: Generalities}
The preheating phase, in general, is episodic with at least three distinct phases. The first phase of preheating is marked by the linear growth of produced field(s) via parametric resonance. The comoving amplitude of the inflaton field remains almost constant throughout this phase while most of the energy of the system is contained in the inflaton field oscillating mostly in the form of kinetic and potential energies. While all other energies are sub-dominant as shown in \Cref{fig:energy_n2}. In the left two figures in \ref{fig:energy_n2}, we have plotted the instantaneous values of different energy components averaged over the lattice points. It is clear from these figures that the interaction energy is negligible compared to the other components throughout the evolution except for a small time interval around the end of the first linear phase. This implies that the occupation number is well defined outside this interval. We can also notice that the gradient energies are negligible well after the initiation of the preheating. They don't became appreciable up until the back-reaction effects kicks in. The characteristic scale marking the end of this phase will be denoted, following the terminology used in\cite{Figueroa:2016wxr}, by the instant $z_{br}$ where \textit{`br'} stands for back-reaction. In practice, we will measure $z_{br}$ as an instant when the comoving inflaton amplitude drops to $95\%$ of its initial value. The second phase is the non-linear regime when the back-reaction of the produced field quanta became appreciable on the dynamics of the inflaton field. As a result, the comoving amplitude of the field starts decreasing appreciably. Inflaton kinetic and potential energies start to decrease resulting in the increase of other energy components. The gradient energy of the inflaton field which represents the growth of the inhomogeneities will also start to increase. This phase ends when the inflaton decay stops. We will denote the end of this phase as $z_{dec}$. Where \textit{`dec'} represents the instant when the inflaton decay ends\cite{Figueroa:2016wxr} and it enters into a stationary regime.

Other interesting aspect of preheating is observed by observing the virialization of the system. The inflaton with a potential $V(\phi) \propto \phi^n$ and the produced field satisfy the following virialization relation.
\begin{align}
 \frac{1}{2}\left\langle \dot{\phi}^2 \right\rangle &= \frac{1}{2}\left\langle \left(\frac{\nabla\phi}{a}\right)^2 \right\rangle + \frac{n}{2} \left\langle V(\phi) \right\rangle + \frac{1}{2} \left\langle g^2\phi^2\chi^2 \right\rangle\\
 \frac{1}{2}\left\langle \dot{\chi}^2 \right\rangle &= \frac{1}{2}\left\langle \left(\frac{\nabla\chi}{a}\right)^2 \right\rangle + \frac{1}{2} \left\langle g^2\phi^2\chi^2 \right\rangle
 \label{eqn:virial}
\end{align}
Where the average is over space and time. We will discuss the virialization of each system separately below. Through out our study we will compare models under consideration for mainly two different values of $\sphi$: one super-Planckian value when the models predictions are similar to their respective power-law chaotic models and one sub-Planckian value when the effects of the scale $\sphi$ is important. 

\begin{figure}[t]
	\centering
	\includegraphics[scale=0.35]{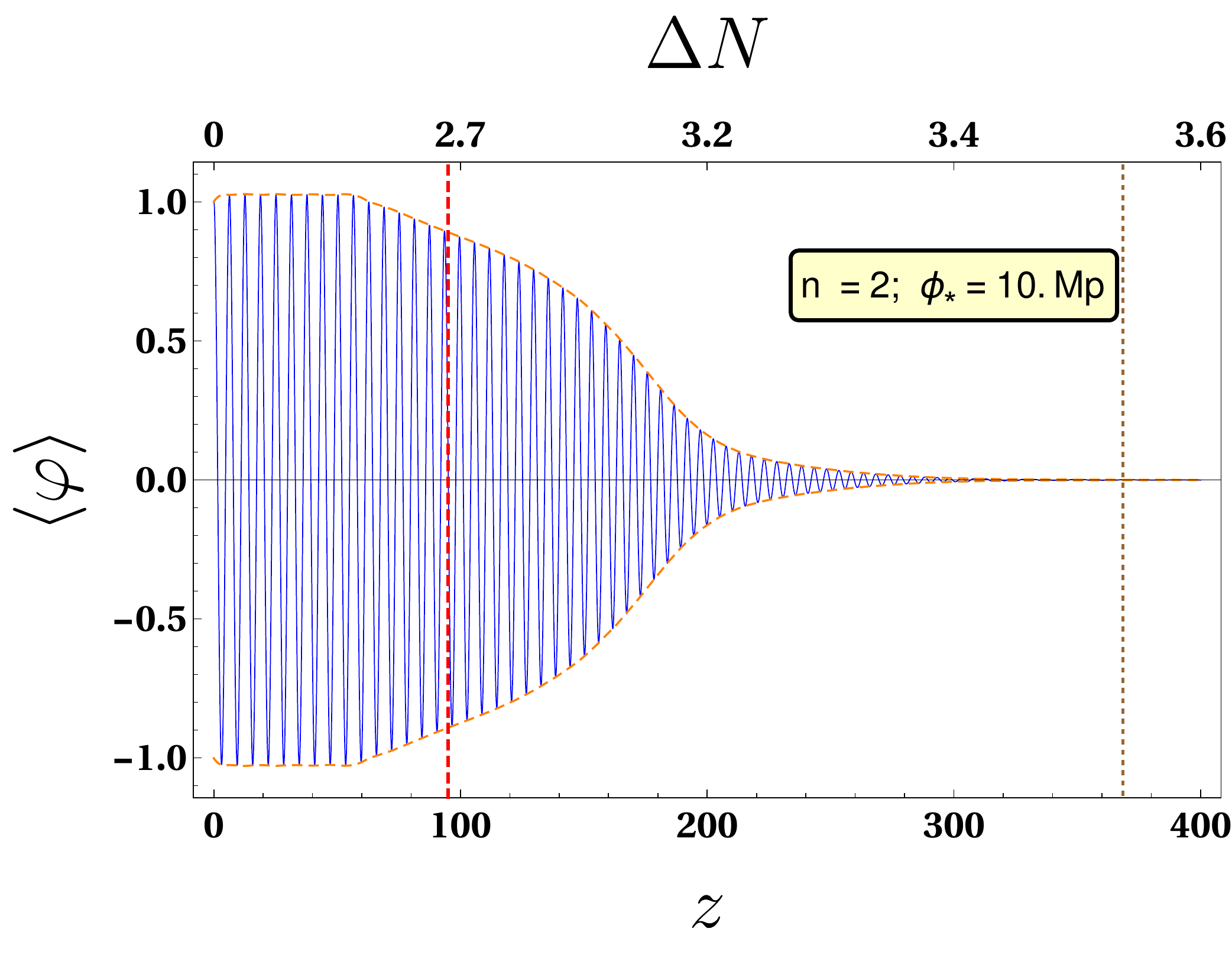}~~~~~~
	\includegraphics[scale=0.35]{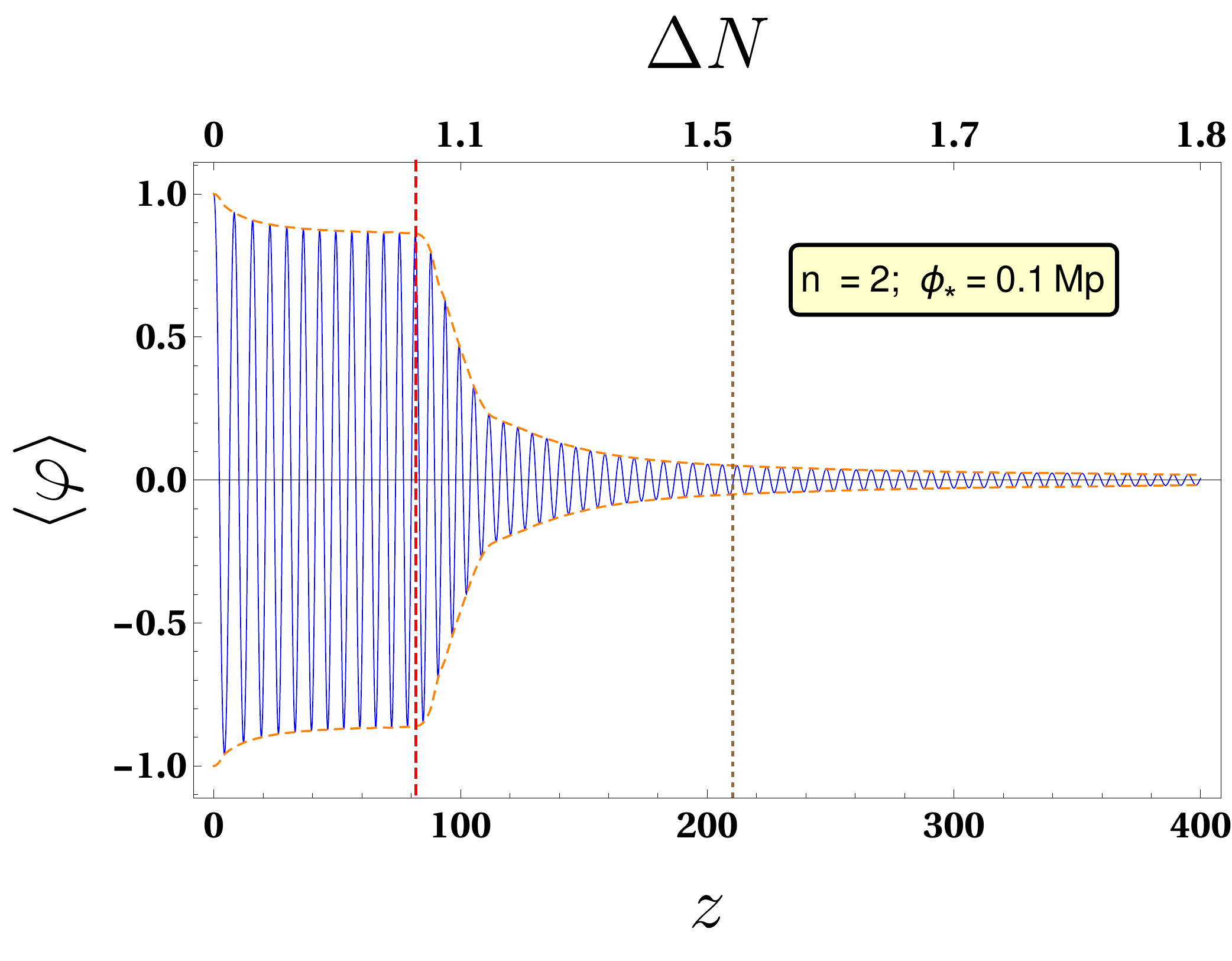}
	\caption{\scriptsize Evolution of volume averaged comoving inflation amplitude   $\varphi=(a/a_i)^{3/2}(\phi/\phi_0)$ for the model potential $n=2$ for $\phi_* = 10M_p$ and $0.1M_p$. The red dashed line shows the instant of back-reaction initiation while the brown dotted line is for the instant of decoupling.}
	\label{phi_n2}
\end{figure}
\begin{figure}[t]
	\centering
	\includegraphics[scale=0.035]{energy_n2A10.pdf}
	\includegraphics[scale=0.035]{frac_energy_n2A10.pdf}\\
	\includegraphics[scale=0.035]{energy_n2A01.pdf}
	\includegraphics[scale=0.035]{frac_energy_n2A01.pdf}
	\caption{\scriptsize The variation of energy components normalized by the initial energy with time and the e-folding number is shown. The plots in the right are the oscillation average of the same plots extending up to the later times. The virialization of the system is also shown in these two plots.}
	\label{fig:energy_n2}
\end{figure}

\begin{table}[t]
	\centering
	\caption{Fractions of energy components at $z=z_{br}$ for $n=2$}\label{tab:frac_br2}
	\begin{tabular}{SSSSSSS} \toprule
		{$\phi_*(M_p)$} & {$\frac{E^K_{\phi}}{E_T}$} & {$\frac{E^P_{\phi}}{E_T}$} & {$\frac{E^K_{\chi}}{E_T}$} & {$\frac{E^I_{\phi \leftrightarrow \chi}}{E_T}$} & {$\frac{E_{\phi}^T}{E_T}$} & {$\frac{E_{\chi}^T}{E_T}$}\\ \midrule
		{$10$} & {$48.0\%$} & {$44.0\%$} & {$3.0\%$} & {$2.0\%$} & {$95.0\%$} & {$5.0\%$}\\\midrule
		{$0.1$} & {$48.0\%$} & {$46.3\%$} & {$2.7\%$} & {$2.1\%$} & {$95.0\%$} & {$5.0\%$}\\ \bottomrule
	\end{tabular}
	\bigskip
	
	\caption{Fractions of energy components at $z=z_{dec}$ for $n=2$}\label{tab:frac_dec2}
	
	\begin{tabular}{SSSSSSSSS} \toprule
		{$\phi_*(M_p)$} & {$\frac{E^K_{\phi}}{E_T}$} & {$\frac{E^P_{\phi}}{E_T}$} & {$\frac{E^G_{\phi}}{E_T}$} & {$\frac{E^K_{\chi}}{E_T}$} & {$\frac{E^G_{\chi}}{E_T}$} & {$\frac{E^I_{\phi \leftrightarrow \chi}}{E_T}$} & {$\frac{E_{\phi}^T}{E_T}$} & {$\frac{E_{\chi}^T}{E_T}$}\\ \midrule
		{$10$} & {$24.5\%$} & {$5.5\%$} & {$14.6\%$} & {$27.6\%$} & {$23.4\%$} & {$4.1\%$} & {$47.0\%$} & {$53.0\%$}\\\midrule
		{$0.1$} & {$29.4\%$} & {$15.4\%$} & {$11.0\%$} & {$22.2\%$} & {$19.1\%$} & {$3.1\%$} & {$57.2\%$} & {$42.7\%$}\\ \bottomrule
	\end{tabular}
	
\end{table}
\subsubsection{$n=2$}
Since we will be considering different inflationary models where different kind of parameterization will be needed for the our lattice simulation. We will describe those in detail for each models. For $n=2$ model, using the default re-scaling scheme of \texttt{LATTICEEASY}, we define the following quantities that will be used for simulation
\begin{align}
\varphi \equiv \frac{\phi}{\phi_0}a^{\frac{3}{2}},\quad X \equiv \frac{\chi}{\phi_0}a^{\frac{3}{2}},\quad z \equiv mt,\quad \vec{z} \equiv m\vec{x}.
\end{align}
where $x^{\mu}\equiv (t,\vec{x})$ is the cosmic time and comoving coordinate. The field equations described in Eqs.\eqref{eq:eoms0} and \eqref{eq:eqchik} now reduces to
\begin{align}
\varphi'' - \frac{1}{a^2}\nabla^2 \varphi -\Bigg[\frac{3}{4}\Big(\frac{a'}{a}\Big)^2 - \frac{3}{2}\frac{a''}{a}\Bigg]\varphi + \frac{\mathcal{Q}}{a^3}\varphi X^2 + \frac{\partial}{\partial \varphi}V(\varphi) &= 0\\
X'' - \frac{1}{a^2}\nabla^2 X -\Bigg[\frac{3}{4}\Big(\frac{a'}{a}\Big)^2 - \frac{3}{2}\frac{a''}{a}\Bigg]X + \frac{\mathcal{Q}}{a^3}\varphi^2 X &= 0.
\end{align}
Where $prime (\prime)$ denotes derivative with respect to $z$ and $\mathcal{Q}$ is the resonance parameter defined as 
\[\mathcal{Q} \equiv \frac{g^2}{m^2}\phi_0^2 \]
We chose the two values of $\phi_{\ast}=(10M_p,0.1M_p).$ To felicitate comparison with previous works, in all our lattice simulation we have set ${\rm m_p = 1}$. PLANCK normalization set the values of the scale $m$ as $ (m_{10},m_{0.1}) = (1.38\times10^{-6} ,~9.38\times10^{-6})$ in unit of ${\rm m_p}$. We have started our simulation when $\varphi'(z_0)=0$ with the initial field values at $(0.191{\rm m_p},~0.0157{\rm m_p})$. The coupling parameter for both values of $\sphi$ is taken to be $g^2=2.2\times10^{-5}.$ We have used a $256^3$ lattice with a comoving edge size $L =(10/m_{10}, 70/m_{0.1})$ respectively for the two values of $\sphi$. The grid spacing corresponding to the above set of parameters are $dx \sim (0.04/m_{10},0.3/m_{0.1})$. Different choices of $L$ are made so that the grid spacing which set the limit to capture the momenta ranges in the simulation is nearly the same in both the cases. The time step for those two cases are $dt = (0.001/m_{10},0.001/m_{0.1})$. The maximum (comoving) momenta that can be possibly excited during the broad parametric resonance in this case will be \eqref{eq:kmax}
\begin{equation}
 k_{max}^2 \sim \mathcal{Q}^{1/2}m^2 .
\end{equation}
Now for a $N^3$ lattice with lattice spacing $L$, the minimum and the maximum (discrete) momenta that can be captured in a lattice simulation are $(2\pi)/L$ and $(\sqrt{3}N\pi)/L$ respectively. Thus in order to capture the correct UV physics, we must ensure that the minimum momenta range above must be well below the value of $k_{max}$. Also, to capture the IR physics for sufficient long time we must also ensure that the maximum momenta is large enough(which call for simulation with large lattice points). We can readily verify that these lattice parameters are adequate to capture both the IR and UV physics for a sufficient time of simulation. The evolution of the comomving inflaton field for two values of $\sphi$ has been shown in fig(\ref{phi_n2}). The back-reaction of the $\chi$ field starts to play role within $\Delta N_{br}\sim (2.7,~1)$ for $\sphi = (10M_p,~0.1M_p)$. It is evident that the duration of parametric resonance phase decreases with decreasing $\sphi$. In the right panels of \Cref{fig:energy_n2}, we have shown the oscillation average of various energy components such that the virialization relations mentions in \Cref{eq:virial_eos} are readily visible. In order to have a quantitative understanding, the fractions of different energy components have been tabulated. From \Cref{tab:frac_br2}, we can clearly see that during the initial phase when particle production is mainly due to parametric resonance, irrespective of $\sphi$ value, only $5\%$ of the total energy component is getting transferred into the reheating field. However most efficient transfer of energy occurs after the onset of back-reaction until the stationary phase is achieved with complete thermalization at around $\Delta N_{dec}\sim(3.5,~1.5)$ for the two values of $\sphi$. After the end of this back-reaction dominated phase at $z=z_{dec}$, all individual energy component freezes out to a constant value given in the \Cref{tab:frac_dec2}. At this point let us bring to our reader's notice an important characteristic feature in the non-linear regime of thermalization is that during this phase the inhomogeneities start to grow rapidly for both the field and then reach a stationary phase. It would be interesting to understand the fact that as we decrease $\sphi$ the growth of inhomogeneities are less. Therefore, for smaller value of $\sphi$ or in other words for small scale inflation, local inhomogeneity during reheating will be suppressed and it can have interesting effect after the end of reheating. We will see in our subsequent discussions that this behavior of inhomogeneous evolution depending upon $\sphi$ will be similar for other models such as $n=4,6$. Analytic understanding of this phenomena could interesting. 
%It is evident that after the daughter particle energy is the dominant component after the stationary phase for $\sphi=10M_p$ which has a higher value of resonance parameter compared to $\sphi=0.1M_p$ for which inflaton energy is still the dominant energy component of the system after the stationary phase of preheating with the four legged interaction considered here. 
As has been mentioned before, irrespective of the value of $\sphi$, only nearly $50\%$ of the inflation energy density has been transferred to the reheating $\chi$ field before non-perturbative production being completely stopped. Therefore, in order to complete the decay we need to perform perturbative decay separately to obtain the reheating temperature, that we will discuss in the end. For this process three leg interaction $\phi \chi^2$ may be dominant. In our future publication we will consider both four and three leg interactions to examine decay process and thermalization. 

In any case for $n=2$, we should emphasize that our numerical simulation does not give the condition of equal energy distribution among the inflaton and daughter particle in the stationary phase. As mentioned earlier the equation of state of both the field does not become that of the radiation as opposed to other models $n=4,6$ that will be discussed in the subsequent sections.    
%%%Energy fraction table for n=2%%%%%%%%%%%%%%%%%%%%%%%%%
%\noindent\rule{\textwidth}{0.2pt}

%\noindent\rule{\textwidth}{0.2pt}
%%%%Table ends%%%%%%%%%%%%%%%%%%%%%%%%%%%%%%%%%%%%%%%%%%%
%\clearpage
\subsubsection{$n=4$}
\begin{figure}[]
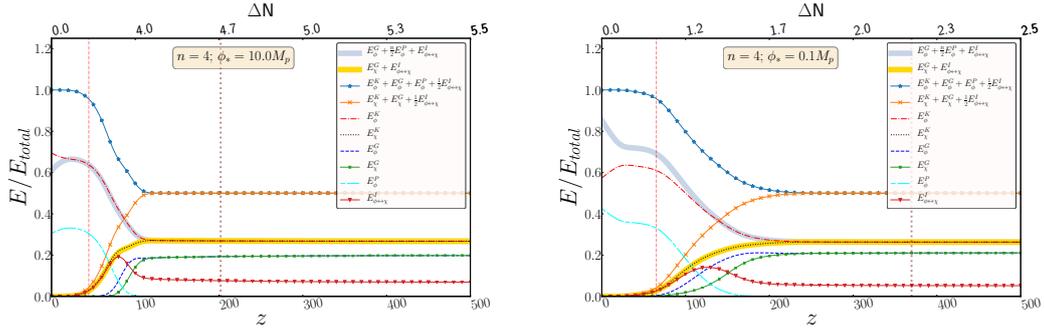

	\centering
	\includegraphics[scale=0.035]{frac_energy_n4A10.pdf}
	\includegraphics[scale=0.035]{frac_energy_n4A01.pdf}
	\caption{\scriptsize The variation of energy components with time and the efolding number is shown. }
	\label{fig:energy_n4}
\end{figure}

For $n=4$ model we define the following dimensionless quantities that will be used for simulation
\begin{align}
\varphi \equiv \frac{\phi}{\phi_0}a^{-1},\quad X \equiv \frac{\chi}{\phi_0}a^{-1},\quad z \equiv \phi_0\sqrt{\lambda}~t,\quad \vec{z} \equiv \phi_0\sqrt{\lambda}~\vec{x}.
\end{align}
where $x^{\mu}\equiv (t,\vec{x})$ is the cosmic time and comoving coordinate. The field equations described in Eqs.\eqref{eq:eoms0} and \eqref{eq:eqchik} reduces to
\begin{align}
\varphi'' - \nabla^2 \varphi - \frac{a''}{a}\varphi + \mathcal{Q}\varphi X^2 + \frac{\partial}{\partial \varphi}V(\varphi) &= 0\\
X'' - \nabla^2 X  - \frac{a''}{a}X + \mathcal{Q}\varphi^2 X &= 0.
\end{align}
The resonance parameter is defined as 
\[\mathcal{Q} \equiv \frac{g^2}{\lambda} \]
For this case too, we chose two values of $\phi_{\ast}=(10M_p,0.1M_p).$ From PLANCK normalization the value of the dimensionless parameter $\lambda$  turned out to be ($4.2\times10^{-13},~4.7\times10^{-8}$). At the end of inflation the field values assumes $(0.342 {\rm m_p},~0.0177 {\rm m_p})$. The coupling parameter for both the run is chosen to be $g^2=2.3\times10^{-5}.$ We have used a $256^3$ lattice with a comomving edge size $L = (10/\mathcal{B}_{10},180/\mathcal{B}_{0.1})$. The grid spacing in these case are $\dd{x} \sim (0.04/\mathcal{B}_{10}, 0.7/\mathcal{B}_{0.1})$. The time step we chose in these cases are $dt = (0.0001/m_{10},0.001/m_{0.1})$. Using \eqref{eq:kmax}, the maximum (comoving) momenta that could be excited here is found to be
\begin{equation}
 k_{max}^2 \sim \mathcal{Q}^{1/2}\phi_0^2\lambda, 
\end{equation} Following the discussion of the $n=2$ case, we can verify that these parameters will capture the necessary momentum ranges for a sufficient time. As we mentioned and also clearly seen from various plots, the qualitative behavior of decaying inflaton for this case is similar to that of $n=2$ model. From \Cref{fig:energy_n4} we see with the decreasing $\phi_*$, the decay of inflaton become efficient. After the initial linear regime, for this case the back-reaction sets in at around $\Delta N_{br}=(3.2,~0.9)$ e-folding number while the subsequent stationary phase is achieved at $\Delta N_{dec}=(4.7,~2.2)$.
The qualitative behavior of this model will be the same as $n=2$ case. However, from the \Cref{tab:frac_dec4} it is important to see that in the stationary phase the total energy becomes equally distributed between the inflaton and the daughter field.  

%%%Energy fraction table for n=2%%%%%%%%%%%%%%%%%%%%%%%%%
%\noindent\rule{\textwidth}{0.2pt}
\begin{table}[htp]
	
	\centering
	\caption{Fractions of energy components at $z=z_{br}$ for model $n=4$}\label{tab:frac_br4}
	\begin{tabular}{SSSSSSS} \toprule
		{$\phi_*(M_p)$} & {$\frac{E^K_{\phi}}{E_T}$} & {$\frac{E^P_{\phi}}{E_T}$} & {$\frac{E^K_{\chi}}{E_T}$} & {$\frac{E^I_{\phi \leftrightarrow \chi}}{E_T}$} & {$\frac{E_{\phi}^T}{E_T}$} & {$\frac{E_{\chi}^T}{E_T}$}\\ \midrule
		{$10$} & {$62.6\%$} & {$29.2\%$} & {$4.5\%$} & {$3.9\%$} & {$93.8\%$} & {$6.2\%$}\\\midrule
		{$0.1$} & {$63.7\%$} & {$30.4\%$} & {$3.0\%$} & {$2.8\%$} & {$95.0\%$} & {$5\%$}\\ \bottomrule
	\end{tabular}
	\bigskip
	
	\caption{Fractions of energy components at $z=z_{dec}$ for model $n=4$}\label{tab:frac_dec4}
	
	\begin{tabular}{SSSSSSSSS} \toprule
		{$\phi_*(M_p)$} & {$\frac{E^K_{\phi}}{E_T}$} & {$\frac{E^P_{\phi}}{E_T}$} & {$\frac{E^G_{\phi}}{E_T}$} & {$\frac{E^K_{\chi}}{E_T}$} & {$\frac{E^G_{\chi}}{E_T}$} & {$\frac{E^I_{\phi \leftrightarrow \chi}}{E_T}$} & {$\frac{E_{\phi}^T}{E_T}$} & {$\frac{E_{\chi}^T}{E_T}$}\\ \midrule
		{$10$} & {$26.9\%$} & {$-$} & {$19.3\%$} & {$27.0\%$} & {$19.3\%$} & {$7.6\%$} & {$50.0\%$} & {$50.0\%$}\\\midrule
		{$0.1$} & {$26.3\%$} & {$0.2\%$} & {$21.0\%$} & {$26.3\%$} & {$21.0\%$} & {$5.2\%$} & {$50.0\%$} & {$50.0\%$}\\ \bottomrule
	\end{tabular}
	
\end{table}

%\noindent\rule{\textwidth}{0.2pt}
%%%%Table ends%%%%%%%%%%%%%%%%%%%%%%%%%%%%%%%%%%%%%%%%%%%

\subsubsection{$n=6$}
For this model the re-scaled variables are defined as
\begin{align}
\varphi \equiv \frac{\phi}{\phi_0}a^{\frac{3}{4}},\quad X \equiv \frac{\chi}{\phi_0}a^{\frac{3}{4}},\quad z \equiv \frac{\phi_0^2}{m}t,\quad \vec{z} \equiv \frac{\phi_0^2}{m}\vec{x}.
\end{align}
The field equations described in Eqs.\eqref{eq:eoms0} and \eqref{eq:eqchik} for $n=6$ reduces to
\begin{align}
\varphi'' - a\nabla^2 \varphi - \Bigg[\frac{3}{4}\frac{a''}{a}- \frac{3}{16}\Big(\frac{a'}{a}\Big)^2\Bigg]\varphi + \mathcal{Q} a^{\frac{3}{2}} \varphi X^2 + \frac{\partial}{\partial \varphi}V(\varphi) &= 0\\
X'' - a\nabla^2 X - \Bigg[\frac{3}{4}\frac{a''}{a}- \frac{3}{16}\Big(\frac{a'}{a}\Big)^2\Bigg]\varphi + \mathcal{Q} a^{\frac{3}{2}}\varphi^2 X &= 0.
\end{align}
Where the resonance parameter turned out as 
\[\mathcal{Q} \equiv \frac{g^2m^2}{\phi_0^2} \]
As opposed to the other two models $n=2,4$, for the this case we have to chose $\phi_{\ast}=(1M_p,0.5M_p)$ to make the lattice input parameters $g^2m^2=(200,~9)$ within the reach of simulation with a practical time step(with our current computational resource). It has been found that the model with $n=6$ takes the longer time for simulation which further increases with increasing value of the coupling parameter $g^2$. With our present computational resource as well as for optimum run time, we are able to simulate for the values of the input parameter $g^2m^2 \textless 1000$. The value of the scale $m$ from WMAP normalization becomes ($2.7\times10^{4},~5.6\times10^{3}$) in unit of ${\rm m_p}$ that will be our one set of input parameters. After the end of inflation the initial field values are taken to be $(0.226{\rm m_p},~0.138{\rm m_p})$ and the coupling parameter for both $\phi_*$ value is chosen as $g^2=2.8\times10^{-7}.$ In this case too we have used a $256^3$ lattice with a comoving edge size $L = (250/\mathcal{B}_{1},250/\mathcal{B}_{0.5})$. The grid spacing in these case are $\dd{x} \sim (0.04/\mathcal{B}_{1}, 0.7/\mathcal{B}_{0.5})$. The time steps for the two cases are set as $dt = (0.0001/\mathcal{B}_{1},0.0001/\mathcal{B}_{0.5})$. For this case \eqref{eq:kmax} yields the maximum momenta to be
\begin{equation}
 k_{max}^2 \sim \mathcal{Q}^{1/2}\frac{\phi_0^4}{m^2} .
\end{equation} Due to increases complexity of the terms in the potential and its derivatives in this case, it has been found that it take longer simulation time for the same program time compared to the previous two cases. With the above lattice parameters we can check that the simulation captures the necessary UV physics. The IR physics is also captured sufficiently within simulation period. We have run the simulation for program time $z=300$ in these cases within which the decay of inflaton is found to reach the desired stationary limit. However, for long time simulation we need larger lattice points for sufficient IR resolution. The simulation results show that the non-linear back-reaction regime starts at around  $\Delta N_{br}=(3.5,~3.0)$ and then finally the equilibrium condition is achieved at $\Delta N_{dec}=(4.5,~3.8)$ for the two values of $\sphi$. In this case too, the energy will be distributed equally to both inflaton and daughter field as opposed to $n=2$ case as seen in \Cref{tab:frac_dec6}.
\begin{figure}[]
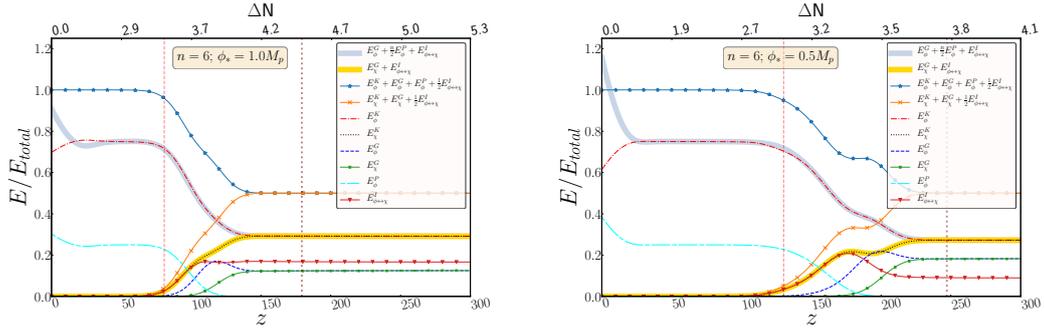

	\centering
	\includegraphics[scale=0.035]{frac_energy_n6A1.pdf}
	\includegraphics[scale=0.035]{frac_energy_n6A05.pdf}
	\caption{\scriptsize Evolution of various energy components for $n=6$ model. }
	\label{fig:energy_n6}
\end{figure}

%%%Energy fraction table for n=2%%%%%%%%%%%%%%%%%%%%%%%%%
%\noindent\rule{\textwidth}{0.2pt}
\begin{table}[htp]	
	\centering
	\caption{Fractions of energy components at $z=z_{br}$ for model $n=6$}\label{tab:frac_br6}
	\begin{tabular}{SSSSSSSS} \toprule
		{$\phi_*(M_p)$} & {$\frac{E^K_{\phi}}{E_T}$} & {$\frac{E^P_{\phi}}{E_T}$} & {$\frac{E^K_{\chi}}{E_T}$} & {$\frac{E^G_{\chi}}{E_T}$} &  {$\frac{E^I_{\phi \leftrightarrow \chi}}{E_T}$} & {$\frac{E_{\phi}^T}{E_T}$} & {$\frac{E_{\chi}^T}{E_T}$}\\ \midrule
		{$1$} & {$71.5\%$} & {$22.8\%$} & {$2.7\%$} & {$0.3\%$} & {$2.7\%$} & {$95.0\%$} & {$5.0\%$}\\\midrule
		{$0.5$} & {$70.3\%$} & {$22.7\%$} & {$3.4\%$} & {$0.2\%$} & {$3.4\%$} & {$95.0\%$} & {$5.0\%$}\\ \bottomrule
	\end{tabular}
	\bigskip
	
	\caption{Fractions of energy components at $z=z_{dec}$ for model $n=6$}\label{tab:frac_dec6}
	
	\begin{tabular}{SSSSSSSS} \toprule
		{$\phi_*(M_p)$} & {$\frac{E^K_{\phi}}{E_T}$} &  {$\frac{E^G_{\phi}}{E_T}$} & {$\frac{E^K_{\chi}}{E_T}$} & {$\frac{E^G_{\chi}}{E_T}$} & {$\frac{E^I_{\phi \leftrightarrow \chi}}{E_T}$} & {$\frac{E_{\phi}^T}{E_T}$} & {$\frac{E_{\chi}^T}{E_T}$}\\ \midrule
		{$1$} & {$29.2\%$}  & {$12.4\%$} & {$29.2\%$} & {$12.4\%$} & {$16.8\%$} & {$50.0\%$} & {$50.1\%$}\\\midrule
		{$0.5$} & {$27.3\%$} & {$18.2\%$} & {$27.3\%$} & {$18.2\%$} & {$9.0\%$} & {$50.0\%$} & {$50.0\%$}\\ \bottomrule
	\end{tabular}
	
\end{table}

%\noindent\rule{\textwidth}{0.2pt}
%%%%Table ends%%%%%%%%%%%%%%%%%%%%%%%%%%%%%%%%%%%%%%%%%%%

\subsection{\label{sec:eos}Equation of state}

\begin{figure}[]
	\centering
	\includegraphics[scale=0.3]{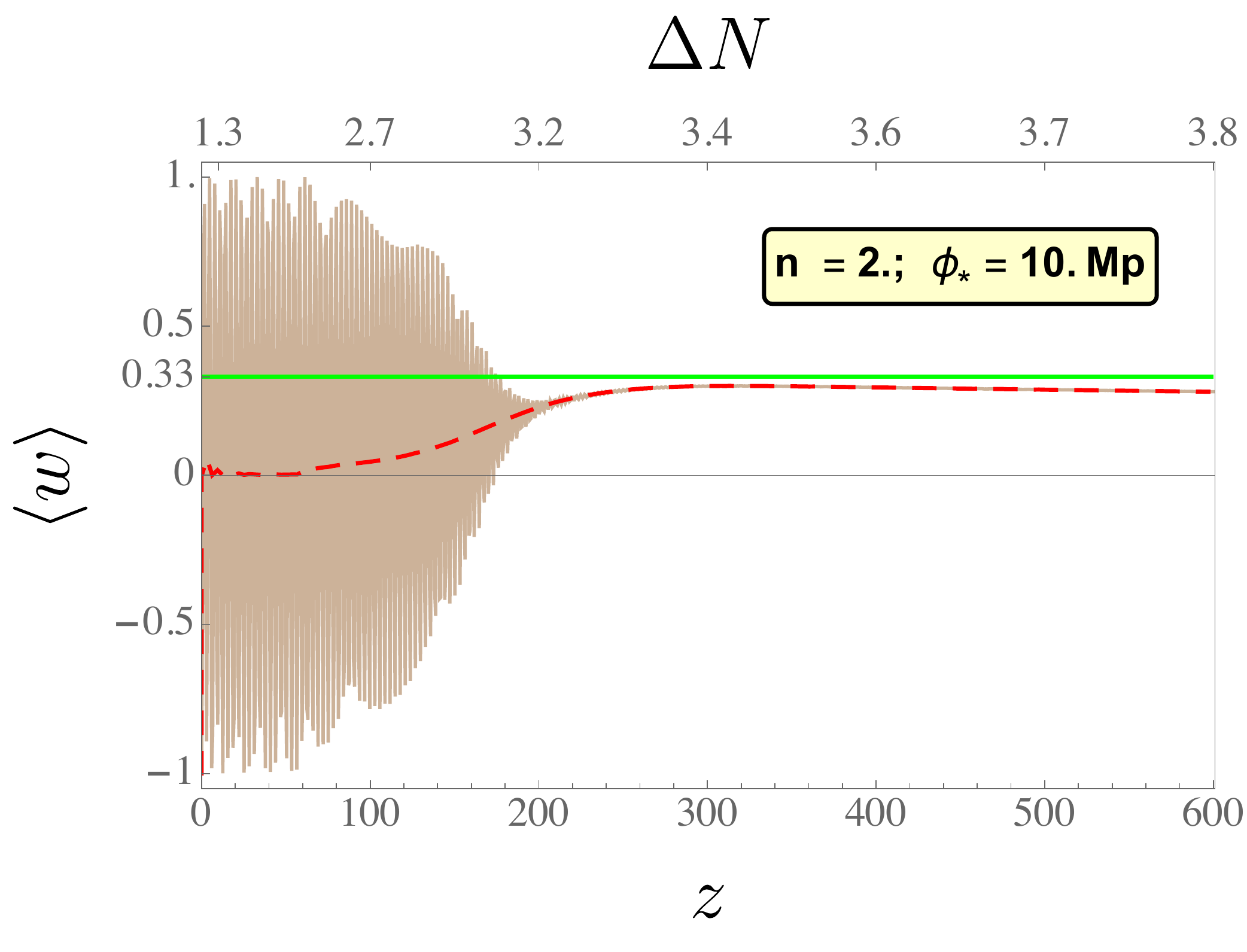}~~~~~~
	\includegraphics[scale=0.3]{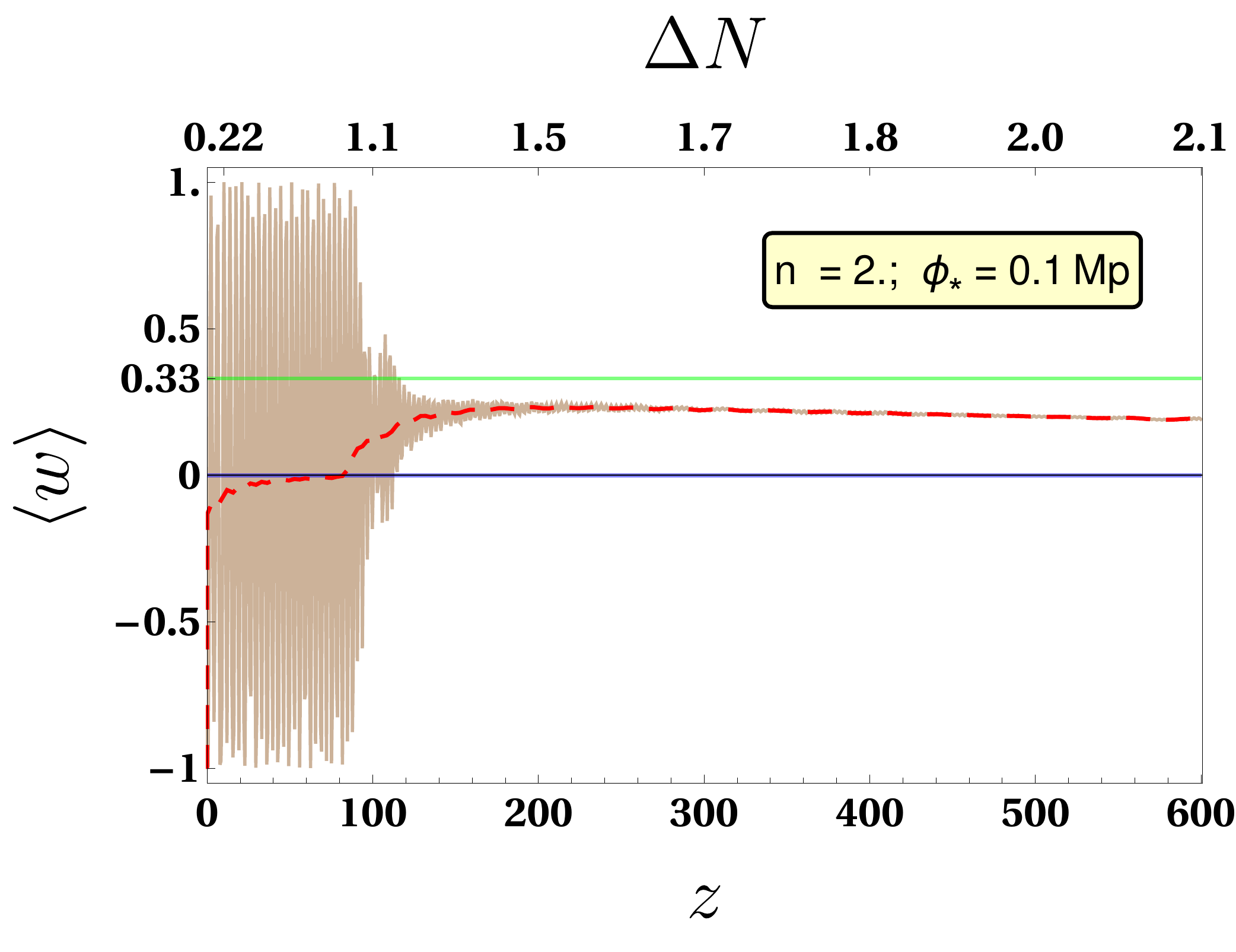}
	\caption{\scriptsize The variation of equation of state with time(or alternatively the e-folding number) for the two values of the scale $\sphi$. The solid blue lines are the instantaneous value of the equation of state while the red dotted line is the averaged value over a period of inflaton oscillation. Time is measured in unit of their respective scale $m$ hence to facilitate a comparison between different models, the efolding number is shown in the upper panel. }
	\label{fig:eos_n2}
\end{figure}

\begin{figure}[]
	\centering
	\includegraphics[scale=0.3]{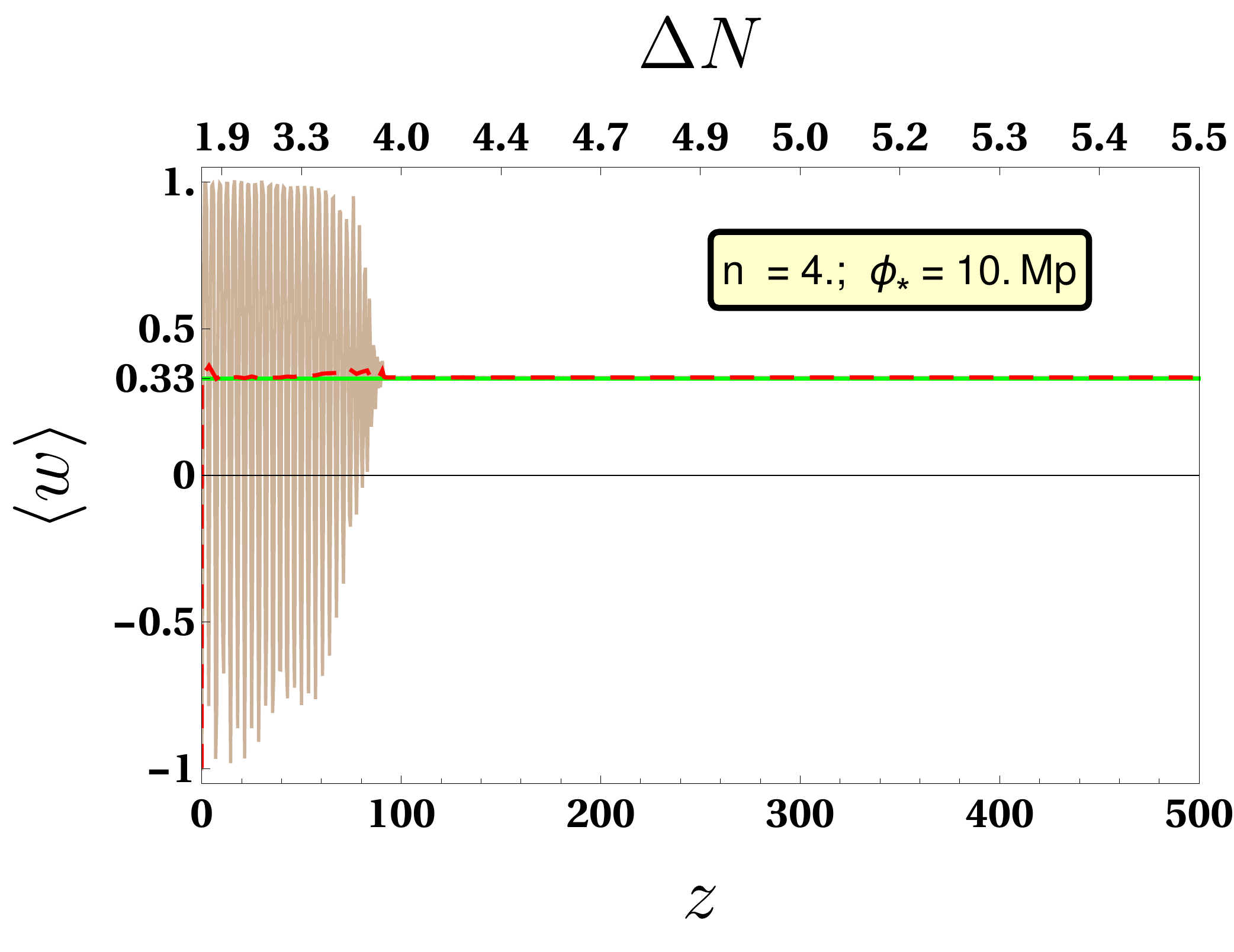}~~~~~~
	\includegraphics[scale=0.3]{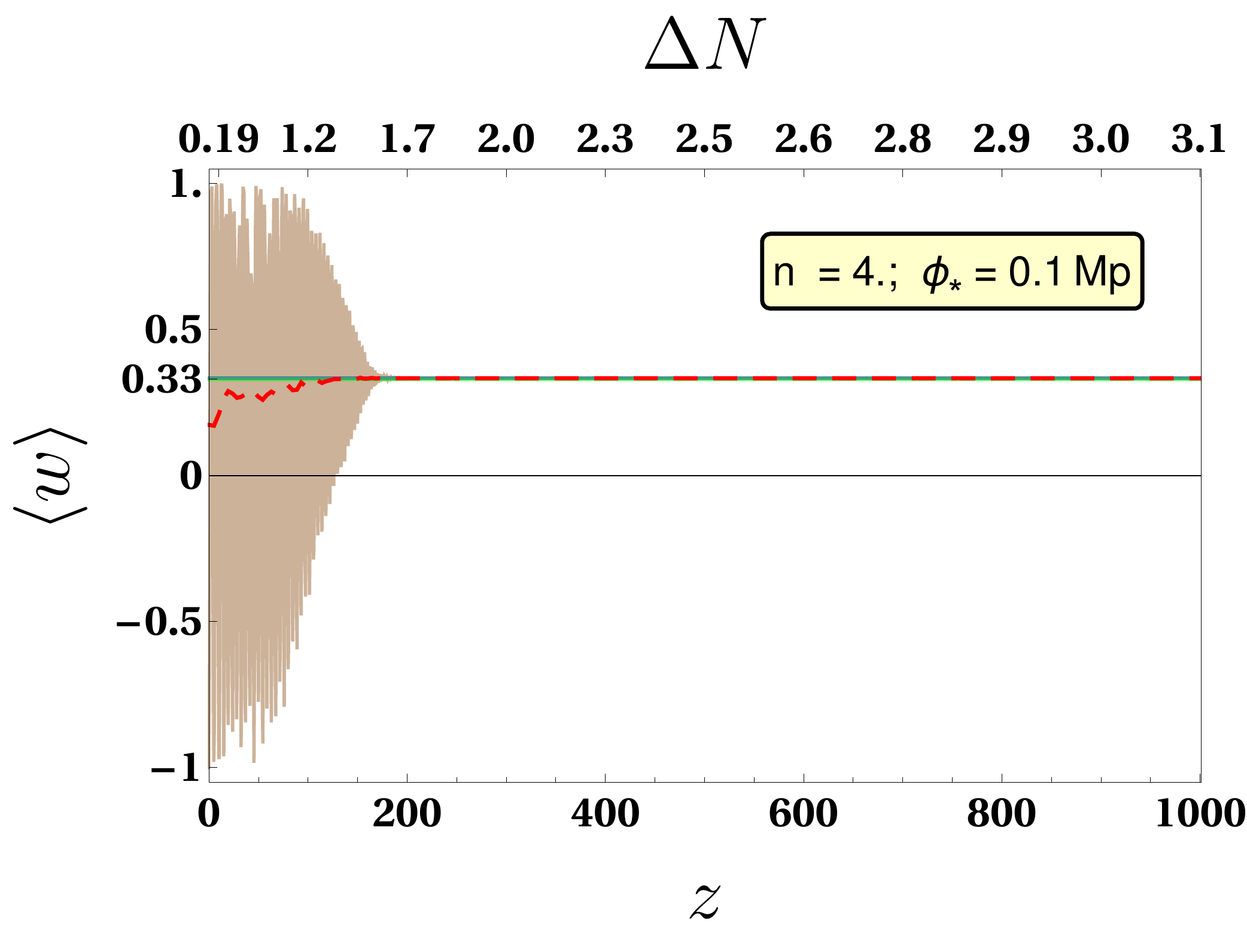}
	\caption{\scriptsize The variation of equation of state with time(or alternatively the e-folding number) for the two values of the scale $\sphi$. The solid blue lines are the instantaneous value of the equation of state while the red dotted line is the averaged value over a period of inflaton oscillation.}
	\label{fig:eos_n4}
\end{figure}

\begin{figure}[]
	\centering
	\includegraphics[scale=0.3]{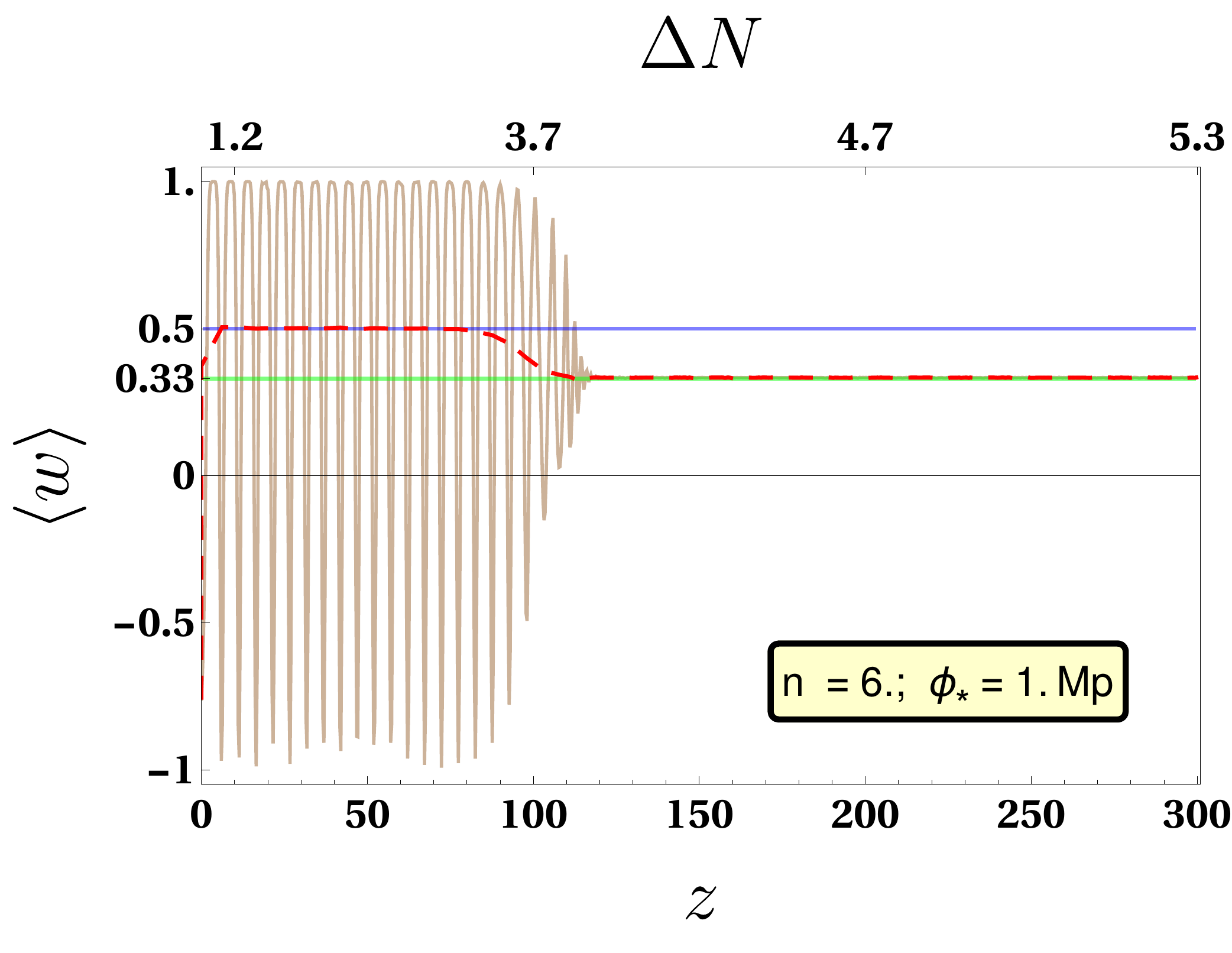}~~~~~~
	\includegraphics[scale=0.3]{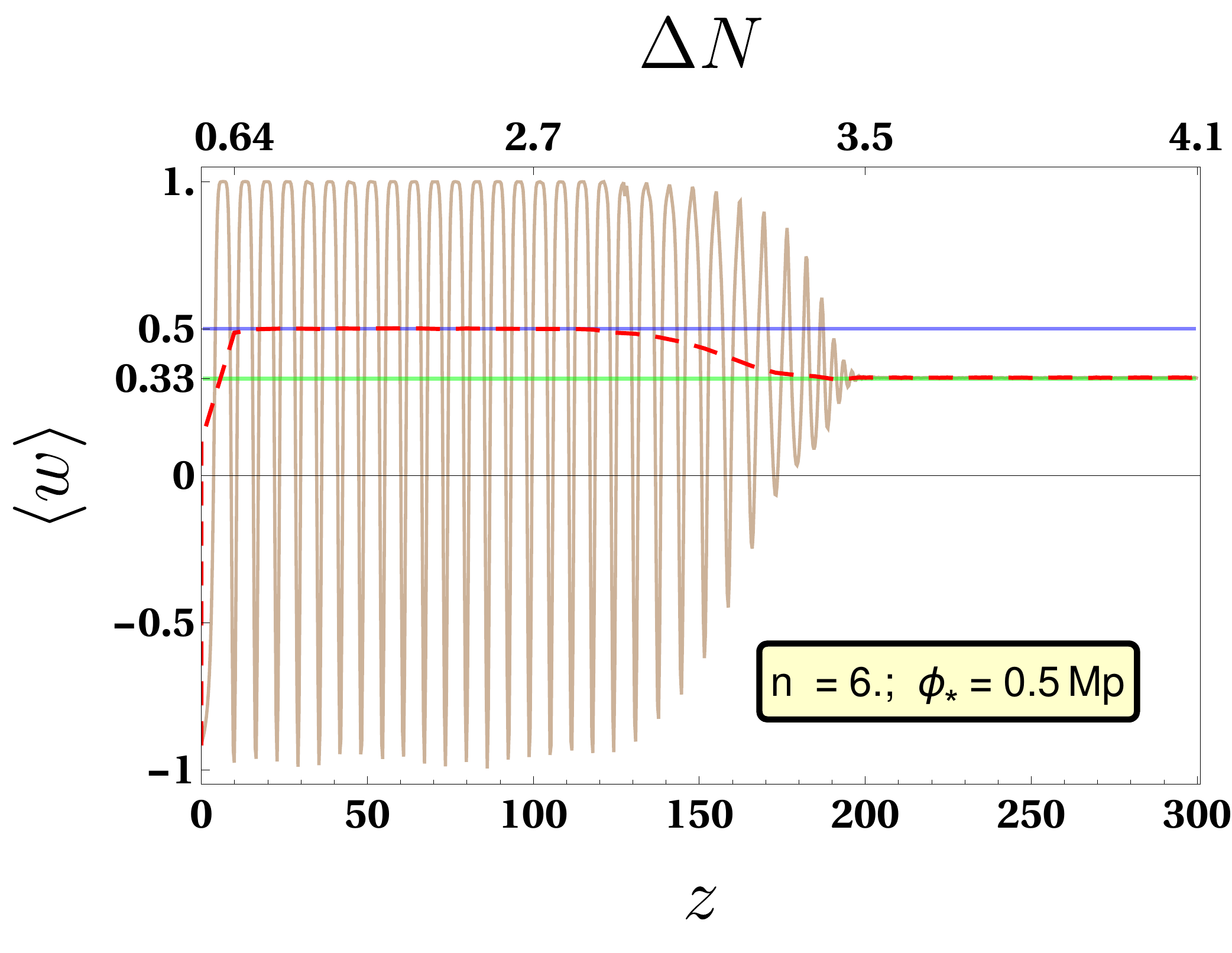}
	\caption{\scriptsize The variation of equation of state with time(or alternatively the e-folding number) for the two values of the scale $\sphi$. The solid blue lines are the instantaneous value of the equation of state while the red dotted line is the averaged value over a period of inflaton oscillation.}
	\label{fig:eos_n6}
\end{figure}

\begin{figure}[]
	\centering
	\subfigure[$n=2$, $\phi_* = 10M_p$]{\includegraphics[scale=0.4]{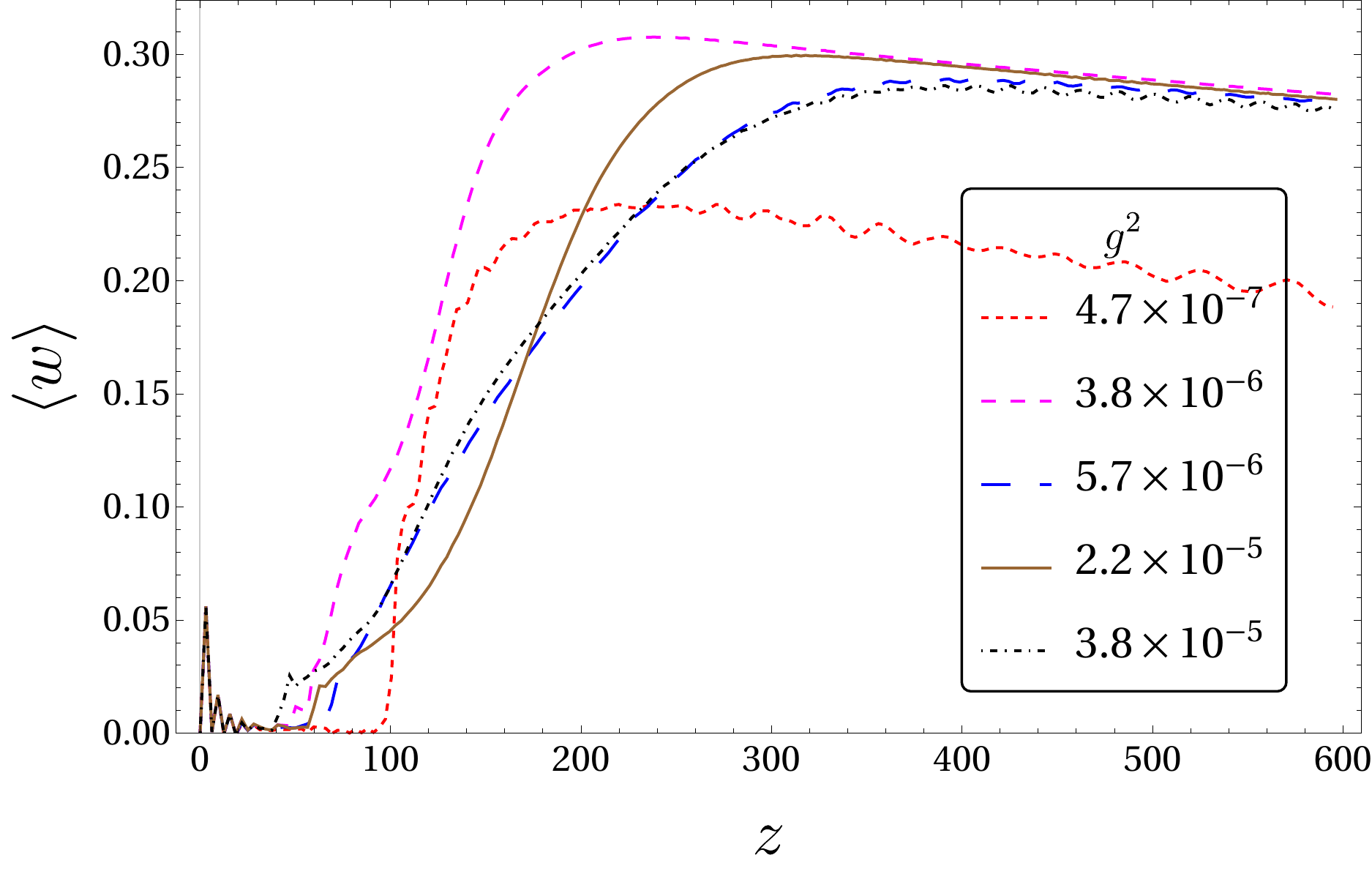}}
	\subfigure[$n=2$, $\phi_* = 0.1M_p$]{\includegraphics[scale=0.4]{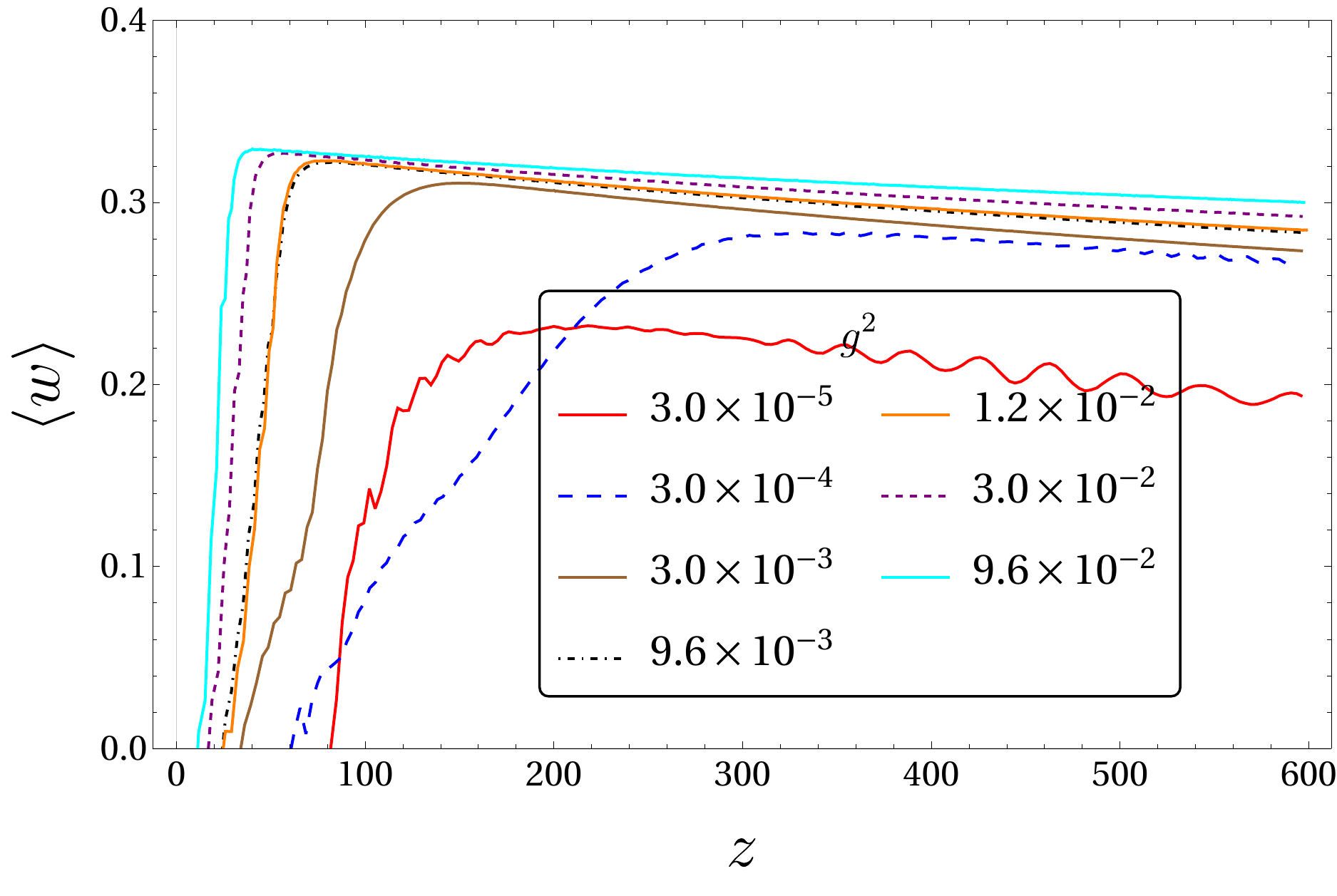}}
	\caption{\scriptsize Variation of equation of state with $g^2$ for model $n=2$ for the two values of $\sphi$. It is evident that indefinitely increasing $g^2$ will not bring radiation domination for $n=2$.}
	\label{fig:eos_n2all}
\end{figure}

The equation of state is one of the most important parameters to study during preheating. The equation of state in this context is defined as:
\be
w = \frac{p}{\rho} = \frac{\frac{1}{2}\dot{\phi}^2 + \frac{1}{2}\dot{\chi}^2 - \frac{1}{3}\frac{(\nabla \phi)^2}{2a^2} - \frac{1}{3}\frac{(\nabla \chi)^2}{2a^2} - V(\phi, \chi)}{\frac{1}{2}\dot{\phi}^2 + \frac{1}{2}\dot{\chi}^2 + \frac{(\nabla \phi)^2}{2a^2} + \frac{(\nabla \chi)^2}{2a^2} + V(\phi, \chi)\hfill}
\label{eq:eos}
\ee
For a homogeneous inflaton condensate oscillating in a potential $V(\phi)\propto \phi^n$, using the virial theorem, the equation state is given by\cite{Mukhanov:2005sc,Turner:1983he} $w=(n-2)/(n+2)$. The inflaton initially oscillates with this equation of state till the other components of the total energy such as gradient, interaction energy became significant and resulting in fragmentation. Ignoring the interaction energy, and using the virial relations given in \Cref{eqn:virial} the definition of $w$ in \ref{eq:eos} reduces to
\begin{align}
w &= \frac{1}{3} + \frac{(n-4)}{6}\frac{1}{ \frac{n+2}{4}+ \frac{\langle (\nabla \phi)^2/2a^2\rangle}{\langle V(\phi) \rangle} + \frac{\langle (\nabla \chi)^2/2a^2\rangle}{\langle V(\phi) \rangle} + \frac{3}{2}\frac{\langle  V_I \rangle}{\langle V(\phi) \rangle} }
\label{eq:virial_eos}
\end{align}
Using the values of average energies of different components listed in \Cref{tab:frac_dec2,tab:frac_dec4,tab:frac_dec6}, we found that $w\to0.2$ for $n=2$ and $w\to1/3$ for $n=4,~\text{and}~6$ in the stationary phase. The results of the simulation are plotted in \Cref{fig:eos_n2,fig:eos_n4,fig:eos_n6}. We have plotted the instantaneous equation of state(brown curves) as well as the average value over a period of one inflaton oscillation(red dashed curves). The green line shows the homogeneous Inflation equation of state. The features we note from these figures for different models are
\begin{itemize}
	\item[(i)] For $n=2$ models, the equation of state make a transition from $w=0$ to $w\sim0.2-0.3$ within a few e-folding numbers. However, the equation of state never reaches the radiation like equation of state($w=1/3$). This behavior has also been noted in \cite{Podolsky:2005bw} for the $m^2\phi^2$ model. Similar behavior is expected here as our model boils down to usual power law potential near the minimum at $\phi=0$. We also observed that reducing the scale $\sphi$ does not improve the scenario except reducing the duration of preheating phase. We have found that with longer simulation equation of state starts decreasing after reaching a maximum $w\sim0.3$ and finally settles to $w\to0.1$. The reason for this behavior is that for $n=2$ model, which is identical to $m^2\phi^2$ around the minimum, the massive inflaton component although may remain under-abundant during preheating eventually rise up to dominate after the inflation decay ceases. It has also been found\cite{Podolsky:2005bw} that $w(t)$ depends non-monotonically on the resonance parameter $q$ (or alternatively on the coupling $g^2$) for higher value of $\phi_*$. However it can be seen that increasing $g^2$ will not help us reaching the radiation domination as depicted in figure (\ref{fig:eos_n2all}). However, interesting observation can be made on the dependence of $\sphi$. For sub-Planckian value of $\sphi = 0.1{\rm M_p}$ the equation of state $w$ changes almost instantaneously from zero to its maximum value before the system could reach the thermal equilibrium and apparently it also depends monotonically on the coupling parameter $g^2$. Therefore, lowering the value of $\sphi$ makes the intermediate or the turbulent phase more efficient.    
		
	\item[(ii)] For $n=4$ model, the homogeneous condensate itself oscillates with $w=1/3$ at the onset of preheating. The simulation results shows that it retains the radiation like equation of state throughout the simulation period. Decreasing $\sphi$ have the same qualitative behavior, with an important difference compared to $n=2$ model is that intermediate or turbulent regime occurs for longer time. Therefore the evolution of $w(t)$ do not provide any additional information about the thermalization process in this case.
	
	\item[(iii)] In the case of model with $n=6$, the homogeneous condensate has equation of state $w=1/2$. Here the equation of state makes a transition from $w=1/2$ to $w\to1/3$ for both values of $\sphi$.

\end{itemize}

%\subsection{Energy Components}

\subsection{\label{sec:occupation}Occupation Numbers}
\begin{figure}[]
	\centering
	\subfigure[$n=2$, $\phi_* = 10M_p$]{\includegraphics[scale=0.45]{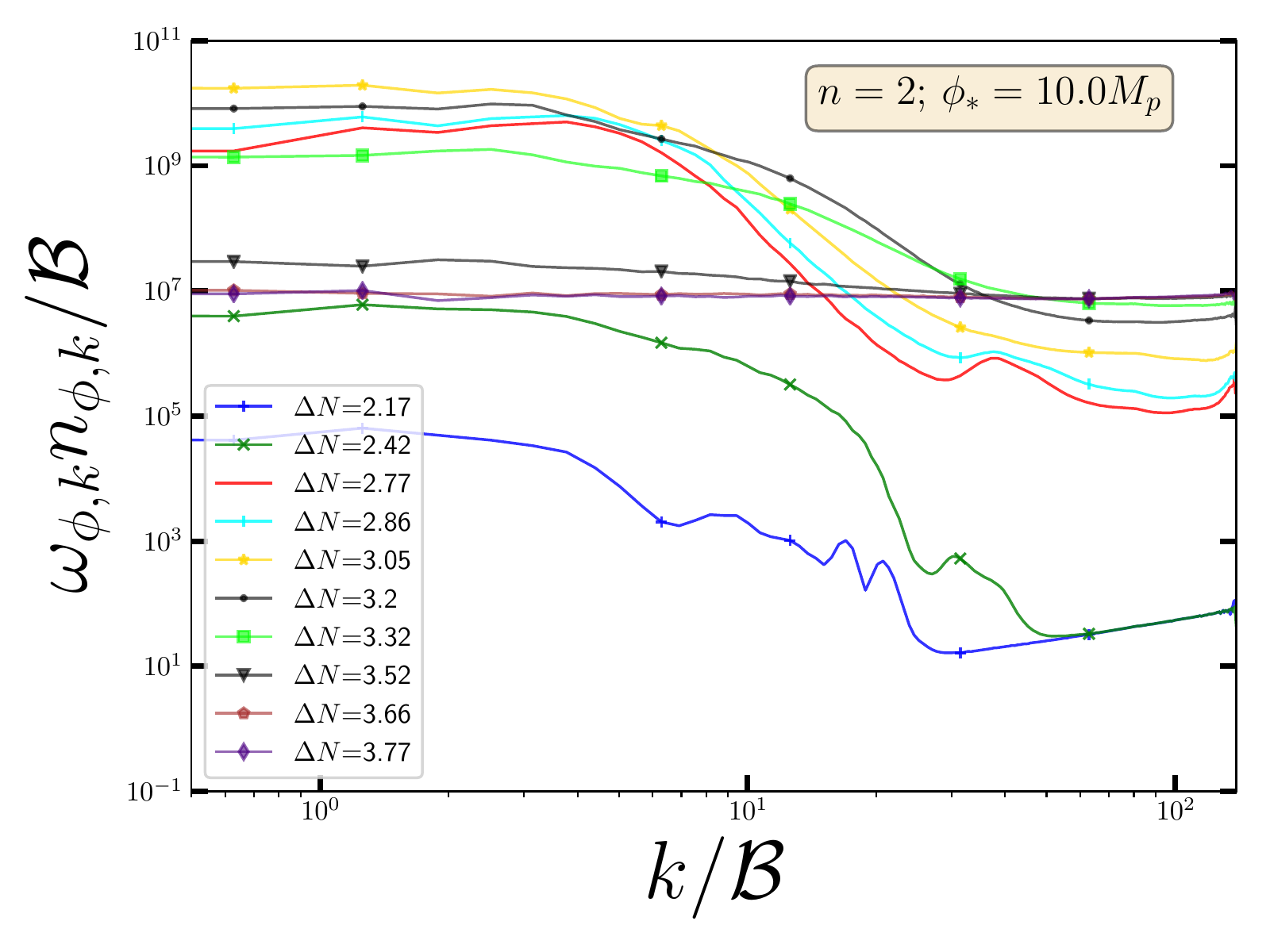}}
	\subfigure[$n=2$, $\phi_* = 0.1M_p$]{\includegraphics[scale=0.45]{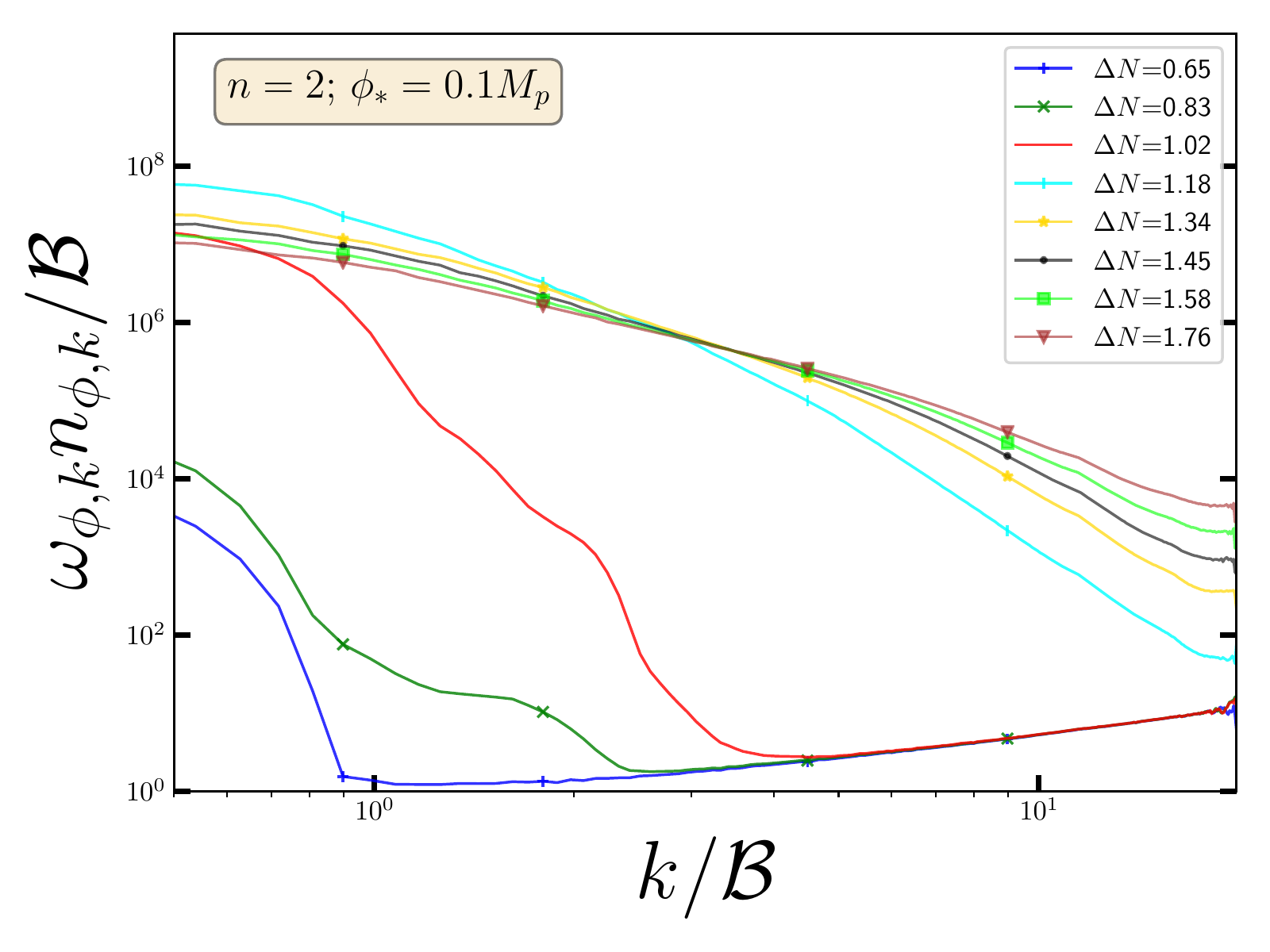}}
	\caption{\scriptsize The evolution of the combination $\omega_{\phi, k}n_{\phi, k}$ as a function of momentum k at different e-folding numbers for $n=2$ for the two values of $\phi_{\ast}$}
	\label{fig:Phikwk_n2}
\end{figure}

\begin{figure}[]
	\centering
	\subfigure[$n=2$, $\phi_* = 10M_p$]{\includegraphics[scale=0.45]{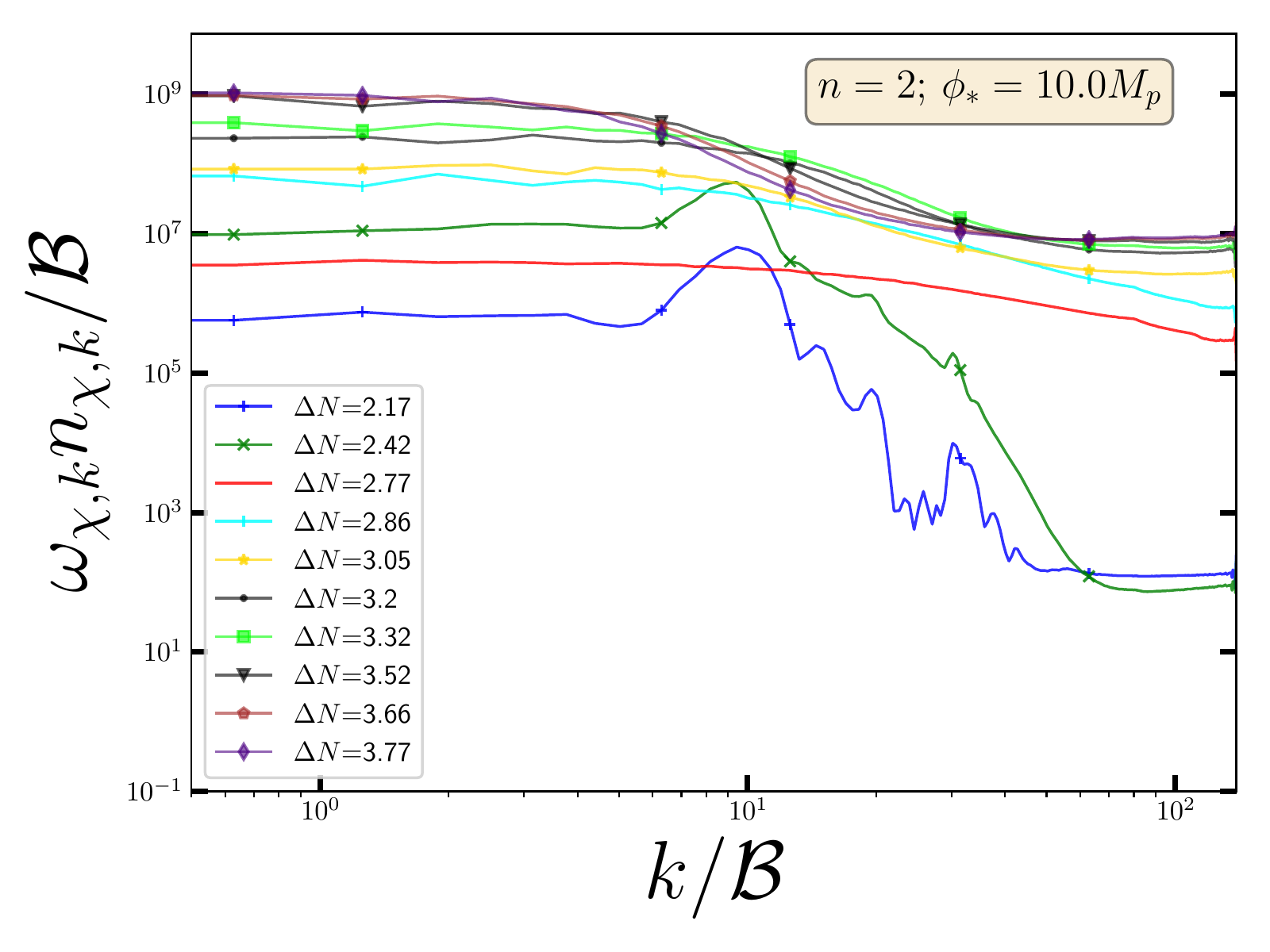}}
	\subfigure[$n=2$, $\phi_* = 0.1M_p$]{\includegraphics[scale=0.45]{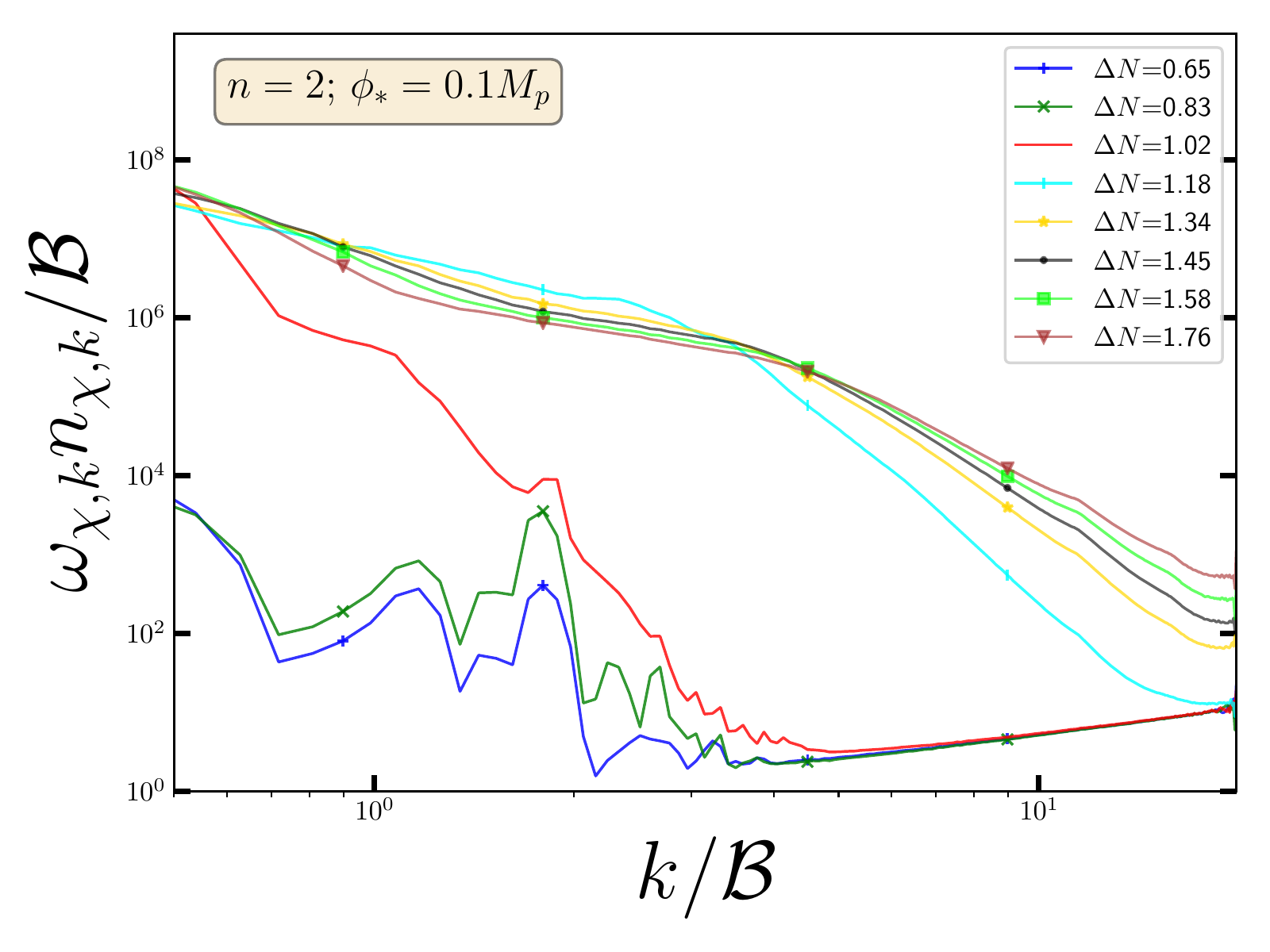}}
	\caption{\scriptsize Same plot as Fig.(\ref{fig:Phikwk_n2}) for the $\chi$ field.}
	\label{fig:Chikwk_n2}
\end{figure}

\begin{figure}[]
	\centering
	\subfigure[$n=4$, $\phi_* = 10M_p$]{\includegraphics[scale=0.45]{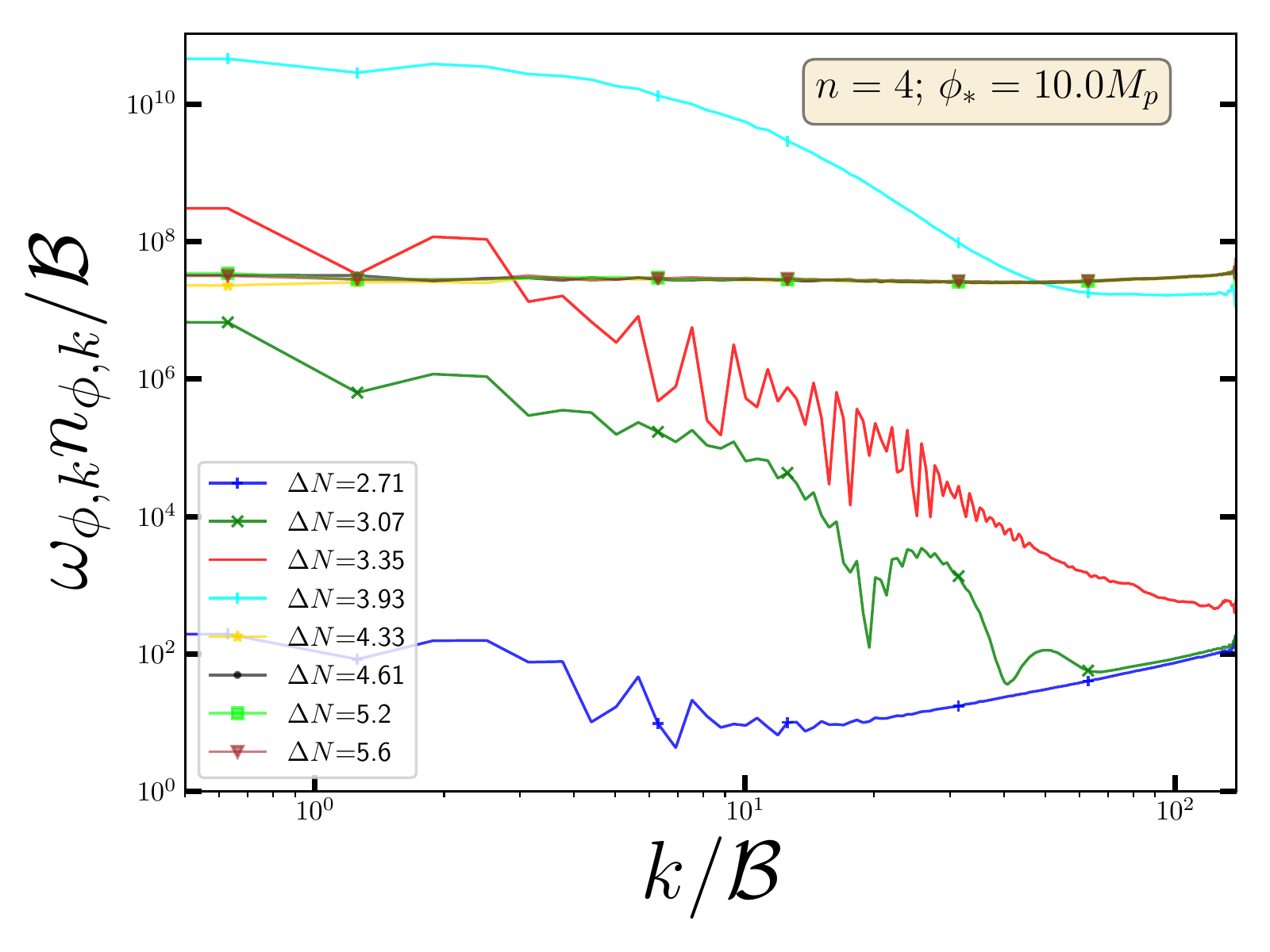}}
	\subfigure[$n=4$, $\phi_* = 0.1M_p$]{\includegraphics[scale=0.45]{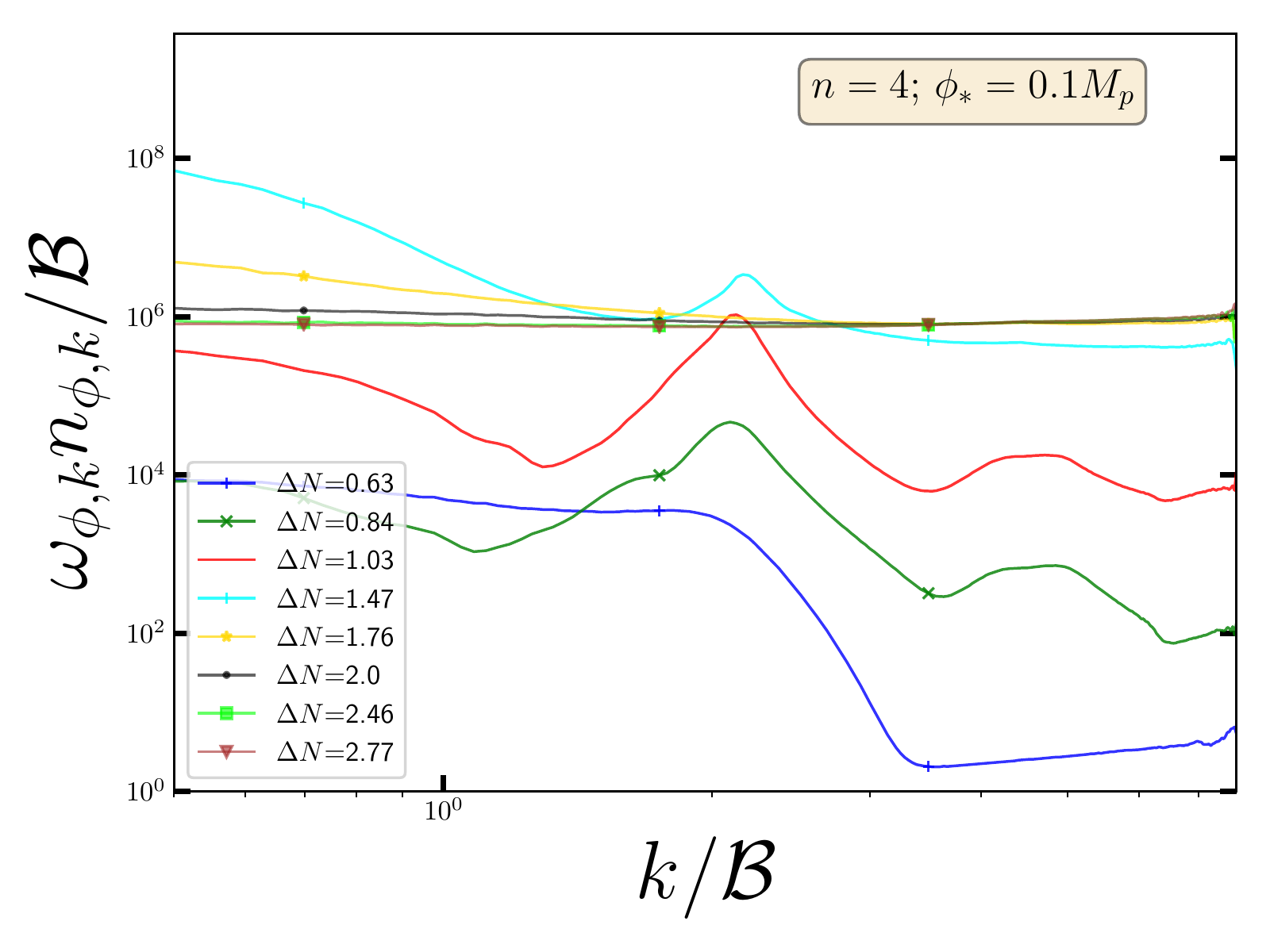}}
	\caption{\scriptsize Same plot as Fig.(\ref{fig:Phikwk_n2}) for the $n=4$}
	\label{fig:Phikwk_n4}
\end{figure}

\begin{figure}[]
	\centering
	\subfigure[$n=4$, $\phi_* = 10M_p$]{\includegraphics[scale=0.45]{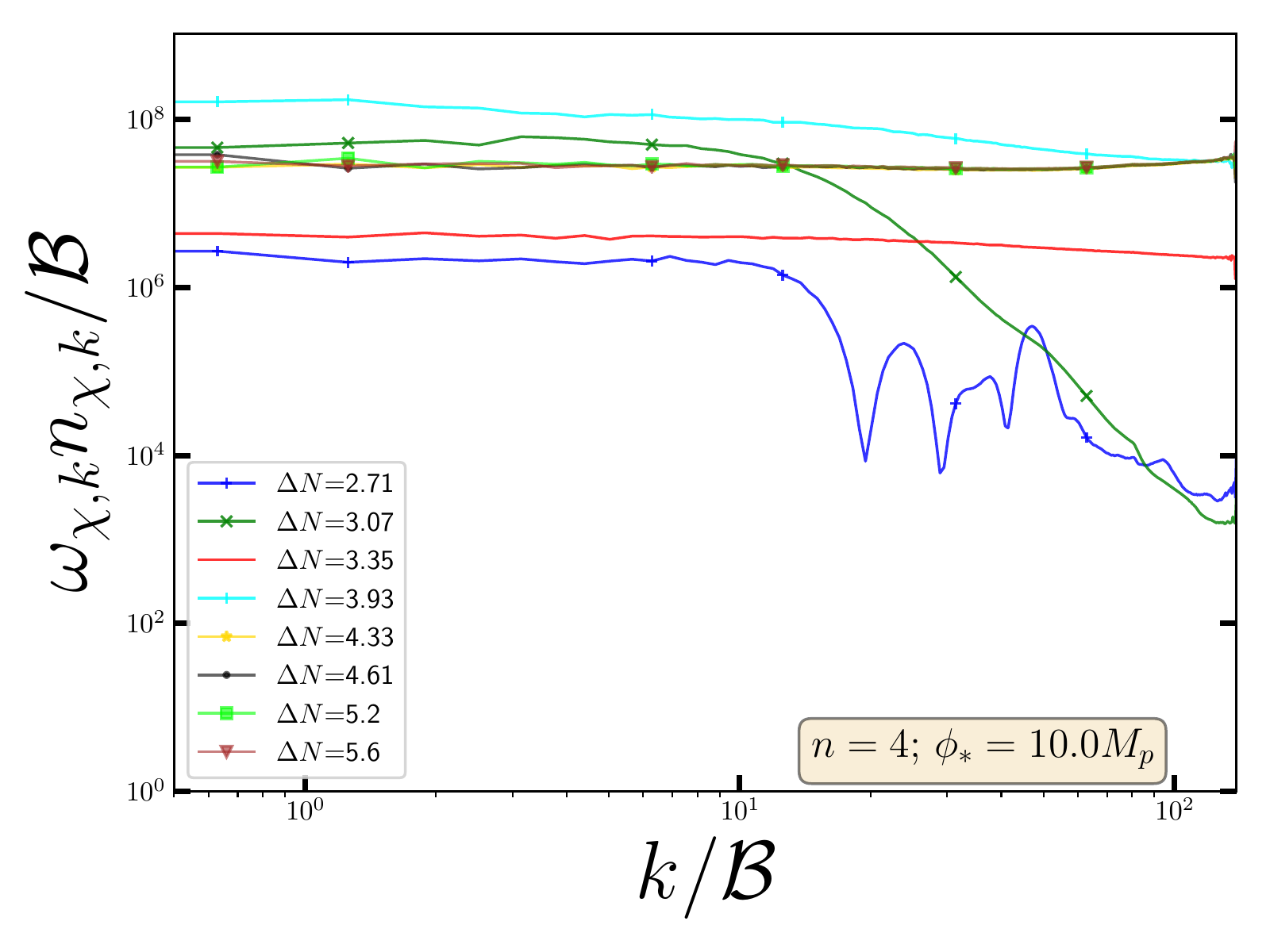}}
	\subfigure[$n=4$, $\phi_* = 0.1M_p$]{\includegraphics[scale=0.45]{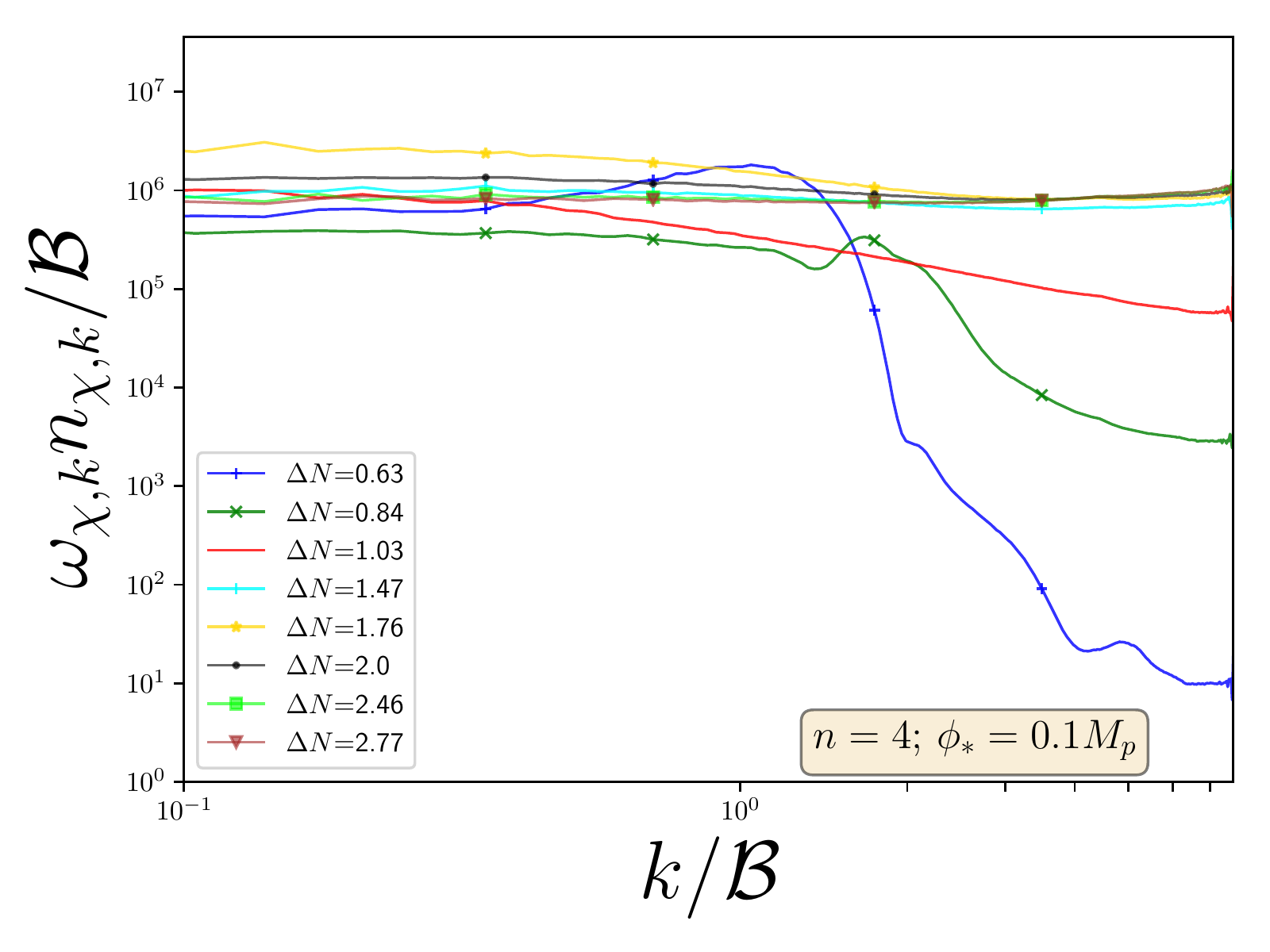}}
	\caption{\scriptsize Same plot as Fig.(\ref{fig:Phikwk_n4}) for the $\chi$ field}
	\label{fig:Chikwk_n4}
\end{figure}

\begin{figure}[]
	\centering
	\subfigure[$n=6$, $\phi_* = 1M_p$]{\includegraphics[scale=0.45]{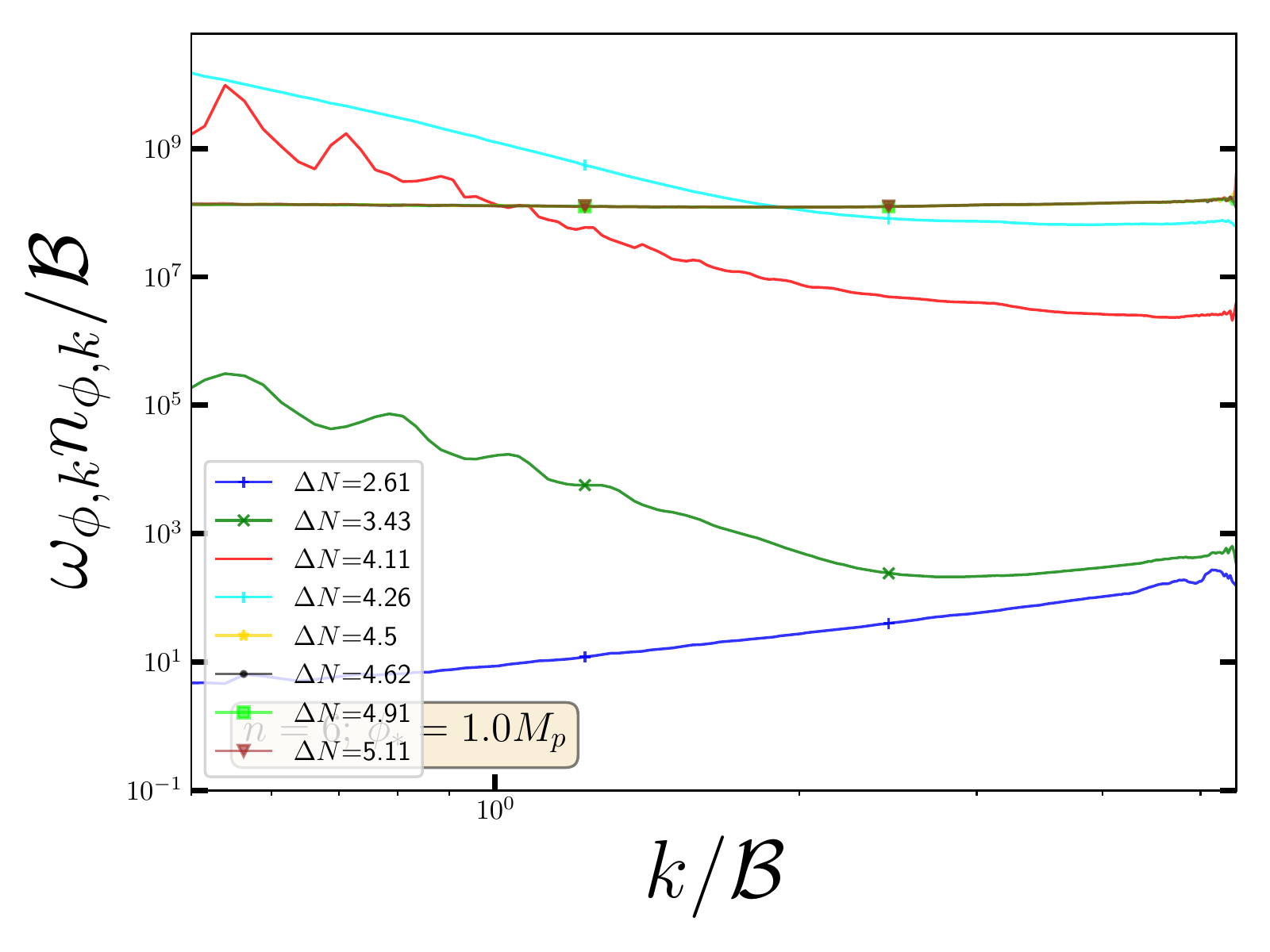}}
	\subfigure[$n=6$, $\phi_* = 0.5M_p$]{\includegraphics[scale=0.45]{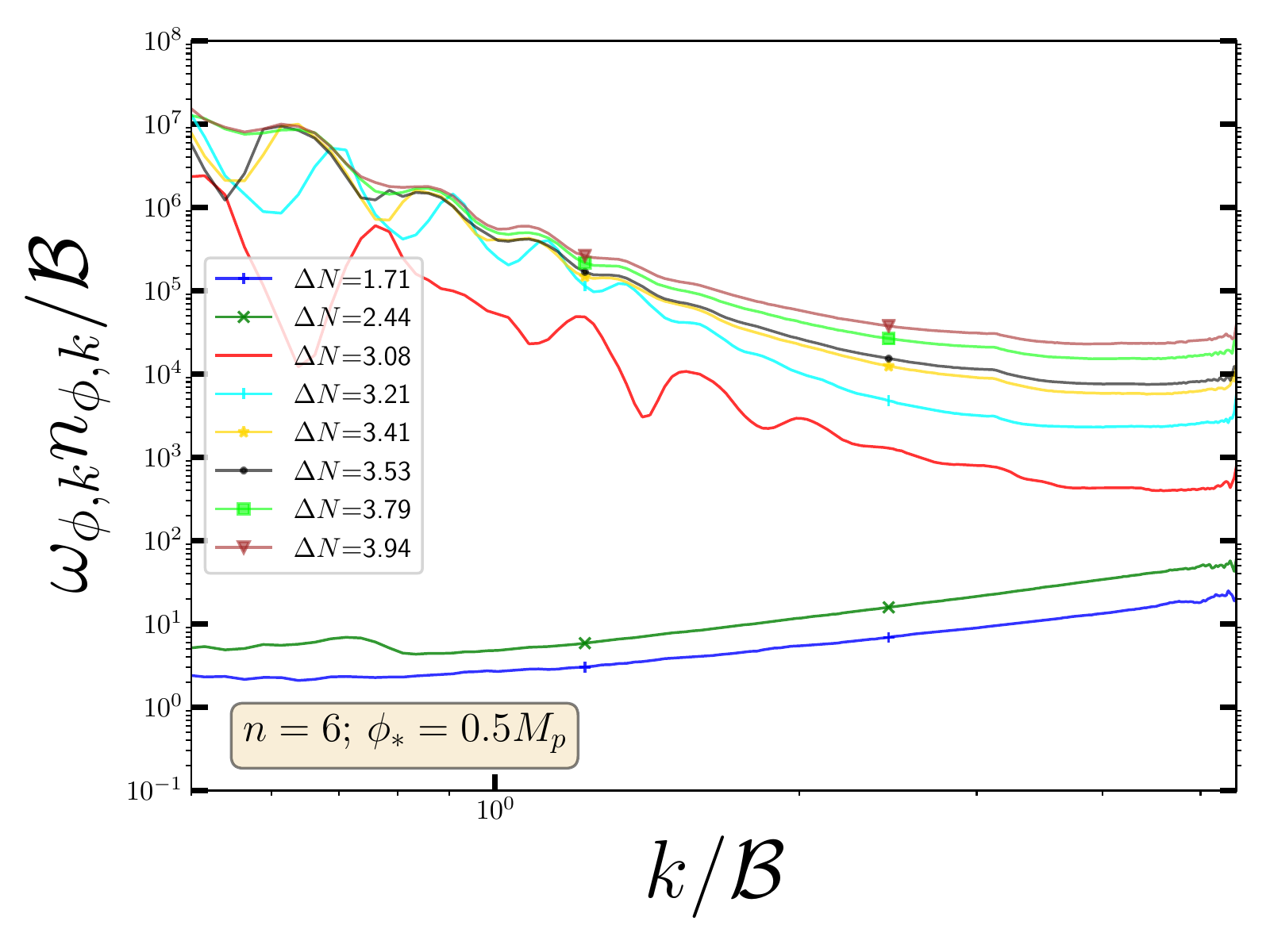}}
	\caption{\scriptsize Same plot as Fig.(\ref{fig:Phikwk_n2}) for the $n=6$. In the inset we have zoomed in the higher values of the UV regime.}
	\label{fig:Phikwk_n6}
\end{figure}

\begin{figure}[]
	\centering
	\subfigure[$n=6$, $\phi_* = 1M_p$]{\includegraphics[scale=0.45]{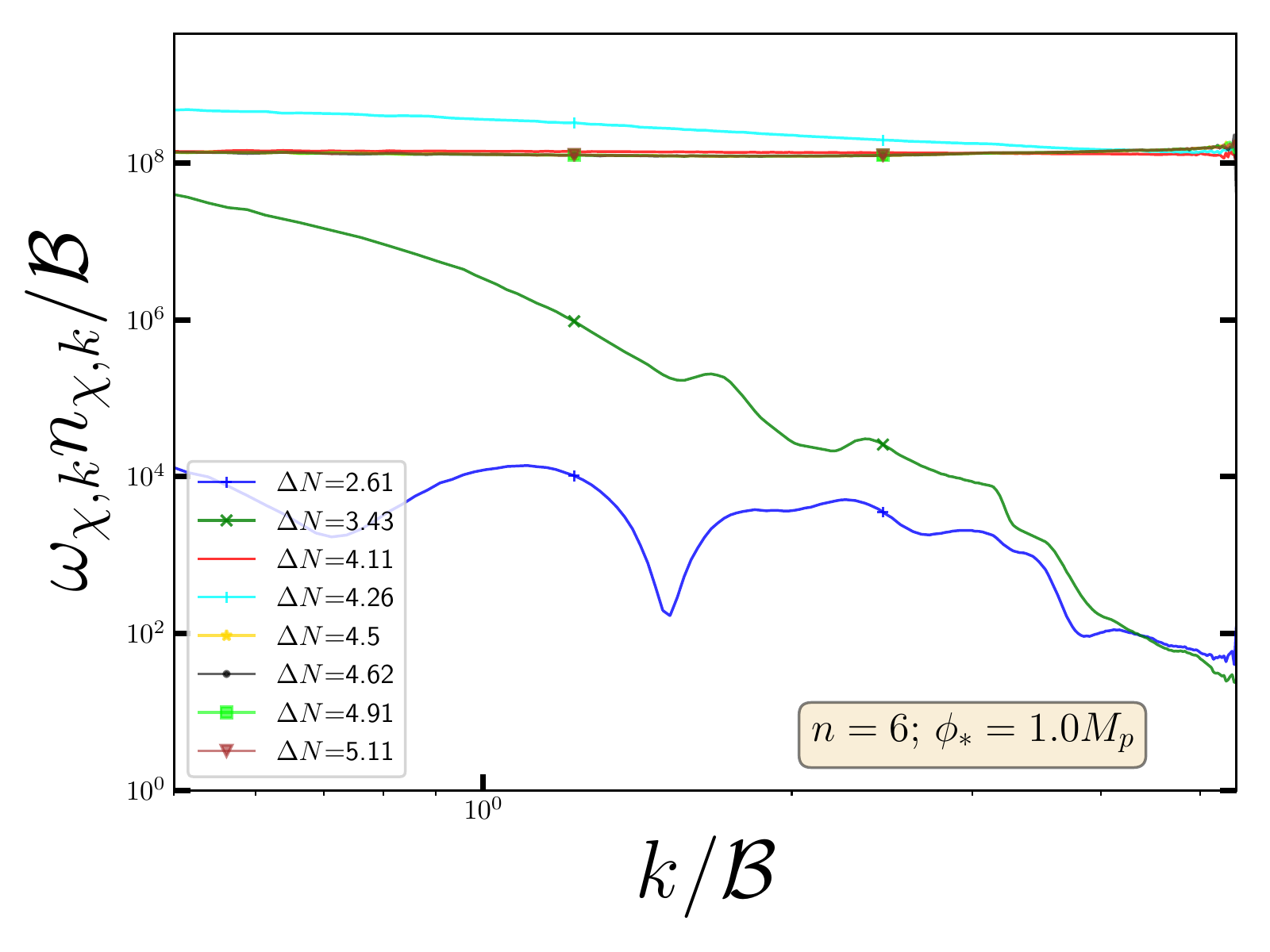}}
	\subfigure[$n=6$, $\phi_* = 0.5M_p$]{\includegraphics[scale=0.45]{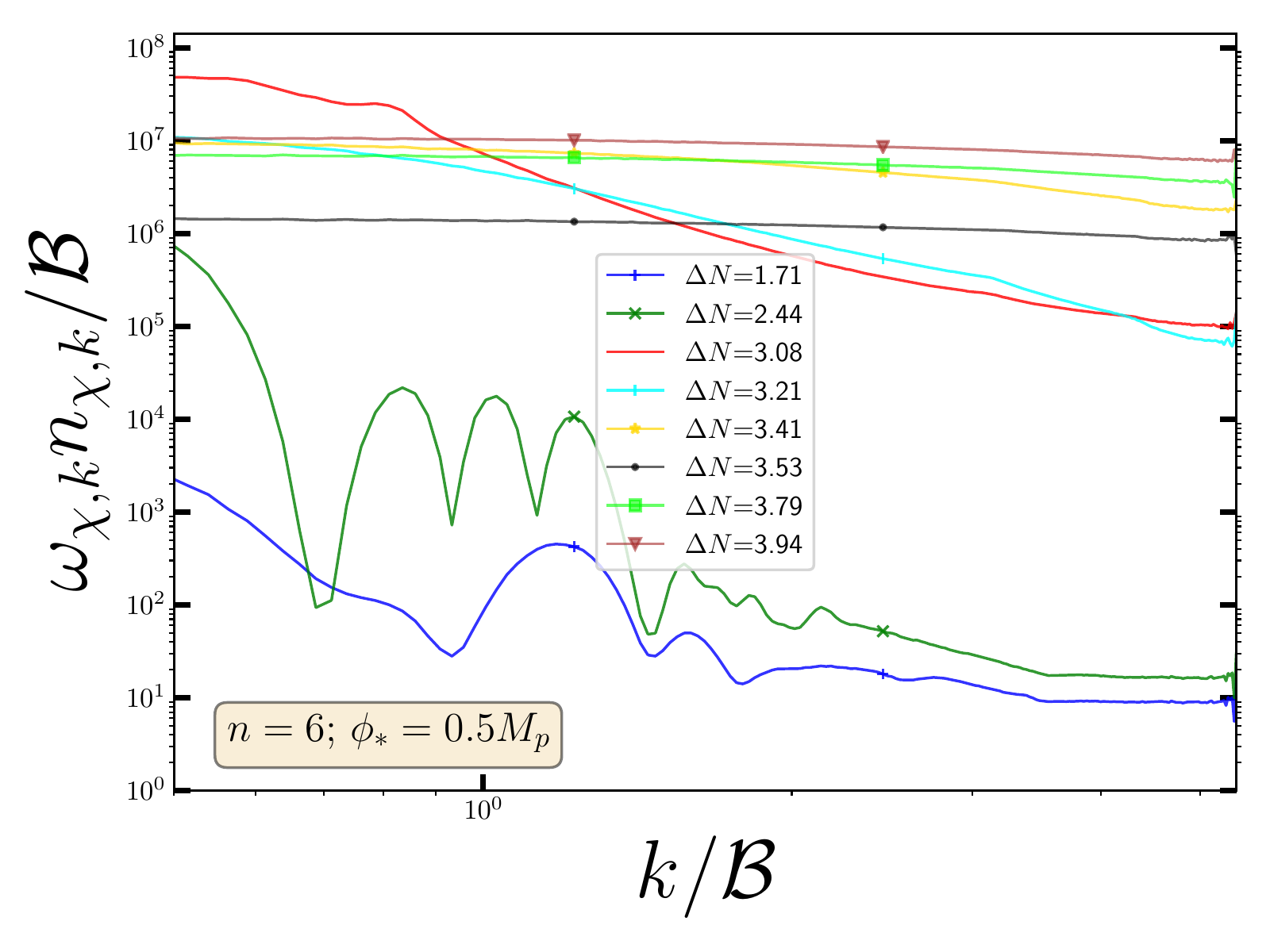}}
	\caption{\scriptsize Same plot as Fig.(\ref{fig:Phikwk_n6}) for the $\chi$ field}
	\label{fig:Chikwk_n6}
\end{figure}

Evolution of occupation number for different modes is an another important parameter that contains important information about the microscopic mechanism of the preheating as well as the details of thermalization process. In general the occupation number will be well defined if the interaction energy is negligible. It is apparent form \cref{fig:energy_n2,fig:energy_n4,fig:energy_n6} that during the whole preheating period, the interaction energy is dominated only for a very brief period of time right after the initial parametric regime where re-scattering effect is important. Therefore, the occupation number($n_k^{\phi},n_k^{\chi}$) of the fields are always well defined except for this small interval. Following\cite{Felder:2007nz,Podolsky:2005bw}, we, therefore, can understand the thermalization process across all possible modes by considering the Rayleigh-Jeans spectrum defined  by the product $n_k\omega_k\simeq T$ in the large occupation number limit which is generally true during preheating. The above Rayleigh-Jeans spectrum can be easily obtained from the following bosonic distribution function defined at finite temperature $T$ as
\begin{equation}
n_k = \frac{1}{exp\left(\frac{\omega_k - \mu}{T}\right) - 1} ,
\label{occupatonb}
\end{equation}
in small chemical potential $\mu$ limit. This implies that rather than plotting the occupation number for a particular mode, we will get a better understanding of the thermalization process by considering the combination $\omega_k n_k$ as a function of comoving wave number $k$ as plotted below in \Cref{fig:Phikwk_n2,fig:Chikwk_n2,fig:Phikwk_n4,fig:Chikwk_n4,fig:Phikwk_n6,fig:Chikwk_n6} for different values $n$. We have, as usual, chosen the previous two values of the controlling scale $\sphi$. The e-folding instants are chosen to cover all the three different stages of preheating discussed earlier.
%
%Next, we note that the energy density of the quanta at momentum $k$ and $k~+~dk$ is $4\pi k^2 \omega_k n_k$.plot the combination $k^2 \omega_k n_k$ as function of $k$.
From the dynamics of occupation number over time, we clearly observes that with increasing $n$, the system thermalize faster. Therefore, it would be interesting to understand this turbulent phase more closely.

For model with $n=2$, during the initial linear regime of parametric resonance, the (infra-red)IR modes are populated first. After the stationary phase, the IR modes shows a greater tendency for thermalization. Nevertheless the overall spectra shows that the thermalization has not been achieved. Decreasing $\sphi$ do not improve the thermalization as we have seen earlier from the study of equation of state. For $n=4$ the spectra after stationary phase evolve towards higher comoving momenta. But the spectra do not show thermalized behavior for $\sphi=10M_p$. Decreasing $\sphi$ shows a greater tendency for thermalization. For models with $n=6$, the spectra is mostly flat after the initial linear stage indicating the achievement of thermalization. 

\begin{figure}[]
	\centering
	\subfigure[$n=2$, $\phi_* = 10M_p$]{\includegraphics[scale=0.4]{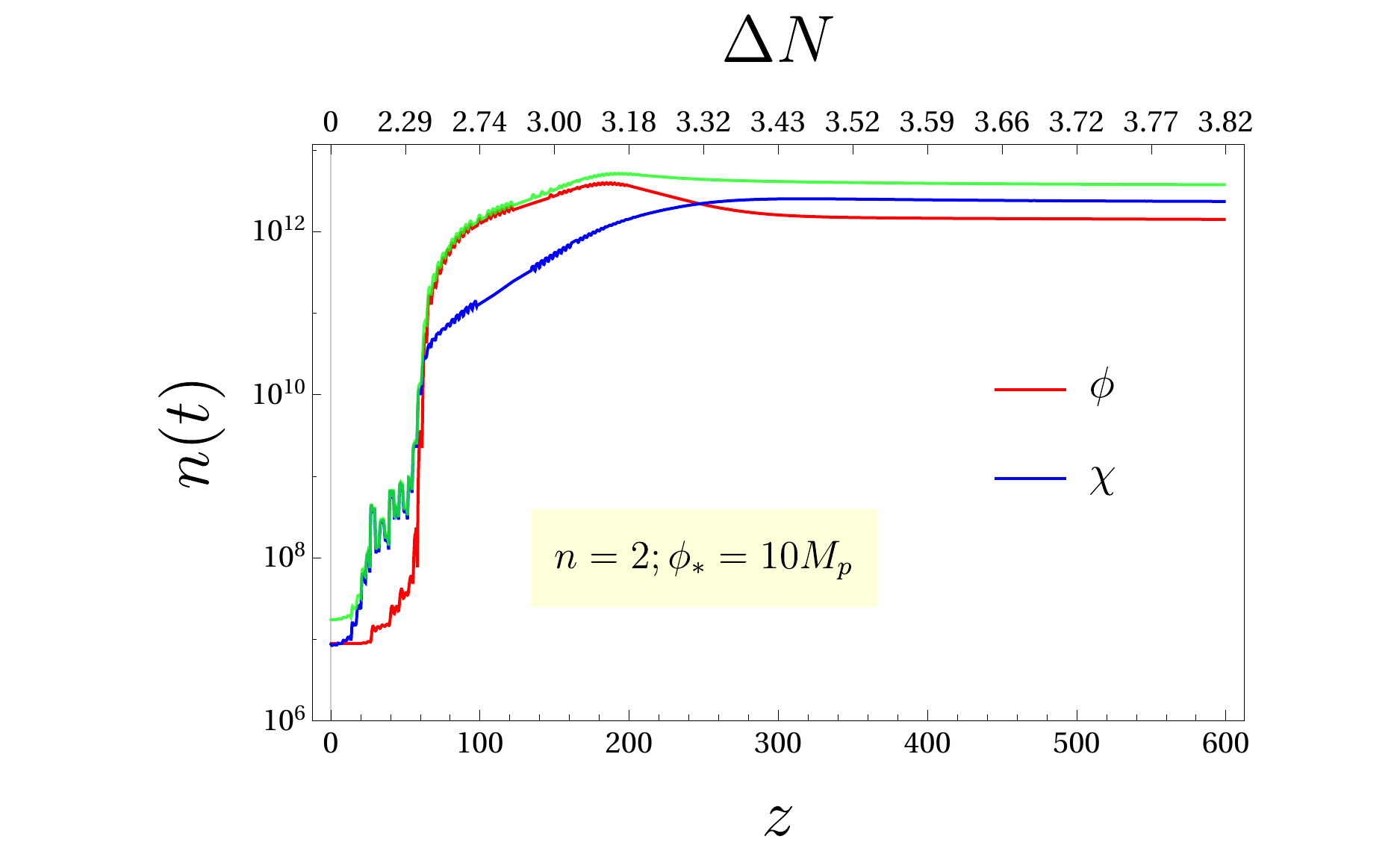}}
	\subfigure[$n=2$, $\phi_* = 0.1M_p$]{\includegraphics[scale=0.4]{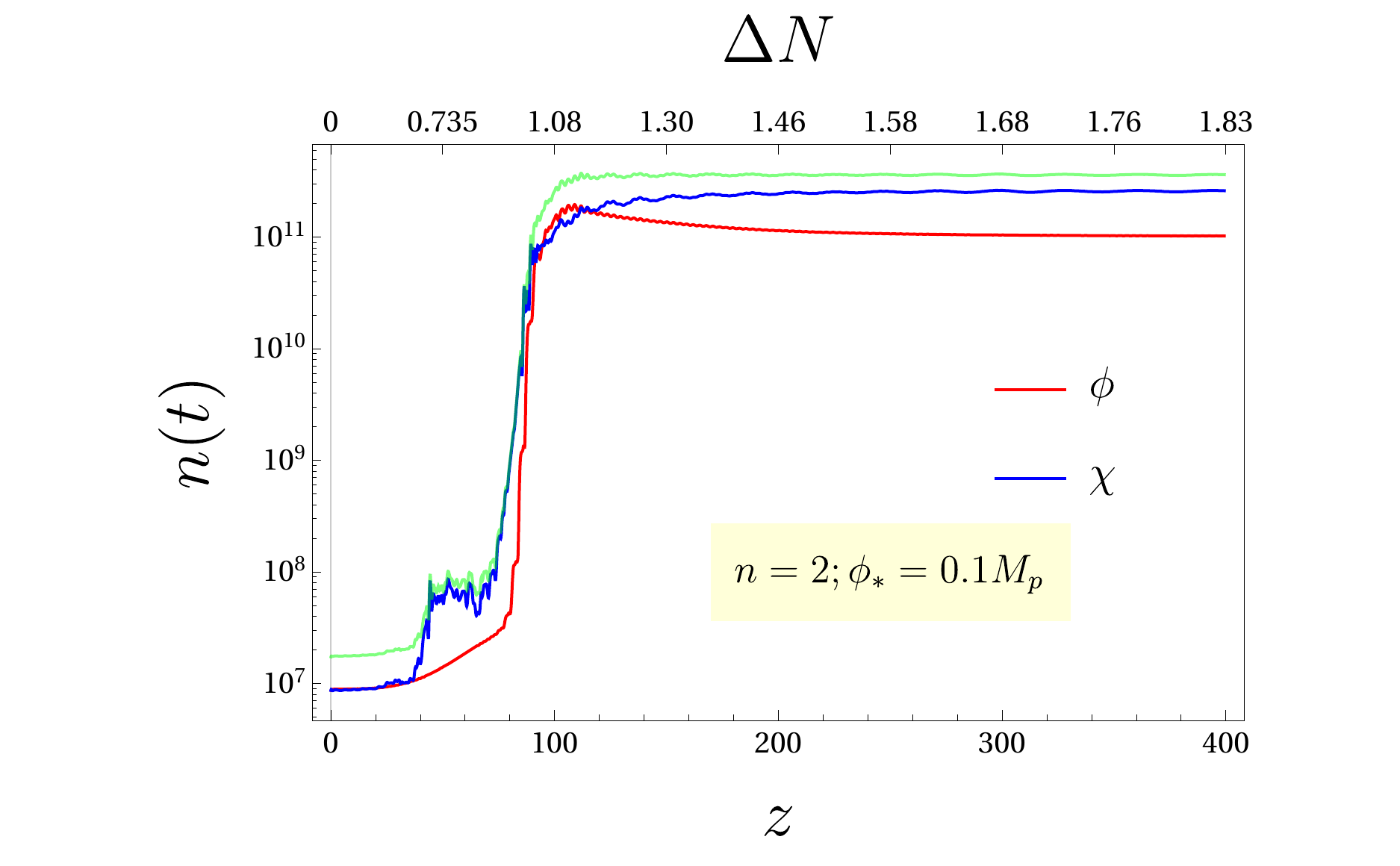}}
	\caption{\scriptsize Total number density for $n=2$ with $\sphi=(10M_p,~0.1M_p)$ Red, blue and green lines corresponds to $n_t^{\phi}$, $n_t^{\chi}$ and $(n_t^{\phi} + n_t^{\chi})$ respectively.}
	\label{fig:ntotal_n2}
\end{figure}

\begin{figure}[!h]
	\centering
	\subfigure[$n=4$, $\phi_* = 10M_p$]{\includegraphics[scale=0.4]{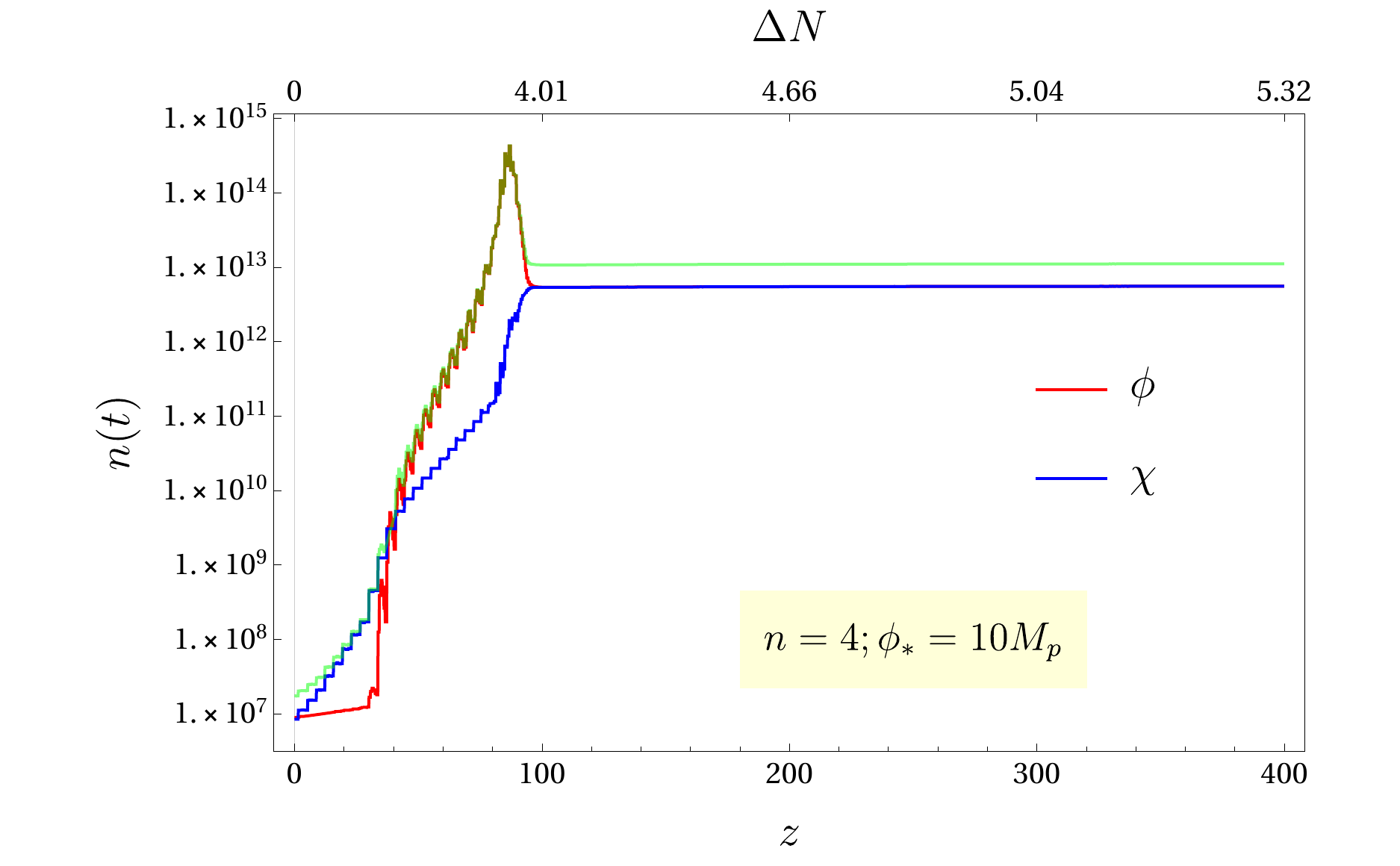}}
	\subfigure[$n=4$, $\phi_* = 0.1M_p$]{\includegraphics[scale=0.4]{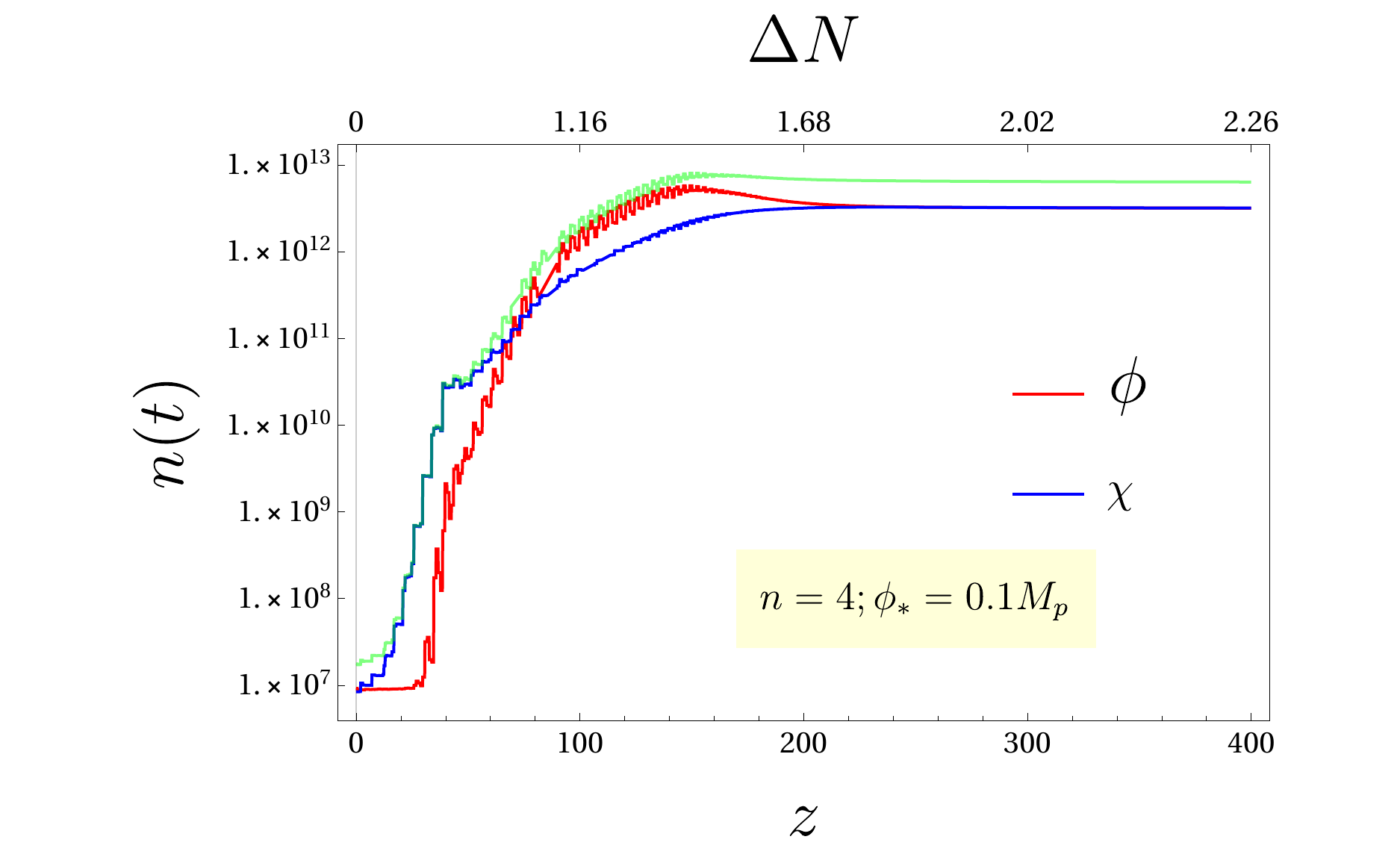}}
	\caption{\scriptsize Total number density for $n=4$ with $\sphi=(10M_p,~0.1M_p)$}
	\label{fig:ntotal_n4}
\end{figure}

\begin{figure}[!h]
	\centering
	\subfigure[$n=6$, $\phi_* = 1M_p$]{\includegraphics[scale=0.4]{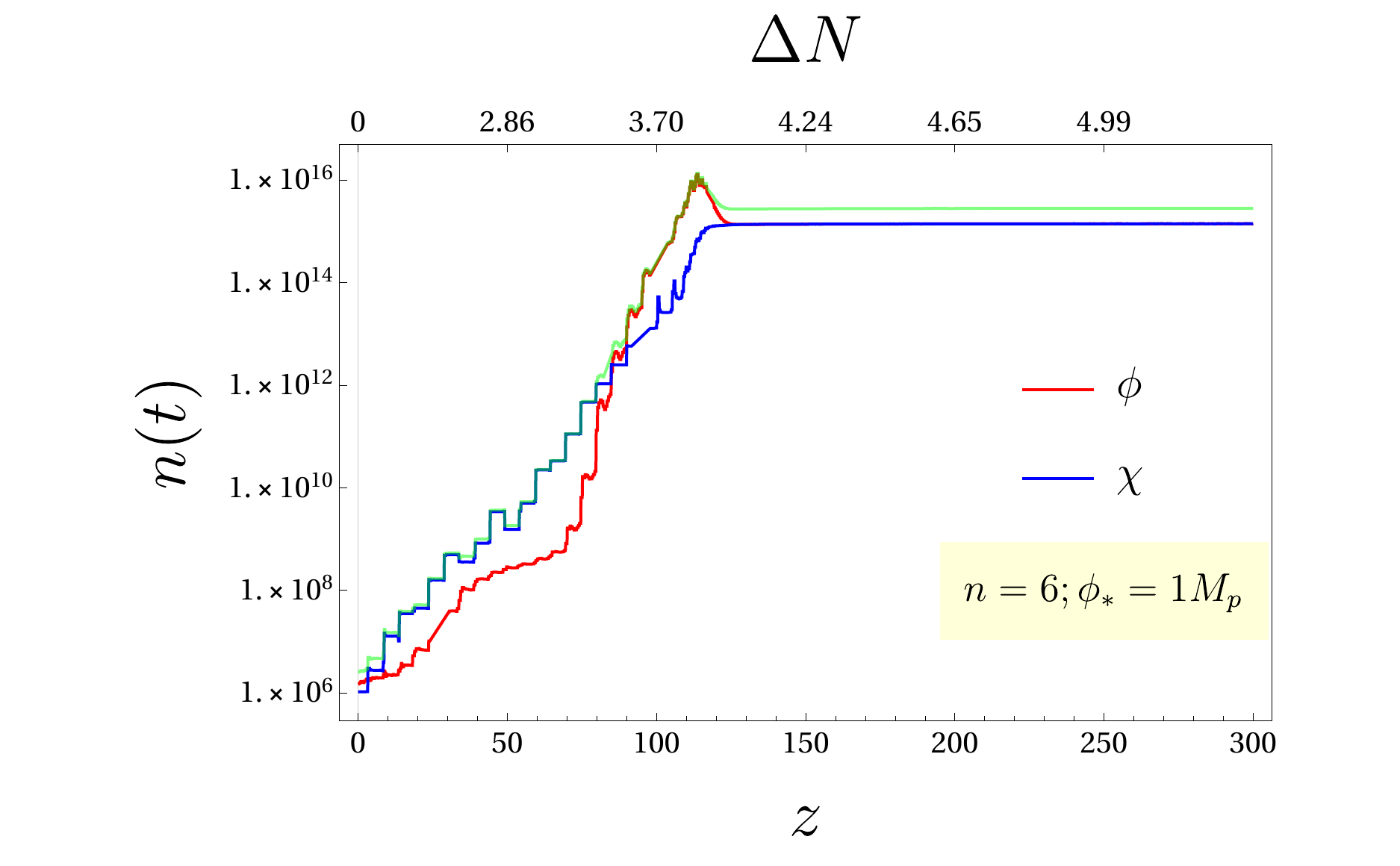}}
	\subfigure[$n=6$, $\phi_* = 0.5M_p$]{\includegraphics[scale=0.4]{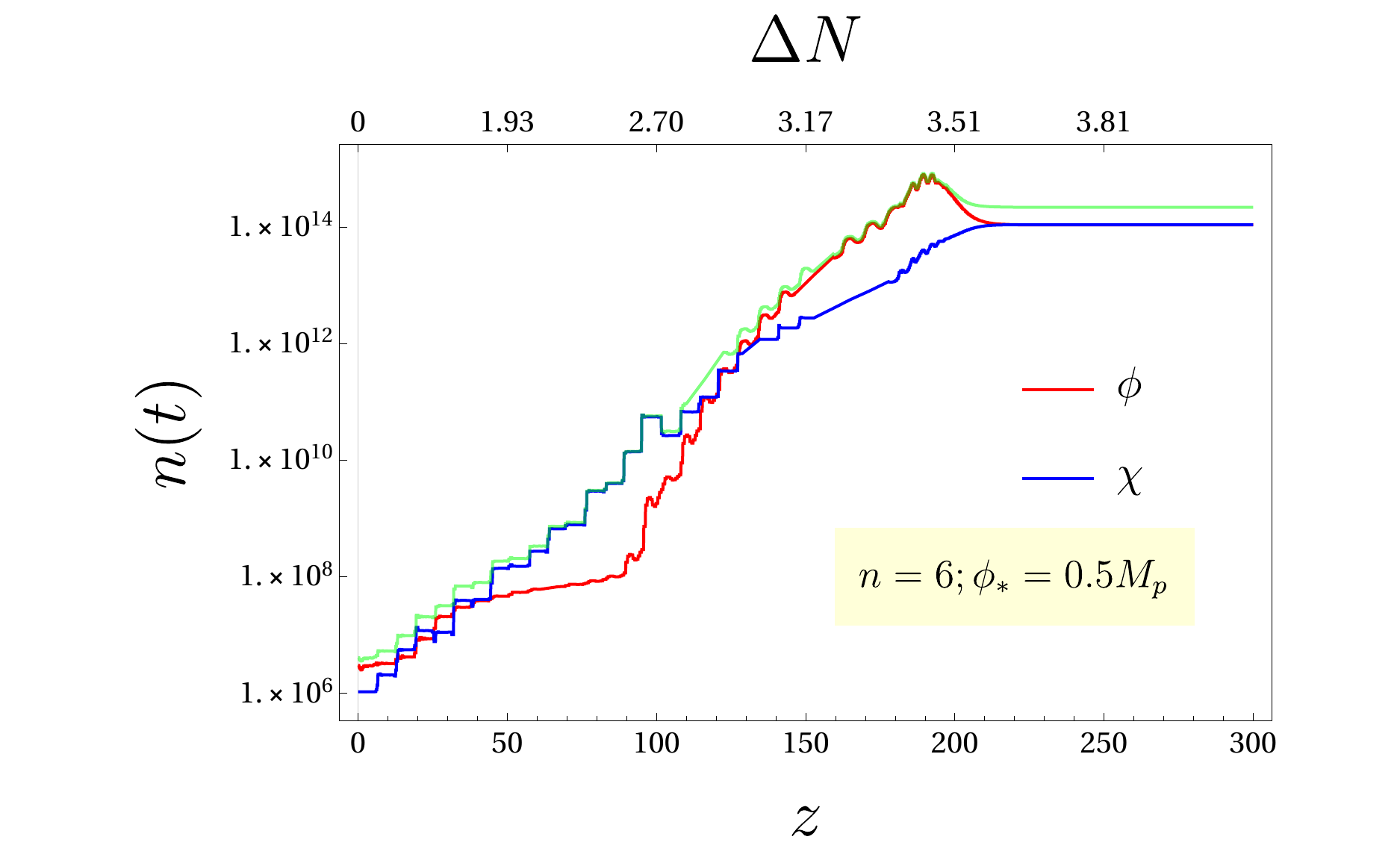}}
	\caption{\scriptsize Total number density for $n=6$ with $\sphi=(1M_p,~0.5M_p)$}
	\label{fig:ntotal_n6}
\end{figure}

Finally, we have plotted the total number densities $n(t)$ of the two fields defined in \Cref{eq:ntotal} for the models in \cref{fig:ntotal_n2,fig:ntotal_n4,fig:ntotal_n6}. Evolution of total number of particles for a particular species encodes the information about the different mechanisms responsible for changing the particle number. For example at small coupling when the particle number is small the perturbative quantum scattering process $(\phi\chi \to \phi\chi)$ conserves the total particle number. This particular phase can be well described by classical wave scattering known as ``weak turbulence'' \cite{Felder:2000hr}. 
The weak turbulence is generally refereed to a phenomena of freely propagating energy cascade when the kinetic description is applicable, and the total particle number remain conserved which is the static phase in our $(n(t)~vs~ z)$ curve. However, before this static phase, in the intermediate regime 
	when the particle number is very large during preheating, despite small coupling, the higher order interactions generically dominates, and the system stays in strong turbulence regime where total particle number will not be conserved. All   
	 these phenomena has already been widely studied in the context of reheating dynamics \cite{Micha:2002ey,Micha:2004bv,Felder:2000hr}, whose late time behavior is characterized by a turbulent and self-similar evolution of distribution functions towards equilibrium.

The particle numbers initially increases exponentially due to parametric resonance which is followed by a gradual decrease in the turbulent regime and finally settling to an asymptotic flat regime during the stationary phase. The gradual decrease is due to the fact that during the turbulent regime the occupation number shifts from low to higher momentum thereby decreasing the overall number. 
For our two non-linearly coupled system, when the same is in the non-linear regime where mode-mode coupling $X X \phi \rightarrow \int dk' dk'' X_{k'} X_{k''}  \phi_{k-k'-k''}$ in the momentum space becomes non-negligible, energy flow towards UV modes should also have compensating opposite flow for $k\rightarrow k'+k''$ decay channel. \cite{Felder:2000hr,Micha:2002ey,Micha:2004bv}. However, as is obvious from the numerical constraints coming from the finite size 3D lattice box, such a flow towards IR modes are absent in the lattice simulation due to the finite size of the box. For, $n=2$, the total number for the two fields do not became identical at the end of stationary phase. For, $n=4$ decreasing $\sphi$ make the two spectra identical. For $n=6$ the total number for the two fields reaches the same value after the stationary phase for both the value of $\sphi$. This feature is consistent with our previous conclusion that for $n=2$ model, the system is not fully thermalized at the end of preheating.

%\subsection{Inflaton fragmentation}

\subsection{\label{sec:pert}Perturbative reheating and constraints from CMB}
The purpose of the reheating phase is to create the right initial conditions for the standard big-bang. Therefore, after the end of reheating the final dominating energy component should be radiation with the characteristic equation of state $w=1/3$. As we have seen for models with $n>2$, the preheating is sufficient to obtain the final equation of state of the system to be radiation like. However, it is important to remember that final state of preheating is a combination of two radiation like fluids with approximately $50\%$ of it is the inflaton field itself. Therefore, even though it behaves like a radiation, in order to connect with the CMB, one needs to consider further decay of inflaton into radiation by perturbative processes. In any case for $n=2$ models we have seen that the final equation of state is not that of radiation. For long time simulation we have seen that $w \to 0.1$ for $n=2$ model. As we have mentioned that we need further perturbative decay of inflaton to complete the reheating and set the correct initial conditions for the big-bang. We have already mentioned before that to obtain the radiation domination, a thee-legs interaction such $\phi\chi^2$ will be important. The full numerical lattice simulation considering both the interaction term will be considered in our subsequent paper. At this point we may attempt an alternative approach that will also help us to connect the reheating phase with present CMB data. We will follow the formalism developed in our recent papers \cite{Maity:2017thw,Maity:2018dgy} generalizing the works in \cite{Martin:2010kz,Dai:2014jja} for decaying inflaton. For recent studies on CMB constraints on reheating phase see\cite{Cook:2015vqa,Ellis:2015pla,Ueno:2016dim,Eshaghi:2016kne,DiMarco:2017zek,Bhattacharya:2017ysa,Drewes:2017fmn,DiMarco:2017zek,Cabella:2017zsa,DiMarco:2018bnw}. In our following analysis, we will assume that the decay of total energy density $\rho_t$ whose initial value will be identified with the value obtained after the end of preheating is via a phenomenological decay term. The perturbative reheating due to such a decay term was first considered in the initial treatments of reheating\cite{Albrecht:1982mp,Dolgov:1982th,Abbott:1982hn,Traschen:1990sw}. For $n=2$, we take the value equation of state at the onset of stationary phase which is $w_{dec} \sim 0.22$. Now introducing the phenomenological decay term for total energy decay we may write
\begin{align}
 \dot{\rho}_{t} + 3H(1+w_{dec})\rho_{\phi} + \Gamma\rho_{t} &= 0,\nno\\
\dot{\rho}_R + 4H\rho_R  - \Gamma\rho_{t}&=0,
\label{eq:rho_phi}
\end{align}
where $\rho_R$ is the additional radiation energy density. 

%For example the decay width can be effectively identified with a three-point interaction of the form $2g^2\sigma\phi\chi^2$ with a tree-level inflaton mass $m_{\phi}$ is found to be\cite{Peskin:1995ev}
% \begin{equation}
%  \Gamma_{\phi\to\chi\chi} = \frac{g^4\sigma^2}{8\pi m_{\phi}}
% \end{equation}

In terms of following dimensionless variables 
\begin{align}
\Phi = \frac{a^{3(1+w_{\rm dec})}\rho_{t}}{a_I^{3(1+w_{\rm dec}}) \rho_I};~~~~~R = \frac{a^4\rho_R}{a_I^3\rho_I} ,
\label{eq:rescaled_vars}
\end{align}
the above equations for energy densities transformed into 
\begin{align}
\Phi'(N) + \frac{\Gamma}{\mathbb{H}}\Phi(N) &= 0,\\
R'(N) - \frac{\Gamma}{\mathbb{H}}e^{(1-3w_{dec})N}\Phi(N) &= 0 ,
\label{eq:Boltzmann1}
\end{align}
where ``prime'' is taken with respect to e-folding ``N'' which is obtained from the Hubble equation 
\begin{equation}
3M_p^2\mathbb{H}^2 = \rho_I\left[\frac{\Phi(N)}{e^{3(1+w_{dec})N}} + \frac{R(N)}{e^{4N}} \right] 
\end{equation}
This equation could be easily solved for a given inflaton decay term.

As mentioned, one of our main goal is to understand the direct constraints coming from the CMB. For a given cosmological scale $k = a_kH_k$ which exits the horizon inflation with scale factor $a_k$ and re-enters the horizon with the scale factor $a_0$ at the present time, satisfies the following relation,
%\begin{align}
%ln\left(\frac{k}{a_k H_k}\right) = ln\left(\frac{a_{end}}{a_k}\frac{a_{pre}}{a_{end}}\frac{a_{re}}{a_{pre}a_0}\frac{a_0}{a_{re}}\frac{k}{H_k}\right) = 0,
%\label{eq:scales}
%\end{align}
\begin{equation}
 N_{k} +  N_{\rm pre} + N_{\rm re}^{\rm pert} + ln\left(\frac{a_0}{a_{\rm re}} \right) + ln\left(\frac{k}{a_0 H_k}\right) = 0,
\label{eq:rel0}
\end{equation}
where $(N_k, N_{\rm pre}, N_{\rm re}^{\rm pert})$ are the inflationary, preheating and perturbative reheating e-folding number respectively. $a_{re}$ is the scale factor after the end of perturbative reheating. Assuming that the entropy is preserved after reheating implies
\begin{align}
g_{re} T_{re}^3 = \left( \frac{a_0}{a_{re}}\right)^3\left(2T_0^3 + 6\times\frac{7}{8}T_{\nu,0}^3 \right) ,
\label{eq:rel1}
\end{align}
where, $T_0 = 2.725K$ is the present CMB temperature, $T_{\nu,0} \sim (4/11)^{1/3}T_0$ is the present neutrino temperature and $g_{re}$ is the effective degrees of freedom at reheating. Using the usual definition of radiation temperature, $T_{re} = ({30}/{g_{re}\pi^2})^{1/4}\rho_I^{1/4}\left({R_{re}/{e^{4N_{re}}}}\right)^{1/4}$
one arrives at the following relation two equation for reheating temperature $T_{re}$ and $N_{re}$ in terms of other known parameters,
\begin{align}
T_{re} =& \left( \frac{43}{11g_{s,re}}\right)^{1/3}\left( \frac{a_0T_0}{k}\right)H_k e^{ N_k}e^{ N_{pre}} e^{ N_{re}^{pert}}  \nonumber\\
N_{re}^{pert} =& \ln \left(\frac{a_{re}}{a_{pre}} \right) .
\label{eq:temperature1}
\end{align}
Where, $a_{pre}$ is the scale factor after the end of preheating. 
We will solve equation \ref{eq:Boltzmann1} keeping the effective decay width as free parameter. The value of the effective decay width is chosen such that $N_{re}^{\rm pert}(\Gamma), T_{re}(\Gamma)$ satisfies Eqs.\ref{eq:temperature1}.
% \begin{figure}[t]
% 	\centering
% 	\subfigure[$n_s~Vs~\mathrm{N}_{\mathrm{re}}$]{\includegraphics[scale=0.6]{nsVsNren2}}
% 	\subfigure[$n_s~Vs~\mathrm{N}_{\mathrm{re}}$]{\includegraphics[scale=0.6]{nsVsTren2}}
% 	\caption{\scriptsize Evolution of $N_{\rm re}(=N_{\rm pre} + N_{\rm re}^{\rm pert})$ and $T_{\rm re}$ with primordial spectral tilt. The dashed lines corresponds to the standard approach when the reheating phase is parametrized by an equation of state parameter $w_{\rm eff} = 0$ while the solid lines are from the calculation where the initial efolding number is from preheating($N_{\rm pre}$).}
% 	\label{fig:decay}
% \end{figure}

\begin{figure}[t]
	\centering
	\subfigure[$n=2~\sphi=10{\rm M_p}$]{\includegraphics[scale=0.3]{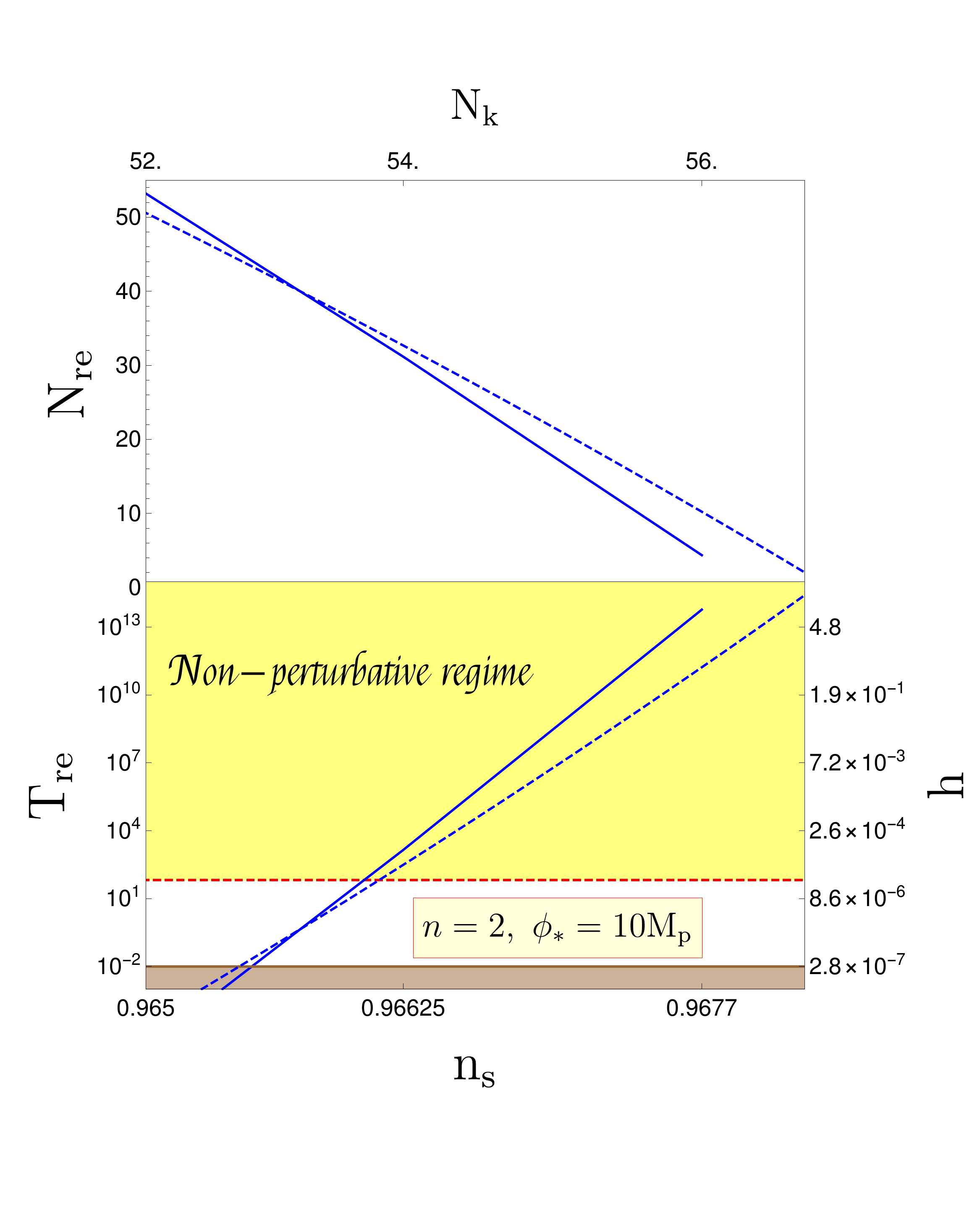}}
	\subfigure[$n=2~\sphi=0.1{\rm M_p}$]{\includegraphics[scale=0.3]{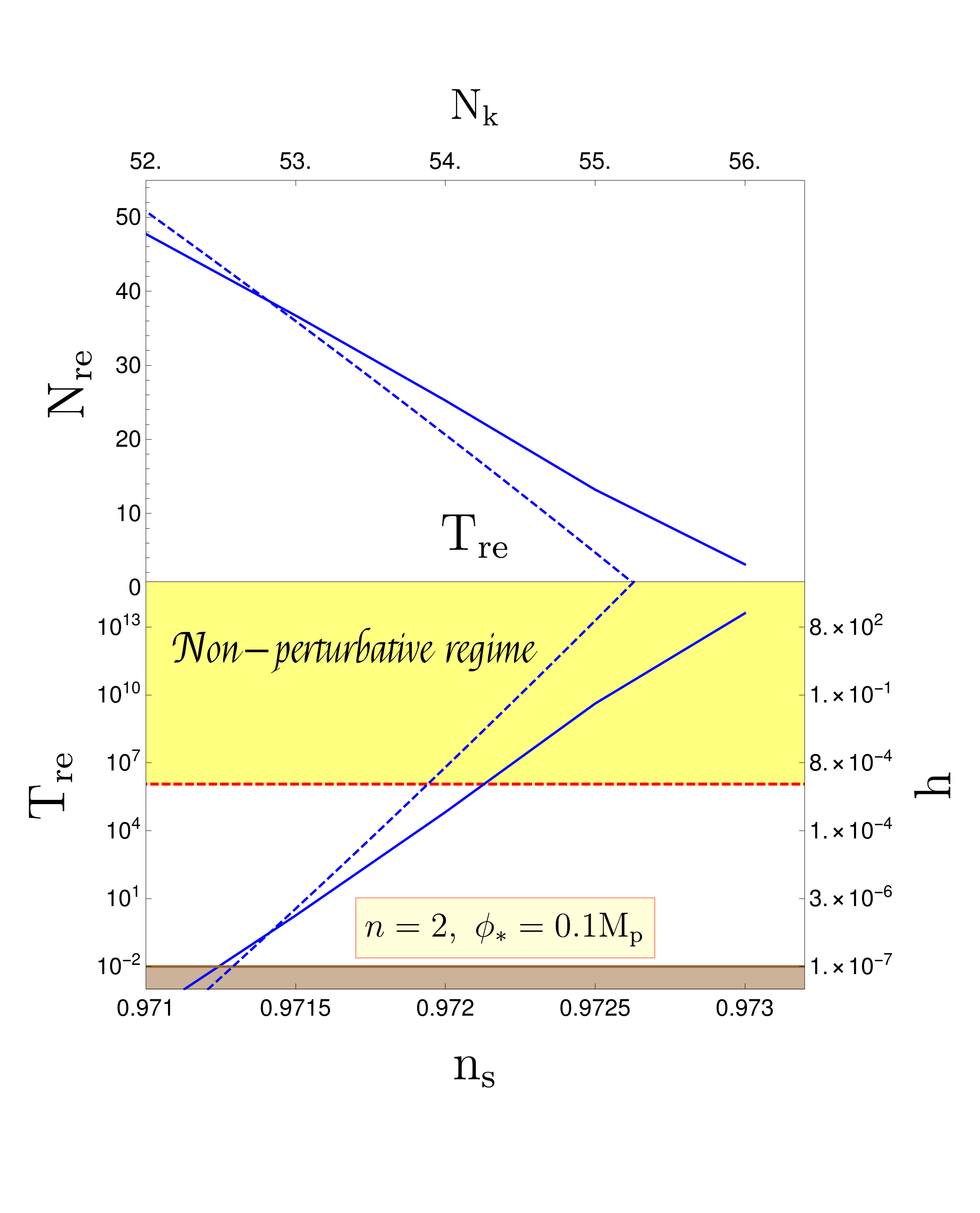}}
	\caption{\scriptsize Evolution $N_{\rm {re}}$ and $T_{\rm {re}}$ with $n_s$ for $n=2$ with two values of $\sphi$ as before. The shaded region is the region when the value of $h$ such that the non-perturbative effects will be important. The dashed lines are the predictions from the conventional approach when the expansion during reheating phase is parameterized by an effective equation of state parameter $w_{eff} =(0.195,~ 0.251)$ for $\phi_* =(10, 0.1) M_p$. The lower brown shaded region is below $10\rm{ MeV}$ that is excluded from the constrains of big bang nucleosynthesis.}
	\label{fig:nsnretre}
\end{figure}

The preheating and consequently the initial conditions for the perturbative reheating dynamics modeled by eqs.\ref{eq:rho_phi} (fraction of inflaton energy density) turned out to be largely independent of the inflationary parameters $(n_s,N_k)$ for a particular model. For $n=2$ model, this important fact fixes the value of $N_{pre} \sim 1.5$ for $\phi_*=0.1$, and $N_{pre} \sim 3.5$ for $\phi_*=10$. Therefore, subsequent dynamics will fix the value of $N_{re} = N_{pre}+ N_{re}^{pert}$ supplemented by the conditions eqs.\ref{eq:temperature1}. Since the reheating temperature is an exponential function of e-folding number, small change in $N_k$ or $n_s$ significantly effects the value of $(N_{re}, T_{re})$. 
In fig.\ref{fig:nsnretre}, we have shown the dependence of $N_{\rm re}$ and $T_{\rm re}$ on the spectral index $n_s$ for the two values of $\sphi$. The dashed lines are the results from the conventional reheating constraint analysis in which complete dynamics of reheating phase is parameterized by an effective equation of state \cite{Martin:2010kz,Dai:2014jja}(for $n=2$). The solid lines is the result from our calculation. We can clearly see the significant difference in  $(T_{re},N_{re})$ for a particular value of $n_s$. The total reheating phase is parametrized by the sum of e-folding numbers due to preheating $N_{\rm pre}$ and the perturbative reheating $N_{\rm re}^{\rm pert}$. In our case, the instantaneous reheating is thus not possible as for $N_{\rm re}^{\rm pert} \to 0$, $N_{\rm re} \to N_{pre}$. Most importantly  preheating dynamics restricts the value of $n_s$ within a very narrow range of $0.971 < n_s < 0.973$ for $\phi_*=0.1~ M_p$. For $\phi_*=10~M_p$, the range approximately is $0.9658 < n_s < 0.9678$. For both the cases the reheating temperature can take a wide range of values with a maximum limit to be $T_{re}^{max} \simeq 10^{13}\sim 10^{16}$ GeV. All these constraints are based on our naive solution of Boltzmann equation. 

However, let us remind the reader that inflaton decay constant $\Gamma$ should not take arbitrary value. For a given interacting model it will have a theoretical constraints. Those are shown as shaded region in the $(n_s, T_{re})$ plot for the given interaction discussed below. For illustration, if we consider a particular Yukawa type interaction between the scalar components and a fermion 
\begin{equation}
\mathcal{L}_{\rm int} \supset -h\phi\bar{\psi}\psi. 
\label{eq:yukawa}
\end{equation}
We may have a estimate the range of coupling when the perturbative treatment can be trusted. Although we have considered a specific interaction, the following discussions will same for other type of three point interaction with some quantitative differences. In the present case the decay rate will be given by\cite{Peskin:1995ev}
\begin{equation}
\Gamma = \frac{h^2 m_{\phi}}{8\pi}
\label{eq:gamma}
\end{equation}
where $m_{\phi}$ is the tree-level mass of $\phi$.
However, in order for our discussion to be valid we need to make sure that the effect of this coupling should be insignificant during the preheating phase. For this we note that, the effect of resonance will be appreciable when the decay rate is greater than the Hubble parameter at the beginning of preheating\cite{Kofman:1997yn,Drewes:2017fmn}. i.e., 
\begin{equation}
q_f^2 m_{\phi}~ \textgreater~ H.
\label{eq:qf0}
\end{equation}
Where $q_f$ is the resonance parameter in case of fermionic preheating analogous to the resonance parameter $q$ appeared in \cref{Hilleq1} and for the interaction term in \cref{eq:yukawa} is given by\cite{Greene:1998nh} 
\begin{equation}
q_f = \frac{h^2\phi_0^2}{m_{\phi}^2}
\label{eq:qf1}
\end{equation}
Combining \cref{eq:qf0} and \cref{eq:qf1}, we obtain the condition on the coupling $h$ when the resonance will be effective as
\begin{equation}
h > \Bigg[\frac{V(\phi_0)^{\frac{1}{2}} m_{\phi}^3}{3^{\frac{1}{2}}{\rm M_P}\phi_0^4} \Bigg]^{\frac{1}{4}}
\end{equation}
This lower limit on $h$ in turn will give us a lower limit in inflaton decay constant $\Gamma$. In \cref{fig:nsnretre} the red dashed line correspond to the aforementioned limit in terms of reheating temperature. Therefore, a significant part of our result that is represented by solid blue lines is in the non-perturbative region, which we may not be trusted. However, one can make progress by fitting our results (blue lines) with an effective time-independent equation of state defined during the entire reheating phase following the standard approach in\cite{Martin:2010kz,Dai:2014jja}, with a simple evolution equation of total energy density $\rho =\rho_{0} a^{-(1+3 w_{eff})}$, and the approximate result (dotted blue lines) turned out to be $w_{\rm eff}=(0.195,~0.251)$ for $\sphi=(10{\rm M_p}~0.1{\rm M_p})$ respectively. Now the dotted lines can be valid in the non-perturbative regime shown in the \cref{fig:nsnretre} with an effective equation of state. To this end we must mention that the origin of the decay term in \cref{eq:rho_phi} is phenomenological thus we are free to chose any interaction terms thereby the bound shown in \cref{fig:nsnretre} can be significantly modified due to  different coupling. In this regard it is also worth mentioning that a combination of perturbative and non-perturbative decay may in fact complicate the preheating scenario as considered in \cite{GarciaBellido:2008ab,Repond:2016sol} which the authors termed as \textit{Combined Preheating}. In this work the authors considered the Higgs Inflation. They show that if the gauge bosons are coupled to fermions, they would quickly dissipate into these lighter degrees of freedom. The onset of preheating in this case is delayed(see also\cite{Freese:2017ace}). The analytic treatment of this mechanism has been given in \cite{Bezrukov:2008ut,GarciaBellido:2008ab} for the case of Higgs inflation within the Standard Model. A full lattice simulation describing the limitations of the analytic study modeling the Higgs-gauge boson interactions as scalar interactions is presented in\cite{Repond:2016sol}. We will not consider those scenarios in this work and will take the inflation decay term is due to a phenomenological origin.
 
% For example for $\phi_*=10 M_p$, taking $N_k=55$, the perturbative reheating happens almost instantaneously with reheating temperature $T_{re}=3.1\times10^{13}$ GeV, shown in the left Fig.\ref{fig:decay}.
% Where as for $N_k=50$, the value of $N_{re}$ turns out to be $\sim 8.2$ with the reheating temperature $T_{re}=2.1\times10^{11}$ GeV, shown in the right Fig.\ref{fig:decay}. 
% It turns out that variation of $\sphi =0.1 M_p$ only change the reheating temperature due to change in initial initial energy density, however, the efolding number for completing of perturbative reheating remains nearly the same. For instance, for $\sphi=0.1M_p$ and $\Delta N_k=50$, we find that $T_{re}=4.5\times10^{11}GeV$ while the duration of perturbative reheating after the preheating($\Delta N_{re}^{pert}=8.2$) is same as in the case with $\sphi=10M_p$.

%\clearpage
\section{{\label{sec:conclusion}}Conclusion and Outlook}

In this work, we have studied the preheating dynamics for a specific class of plateau type inflationary potential proposed earlier by us. As we have shown previously, these type of plateau potentials fit will with the cosmological observations. The plateau potential in the present context can be thought of as a generalization of the chaotic power law potentials. Indeed the potential reduces to the form $V(\phi) \propto \phi^n$ around the minimum of the potential. However, we have a scale $\sphi$ in our model that controls the height and width of the potential. In the present work we have explored in detail the effect of this scale on the preheating dynamics for different inflationary model parameterized by $n=(2,4,6)$. Lowering the value of $\sphi$, the system reaches its stationary stage faster. Even though qualitative behavior remains same for all the model. Some important microscopic detail changes which are worth studying in future. In order to do a comparative study, we take the same value of $g^2$ for different $\sphi$. Below we list some of the important findings of our study,
\begin{itemize}
\item[(a)] Given a particular model with fixed $(n, g^2)$, decreasing $\sphi$ makes the energy transfer efficient by reducing the efolding number.
	\item[(b)] However at the end of preheating, the total energy is distributed almost equally among the different fields taking part in the reheating dynamics. This distribution is nearly independent of the all the model parameters. Microscopic details of the dynamics will be dependent upon the parameters. For $n=2$ this energy distribution is little asymmetric mainly due massive inflaton component. 
	\item[(c)] Most importantly for the models $n=4,6$, 
	the final equation of state turns out to be that of radiation $w=1/3$ independent of all the other parameters. Similar observation has been made in the recent works\cite{Lozanov:2016hid} considering the self resonance of the inflaton for similar plateau type inflationary models. For our case resonance will occur for both the fields $(\phi,\chi)$. We believe this is true for any model with $n>2$, and for all those models the final equation of state will be that of the radiation.
	For $n=2$ model, however, dynamics is little different. For this model self resonance is inefficient. The parameter $m$ plays the role of mass of the inflaton. Therefore, at the end of preheating, inflaton behave like a non-relativistic massive particle with a little asymmetric energy distribution with the relativistic $\chi$-particle. Therefore, the total equation state never reaches to that of radiation in the stationary phase. 
	\item[(d)] From the microscopic point of view, we found that the transition from homogeneous inflaton equation of state $w=(n-2)/(n+2)$ to that of the saturated preheating equation of state $w_{\rm dec}$ depends on $(n,\sphi)$. For $n=2$, with the decreasing $\sphi$ the transition from $w=0$ to its maximum value tend to be instantaneous. Therefore, the system must go through highly turbulent phase.  
	This transition of equation of state occurs in the phase when transfer of energy from inflaton to reheating field is efficient. Therefore, all the interesting non-equilibrium phenomena such as chaos, turbulence, thermalization will happen in this phase. Thus, homogeneous inflaton equation of state plays very important role for those non-equilibrium processes during preheating. Detailed study of this phase will be done in future.
\end{itemize}

It is apparent that the preheating itself is not the full story of the reheating dynamics. As nearly half of the total energy density is stored in the form of inflaton, we need further mechanism to get the complete transfer of inflaton energy to radiation phase, that will set the initial condition for the standard big-bang. Since we need the complete decay of inflaton, three legs interaction\cite{Podolsky:2005bw} may be important. The effect of these interaction in the preheating has been discussed in \cite{Shuhmaher:2005mf,Dufaux:2006ee,Abolhasani:2009nb}. A three legs interaction will also necessitate incorporating the self-interaction term such as $\lambda_{\chi} \chi^4$ that makes the potential bounded from below. Depending upon the value of $\lambda_{\chi}$, most of the energy density may or may not be converted into $\chi$ quanta before the back-reaction effect will became dominant\cite{Dufaux:2006ee} in the preheating dynamics. However, during the initial stage of preheating the four-legs interaction will be dominant over the three-legs. If both the interactions are present, we will have several interesting behavior of the system\cite{Dufaux:2006ee,Abolhasani:2009nb}. We will look into those effects in a separate work. 
In the present work, we have considered the usual perturbative decay of inflaton with a phenomenological decay term. This helped us to connect the reheating phase with CMB. Considering $n=2$ model, our naive analysis of Boltzmann equation limits 
the value of the spectral index  within $0.971 < n_s < 0.973$ for $\phi_*=0.1~ M_p$, and $0.9658 < n_s < 0.9678$ for $\phi_*=10~M_p$. In the above estimation, the reheating temperature is considered to be within $10 \mbox{MeV} < T_{re} < 10^{15} \mbox{GeV}$. Finally we discussed qualitatively on the bound on the reheating temperature by looking at the non-perturbative limit on inflation decay constant $\Gamma$ which is assumed to be originated from the Yukawa interaction between the scalar fields and a fermion field. As we have seen that the upper limit on the Yukawa coupling below which the perturbative analysis can be carried put a maximum bound on the reheating temperature for the models under consideration. The maximum reheating temperature has been found to be $T_{\rm re}\sim 10^2{\rm GeV}$ with  $n_s^{\rm max}\sim 0.966$ for $\sphi=10{\rm M_p}$ and $T_{\rm re}=10^6{\rm GeV}$ with $n_s^{\rm max}=0.972$ for $\sphi=0.1{\rm M_p}$. For higher values of coupling Yukawa coupling $h$ when perturbative analysis can not be trusted, we can obtain the qualitative result considering the effective equation of state $w_{\rm eff}$ that can qualitatively fit our results.   

However important point we should understand that there is no unified description of both the perturbative and non-perturbative reheating process. In our present analysis we have considered those as two separate phenomena connected by the initial condition for the Boltzmann equation after the end of preheating.
Therefore, it would be interesting to understand those in a single framework. Another important issue we have not discussed is related to incomplete decay of inflaton for $n> 2$ models where final equation of state becomes $1/3$ after preheating. For these cases simple Boltzmann description of perturbative reheating becomes untenable for $(T_{re}, N_{re})$. Therefore for such situation how and when the actual radiation domination starts will be an important question to discuss.

%\begin{figure}[h!]
%	\centering
%	\includegraphics[scale=0.3]{slice1.jpg}\\
%\includegraphics[scale=0.12]{slice2.jpg}
%\caption{\scriptsize . }
%\label{slice}
%\end{figure}
\section{Acknowledgement}
We thank our HEP and Gravity group members for their valuable comments and discussions. PS thank Gary Felder for useful discussions and help during various steps of working with \texttt{LATTICEEASY}. We thank Javier Rubio for useful comment on the draft. We thank the referee of critical comments which helped us to improve the quality of the work. We acknowledge the help of computing facilities at Department of Physics, IIT Guwahati.

\providecommand{\href}[2]{#2}\begingroup\raggedright\endgroup

\end{document}